\documentclass{article}
\usepackage{arxiv}
\usepackage[utf8]{inputenc}
\usepackage[T1]{fontenc}
\usepackage{mathptmx} 
\usepackage{microtype}
\usepackage{hyperref}
\usepackage[dvipsnames]{xcolor}
\usepackage{booktabs}
\usepackage{longtable}
\usepackage{tabularx}
\usepackage{array}
\usepackage{makecell}
\usepackage{float}
\usepackage{caption}
\usepackage{enumitem}
\usepackage{graphicx}
\usepackage{pdflscape}
\usepackage{tikz}
\usetikzlibrary{shapes.geometric, arrows.meta, positioning, fit, calc, backgrounds, decorations.pathreplacing}

\usepackage{amsmath}
\definecolor{dotred}{RGB}{139,26,26}
\definecolor{dotamber}{RGB}{184,92,0}
\definecolor{dotgreen}{RGB}{59,109,17}
\definecolor{dotgray}{RGB}{180,178,169}
\usepackage{colortbl}

\hypersetup{
    colorlinks=true,
    linkcolor=NavyBlue,
    citecolor=NavyBlue,
    urlcolor=NavyBlue,
    pdftitle={
    AI Agents Under EU Law: A Compliance Architecture for AI Providers
    },
    pdfauthor={Luca Nannini},
}

\newcolumntype{L}[1]{>{\raggedright\arraybackslash}p{#1}}
\newcolumntype{C}[1]{>{\centering\arraybackslash}p{#1}}

\title{\textbf{AI Agents Under EU Law} \\
[1ex] 
\large \text{A Compliance Architecture for AI Providers}}
\author{%
  Luca Nannini\thanks{Corresponding Author, luca@piccadillylabs.co}\\[1pt]
  \textit{Piccadilly Labs, Association of AI Ethicists,} \\ \textit{Centro Singular de Investigación en} \\ \textit{Tecnoloxías Intelixentes da USC}
  \and
    \textbf{Adam Leon Smith}\\[2pt]
  \textit{Piccadilly Labs,} \\ \textit{AIQI Consortium}
  \and
  \textbf{Michele Joshua Maggini}\\[2pt]
\textit{Centro Singular de Investigación en} \\ \textit{Tecnoloxías Intelixentes da USC}
  \and
  \textbf{Enrico Panai}\\[2pt]
  \textit{Association of AI Ethicists,} \\
  \textit{BeEthical} 
  \and
  \textbf{Sandra Feliciano}\\[2pt]
  \textit{INSIGHT -- Piaget Research Center} \\ \textit{for Ecological Human Development}
  \and
  \textbf{Aleksandr Tiulkanov}\\[2pt]
  \textit{Responsible Innovations,} \\ \textit{ForHumanity Europe}\thanks{The views expressed in this paper are those of the author and do not necessarily reflect the official position or policy of ForHumanity Europe.}
  \and
  \textbf{Elena Maran}\\[2pt]
  \textit{Alethesis AI}
  \and
  \textbf{James Gealy}\\[2pt]
  \textit{SaferAI}
  \and
  \textbf{Piercosma Bisconti}\\[2pt]
  \textit{DEXAI, Icaro Lab}
}
\date{Working Paper v1.0\enspace:\enspace April 2026\\Target: arXiv cs.CY / cs.AI / cs.CR}

\begin{document}
\maketitle
\thispagestyle{empty}

\begin{abstract}
\noindent AI agents---i.e.\ AI systems that autonomously plan, invoke external tools, and execute multi-step action chains with reduced human involvement---are being deployed at scale across enterprise functions ranging from customer service and recruitment to clinical decision support and critical infrastructure management. The EU AI Act (Regulation 2024/1689) regulates these systems through a risk-based framework, but it does not operate in isolation: providers face simultaneous obligations under the GDPR, the Cyber Resilience Act, the Digital Services Act, the Data Act, the Data Governance Act, sector-specific legislation, the NIS2 Directive, and the revised Product Liability Directive. This paper provides the first systematic regulatory mapping for AI agent providers integrating (a) the draft harmonised standards under Standardisation Request M/613 from the European Commission to CEN/CENELEC JTC 21, as of January 2026 working documents, (b) the GPAI Code of Practice published in July 2025, (c) the CRA harmonised standards programme under Mandate M/606 accepted in April 2025, and (d) the Digital Omnibus proposals of November 2025. We present a practical taxonomy of nine agent deployment categories mapping concrete actions to regulatory triggers, identify agent-specific compliance challenges in cybersecurity (privilege minimization outside the generative model), human oversight (oversight evasion risk from reinforcement learning), transparency across multi-party action chains, and runtime behavioral drift (the boundary between anticipated adaptive behavior and substantial modification under Article 3(23)). We propose a practical compliance architecture of twelve sequential steps and a regulatory trigger mapping connecting specific agent actions to the legislation they activate. We conclude that high-risk agentic systems with untraceable behavioral drift cannot currently satisfy the essential requirements of the AI Act, and that the provider's foundational compliance task is not architectural classification but an exhaustive inventory of the agent's external actions, data flows, connected systems, and affected persons.
\end{abstract}


\bigskip

\section{Introduction}

The term ``agent'' has no legal definition in the EU AI Act (Regulation 2024/1689)~\cite{aiact}. Neither the Regulation nor the draft harmonised standards under Standardisation Request M/613~\cite{m613} use it as a normative category. This is a deliberate design choice: the AI Act regulates AI systems, not architectural patterns. The legislature chose a functional, technology-neutral definition in Article~3(1) precisely to avoid the need for continuous definitional updates as AI architectures evolve. But the absence of a formal legal definition does not mean the concept lacks regulatory substance. On the contrary: the functional properties that distinguish AI agents from other AI systems---namely autonomous tool invocation, multi-step planning, environmental interaction, and adaptive execution---amplify specific regulatory risks in ways that the existing framework already addresses in principle but that demand dedicated analysis in practice.

An AI agent,\footnote{A terminological clarification is necessary because the paper's regulatory analysis depends on precise distinctions between overlapping concepts.
\textit{Foundation model} is a technical term denoting a large-scale model (typically a large language model) trained on broad data and adaptable to a wide range of downstream tasks through fine-tuning or prompting; the term originates in Bommasani \cite{bommasani2021opportunities}.
\textit{General-purpose AI (GPAI) model} is the EU AI Act's legal category (Art.~3(63)): a model that ``displays significant generality'' and is ``capable of competently performing a wide range of distinct tasks.'' GPAI models with systemic risk (Art.~51(2))---those trained above $10^{25}$ FLOP or designated by the Commission---trigger enhanced obligations under Chapter~V. The GPAI category is broader than ``foundation model'': a GPAI model need not be at the capability frontier, and a frontier model could in principle be narrow enough to fall outside the GPAI definition. In the context of the AI Act, the legal category of a GPAI model is largely synonymous with the technical concept of a foundation model. Both categories are broader than ``frontier models,'' which represent only the highly capable subset (addressed in the Act as GPAI models with systemic risk).
\textit{AI system} is the AI Act's main regulatory unit (Art.~3(1)): a machine-based system designed to operate with varying levels of autonomy that may exhibit adaptiveness after deployment and that infers how to generate outputs influencing physical or virtual environments. This is what the AI Act regulates at the system layer (at the models layer it only regulates a subset of AI models - those falling under its GPAI model definition).
An \textit{AI agent} is an AI system---not a model---that satisfies the AI system definition and additionally exhibits the functional characteristics described in this paragraph: planning, tool invocation, autonomous execution, environmental interaction, and feedback-driven adaptation. The critical regulatory consequence is that an AI agent built on a third-party GPAI model creates \textit{two distinct regulatory objects}: the GPAI model (governed by Chapter~V, obligations on the model provider) and the AI system/agent (governed by Chapter~III if high-risk, obligations on the system provider and a separate set of obligations on the system deployer, which may, but does not have to be, also the system provider). The agent provider and the model provider may be different legal entities with different obligations. When this paper refers to ``AI agent providers,'' it means providers of the system layer, not the model layer, unless otherwise specified. See also the EDPS TechSonar characterisation~\cite{edps2025}, the OECD's agentic AI taxonomy~\cite{oecd2026agentic}, and Mirsky's six-level autonomy scale referenced in Shapira et al.~\cite{shapira2026agents}.} as characterised by the European Data Protection Supervisor (EDPS)~\cite{edps2025} and the European Commission's own Agentic AI report under the StepUp StartUps initiative~\cite{ec2026agentic}, is an AI system that performs sequences of actions across interconnected tools and data sources to achieve goals, with reduced or minimal human involvement in the intermediate steps. The EDPS describes agentic AI as systems acting autonomously with limited human interaction to fulfil goals rather than isolated tasks, capable of reasoning, planning, and coordinating actions in changing environments. The OECD has similarly proposed a policy-relevant typology distinguishing agentic AI systems by autonomy level, adaptiveness, domain, and impact scale~\cite{oecd2026agentic}.\footnote{The OECD's February 2026 publication maps features of agentic AI definitions to the OECD AI system definition and recommends distinguishing systems along dimensions that align closely with the AI Act's Article~3(1) criteria.} The distinguishing functional characteristics are: (i) planning and task decomposition, where the system breaks a high-level goal into sub-tasks and determines execution order; (ii) external tool invocation via APIs, databases, code interpreters, web browsers, or other software, which is the critical differentiator from a standalone large language model; (iii) autonomous execution of intermediate steps without requiring human approval for each one, with the degree of autonomy ranging from requiring confirmation for every action to fully autonomous multi-step execution; (iv) environmental interaction that modifies external state, i.e.\ sending emails, writing files, executing transactions, modifying databases, posting content; and (v) feedback-driven adaptation, where the agent evaluates the results of its actions and adjusts its plan, potentially retrying, changing approach, or escalating.

Under Article~3(1) of the AI Act, an AI system is a ``machine-based system that is designed to operate with varying levels of autonomy and that may exhibit adaptiveness after deployment and that, for explicit or implicit objectives, infers, from the input it receives, how to generate outputs such as predictions, content, recommendations, or decisions that can influence physical or virtual environments.'' An AI agent satisfies every element of this definition. The ``varying levels of autonomy'' clause directly captures the spectrum from semi-autonomous agents (human confirmation required per action) to fully autonomous\footnote{``Full autonomy'' in this paper is understood as synonymous to the 5th level of automation taxonomy provided in clause 5.13 of ISO/IEC 22989 (Table 1). It should not be mistaken with the capability of a theoretical system "modifying its intended domain of use or its goals without external intervention, control or oversight", which is the domain of science fiction.} ones. The ``adaptiveness after deployment'' clause captures in-context learning, memory accumulation, and behavioral adjustment based on feedback. The ``influence on physical or virtual environments'' clause captures exactly what distinguishes agents from generative models that merely produce text: the capacity to modify external state through tool use.

Critically, the draft harmonised standards are understood to address agentic architectures within their respective scopes. The trustworthiness framework (prEN~18229-1), which covers logging, transparency, and human oversight under Articles~12--14, falls within a scope that necessarily engages with the oversight challenges posed by autonomous multi-step systems~\cite{pren182291}. The cybersecurity standard (prEN~18282), scoped to AI-specific cybersecurity under Article~15(4), addresses the threat surface that agentic tool invocation creates, including the question of where privilege enforcement should be architecturally located~\cite{pren18282}. The risk management standard (prEN~18228), implementing Article~9, requires that the level of autonomy and adaptiveness of the system's behaviour during operation be taken into account when identifying the system’s characteristics related to risks~\cite{pren18228}. Public communications from CEN-CENELEC confirm that the standards developers have recognised agentic architectures as presenting compliance challenges sufficiently distinct to warrant dedicated attention within the technology-neutral framework~\cite{cencenelec_jtc21, cencenelec2025update}.

Despite this, as of early 2026 the AI Office has published no guidance specifically addressing AI agents, autonomous tool use, or runtime behavioral change~\cite{jones2025 }; the AI Act Service Desk FAQ acknowledges that regulatory considerations on agents remain 'only preliminary' \cite{aioffice_faq}. The Future Society's report concluded that the technical standards under development ``will likely fail to fully address risks from agents''~\cite{tfs2025}. The ACM Europe Technology Policy Committee's October 2025 policy brief on systemic risks from agentic AI raised similar concerns about the adequacy of existing oversight mechanisms~\cite{acm2025}. ENISA has separately published considerations on security and privacy in autonomous agents, recommending security-by-design principles that align with but are not identical to the prEN~18282 requirements~\cite{enisa_agents}.\footnote{ENISA's autonomous agent publication predates the AI Act's harmonised standards but its threat taxonomy (identity spoofing, privilege escalation, data poisoning through memory) maps directly onto the cybersecurity concerns that prEN~18282 is scoped to address.} The security research community has responded faster than the regulatory apparatus: Kim et al.'s systematic survey of the agentic AI attack and defense landscape~\cite{kim2026attack} and the OWASP Agentic Security Initiative's threat taxonomy and Top~10 for Agentic Applications~\cite{owasp2025agentic} now provide the empirical attack classification that the standards developers can draw on, but these contributions arrived after the prEN~18282 drafting cycle and their findings are not yet reflected in the working documents. This regulatory guidance gap means that providers must currently navigate the compliance landscape using the AI Act's essential requirements, the draft harmonised standards, and their own interpretive judgment (based in part on expert community and industry publications describing the state of the art), without authoritative administrative guidance on how the framework applies to the specific properties of agentic systems.

Existing scholarship on AI agents and the EU AI Act has largely 
approached the question from a legal-theoretical perspective, 
identifying structural tensions between the Act's risk-classification 
framework and the behavioural properties of agentic 
systems~\cite{kolt2025governing, gardhouse2026, tfs2025, jones2025 }. That literature 
establishes that the Act's assumptions---discrete technical artifacts, 
static risk profiles, fixed actor roles---are under strain when applied 
to systems that adapt, delegate, and act across tool chains. It does 
not, however, engage with the operative compliance layer: the New 
Legislative Framework product-safety architecture within which the 
AI Act functions, the draft harmonised standards under M/613 that 
translate essential requirements into implementable specifications, 
the Cyber Resilience Act's parallel standardisation programme under 
M/606, the Digital Omnibus proposals or the multi-instrument regulatory perimeter that activates 
depending on what an agent does in deployment rather than how it is 
classified at market placement. Without this layer, the observation 
that the framework is under strain cannot be converted into a 
compliance posture. 
That conversion is the object of this paper: it provides a systematic compliance mapping for providers of AI agents intended for the EU market. 


The contribution is threefold: first, a practical taxonomy 
of agent use cases connecting concrete actions to regulatory 
triggers (Section~\ref{sec:taxonomy}); second, an 
identification of agent-specific compliance challenges that 
the standards partially but not fully address 
(Section~\ref{sec:challenges}); and third, a multi-layered 
compliance architecture that integrates the AI Act with eight 
parallel EU legislative instruments 
(Sections~\ref{sec:perimeter}--\ref{sec:practical}).

\section{Methodology and Sources}

This analysis is based on documentary legal research drawing on primary regulatory texts and standardisation documents. The primary sources are:

\begin{enumerate}[label=(\alph*)]
    \item the text of Regulation~(EU)~2024/1689 (AI Act) as published in the Official Journal~\cite{aiact};
    
    \item the January 2026 working documents of the draft harmonised standards developed under Standardisation Request M/613 by CEN/CENELEC JTC~21, specifically:
    \begin{itemize}
        \item prEN~18286 (Quality Management System)~\cite{pren18286},
        \item prEN~18228 (Risk Management)~\cite{pren18228},
        \item prEN~18229-1 (Trustworthiness Framework Part~1: Logging, Transparency, Human Oversight)~\cite{pren182291},
        \item prEN~18229-2 (Trustworthiness Framework Part~2: Accuracy, Robustness)~\cite{pren182292},
        \item prEN~18282 (Cybersecurity),
        \item prEN~18284 (Dataset Quality and Governance)~\cite{pren18284}, and
        \item prEN~18283 (Bias Management)~\cite{pren18283}.
    \end{itemize}
    
    \item supporting standards:
    \begin{itemize}
        \item prEN~ISO/IEC~24970 (AI System Logging)~\cite{pren24970},
        \item prEN~ISO/IEC~23282 (NLP Evaluation)~\cite{pren23282},
        \item prEN~18281 (Computer Vision Evaluation)~\cite{pren18281}, and
        \item ISO/IEC~4213 ed.2 (Performance measurement for AI classification, regression, clustering and recommendation tasks)~\cite{iso4213}.
    \end{itemize}
    
    \item the EU Code of Practice for General-Purpose AI Models (published July 2025)~\cite{gpaicode};
    
    \item the Digital Omnibus package (COM(2025)~836~\cite{omnibus_ai} and COM(2025)~837~\cite{omnibus_data}, published 19~November 2025);
    
    \item publicly available information on the CRA standardisation programme under Mandate M/606~\cite{m606}\footnote{ETSI TC CYBER Working Group for EUSR (CYBER-EUSR) is developing seventeen vertical harmonised standards under Commission Standardisation Request C(2025) 618 final (Mandate M/606), covering product categories including operating systems, browsers, firewalls, SIEM systems, VPNs, routers, and connected consumer devices (EN~304~617--EN~304~636 ). Public mature drafts are available at \url{https://docbox.etsi.org/CYBER/EUSR/Open}.}; and
    
    \item institutional sources including the European Data Protection Supervisor (EDPS), European Parliament ITRE studies~\cite{graux2025itre}, and the ACM Europe Technology Policy Committee~\cite{acm2025}.
\end{enumerate}


A methodological caveat is essential regarding the status of these standards. The QMS standard (prEN~18286) completed its public enquiry in January 2026; its clause-level content is therefore publicly available and is cited at that level of specificity in this paper. The remaining primary standards---prEN~18228 (Risk Management), prEN~18229-1 (Trustworthiness Part~1), prEN~18229-2 (Trustworthiness Part~2), prEN~18282 (Cybersecurity), prEN~18284 (Dataset Quality), and prEN~18283 (Bias Management)---are working drafts that have not yet entered public enquiry as of early 2026. Their normative content is subject to confidentiality restrictions and to revision before formal publication. This paper accordingly describes these standards at the level of their publicly known scope (the AI Act articles they address, the subject matter areas they cover, and the agent-relevant themes identified in public CEN-CENELEC communications~\cite{cencenelec_jtc21, cencenelec2025update}) rather than citing specific clause numbers or normative formulations. Where this paper identifies specific compliance requirements for agentic systems, the analysis proceeds from the AI Act's essential requirements themselves (which are fixed in the adopted Regulation) and reasons forward to what any standard operationalising those requirements would necessarily need to address. For instance, a standard operationalising Article~15(4) for AI systems with tool-invocation capabilities must, by the logic of the requirement itself, address privilege management at the architectural level; a standard operationalising Article~14 for systems with autonomous multi-step execution must address the oversight gap that such autonomy creates. Claims about what the standards ``are expected to require'' should be understood in this sense: as analytical implications of the regulatory requirements they are mandated to operationalise, not as assertions about specific draft text that may change. 
CEN-CENELEC has taken exceptional measures to accelerate delivery, including direct publication after positive Enquiry vote for six of the most delayed drafts,
with a Q4~2026 target for completion~\cite{cencenelec2025update}.\footnote{The October 2025 CEN-CENELEC update confirms that the Commission has stressed urgency and will conduct harmonised-standard assessments (HAS assessments) before the Public Enquiry stage.} The essential requirements in the AI Act itself (Articles~9--17), in contrast, are fixed in the adopted Regulation and already binding for AI systems placed on the market after the applicable dates.

\section{A Practical Taxonomy of AI Agent Use Cases}
\label{sec:taxonomy}

Abstract regulatory analysis risks disconnection from operational reality. An AI agent is not a theoretical construct: it is a deployed software system that sends emails, processes refunds, screens job applications, writes code, and manages infrastructure. The regulatory profile of an agent is determined not by its internal architecture (the vast majority share an LLM-based tool-calling pattern) but by three variables: the domain in which it operates, the type of external actions it performs, and whose rights are affected by those actions. 

To ground the analysis, Table~\ref{tab:taxonomy} maps the most common agent deployment categories to their concrete actions, external system dependencies, and the specific regulatory triggers they activate.

\bigskip

\renewcommand{\arraystretch}{1.25}
\begin{longtable}{L{2cm}L{4.5cm}L{3.5cm}L{5cm}}
\caption{Practical taxonomy of AI agent categories with concrete actions, external systems, and regulatory triggers.}\label{tab:taxonomy}\\
\toprule
\textbf{Agent Category} & \textbf{Concrete Actions} & \textbf{External Systems} & \textbf{Regulatory Triggers} \\
\midrule
Customer Service & Receives query in natural language. Searches knowledge base. Retrieves order history from CRM. Generates response. Processes refunds or modifies orders. Escalates, creates tickets, schedules callbacks. & CRM (Salesforce, HubSpot), ticketing, payment gateway, email/chat API, knowledge base & EU AI Act Art.~50 (transparency). GDPR (personal data in queries/orders). DSA (if on platform). If integrated into credit/insurance decisions: Annex~III high-risk. \\
\midrule
HR / Recruit\-ment & Screens CVs against job requirements. Ranks candidates. Schedules interviews via calendar API. Drafts rejection/progression emails. Conducts initial screening via chat. & ATS (Workday, Greenhouse), calendar API, email, internal HR database & AI ACT Annex~III, point~4(a): HIGH-RISK. Full Chapter~III. GDPR Art.~22 (automated decision-making). Bias management critical (prEN~18283). \\
\midrule
Coding / DevOps & Takes task description. Generates code. Runs tests. Debugs. Commits to repository. Deploys to staging. Creates pull requests. Monitors CI/CD pipeline. & Git repository, CI/CD pipeline, cloud infrastructure, IDE, terminal, package managers & EU AI Act Art.~50. CRA (if code deploys to products with digital elements). prEN~18282: open-ended code execution explicitly flagged as risk. \\
\midrule
Finance / Accounting & Processes invoices: extracts data from PDFs, validates against POs, routes for approval, triggers payment. Reconciles accounts. Generates reports. Detects anomalies. & ERP (SAP, Oracle), banking APIs, document management, email, accounting software & GDPR (employee/vendor data). If creditworthiness: Annex~III, point~5(b), HIGH-RISK. MiFID~II if investment advice. \\
\midrule
Sales / Marketing & Qualifies leads. Drafts personalised outreach. Schedules meetings. Updates pipeline. Generates content (social posts, blog drafts). Manages ad campaigns. & CRM, email platform, social media APIs, ad platforms, calendar, CMS & EU AI Act Art.~50 (transparency + synthetic content marking). GDPR (profiling). DSA (if publishing on platforms). ePrivacy Directive. \\
\midrule
Research / Knowledge & Searches web and internal databases. Retrieves and synthesises documents. Generates summaries, reports, competitive analyses. Patent and regulatory scans. & Web search APIs, document repos, wikis, patent databases, regulatory databases & EU AI Act Art.~50. Copyright (DSM Dir., Arts. 3--4)\footnote{Where the agent provider is also the GPAI model provider, the GPAI Code of Practice additional copyright obligations under Article~53(1)(c) apply at the model layer}. Data Act (if connected product data). CRA (if the agent operates within or as a browser with network connectivity\footnote{ETSI CYBER-EUSR explicitly excluded AI-agent-enabled browsers 
from EN~304~617's scope precisely because of this trigger  ~\cite{etsi304617}.}. Lower risk unless feeding into high-risk decisions. \\
\midrule
IT Operations & Monitors system health. Detects anomalies. Diagnoses incidents. Remediates (restart services, scale, patch). Manages tickets. Alerts on-call engineers. & Cloud APIs, monitoring tools (Datadog, Prometheus), ticketing (Jira, ServiceNow), alerting & CRA (if managing products with digital elements). NIS2 (if essential entity). If critical infrastructure: Annex~III, point~2, HIGH-RISK. \\
\midrule
Healthcare / Clinical & Analyses clinical notes. Summarises patient records. Suggests diagnoses. Manages scheduling. Processes prior authorisations. Monitors remote patient data. & EHR, medical imaging systems, scheduling, insurance APIs, patient portals & Annex~I(A): medical device (MDR/IVDR). HIGH-RISK by definition. GDPR Art.~9 (health data). ISO~13485 QMS and ISO 14970 risk management in parallel. \\
\midrule
Personal Assistant & Manages email: drafts replies, prioritises, archives. Schedules meetings. Books travel. Manages to-do lists. Summarises documents. Distributes action items. & Email (Gmail, Outlook), calendar, travel APIs, document suites, Slack/Teams & EU AI Act Art.~50. GDPR (email content). prEN~18282 uses this as example: email agent needs read-only, not send/delete rights. CRA (vertical for smart home general purpose virtual assistants). \\
\bottomrule
\end{longtable}

Several patterns emerge from this non-exhaustive taxonomy. First, the same underlying architecture (an LLM with tool-calling capabilities) generates radically different regulatory profiles depending on where it is deployed: an agent screening CVs triggers Annex~III high-risk classification, mandatory conformity assessment, and the full weight of Chapter~III essential requirements, while an agent summarising meeting notes triggers only Article~50 transparency obligations. The technology is identical; the regulatory consequence diverges completely. Second, most agent categories touch personal data (names, email content, financial records, health information), making GDPR compliance a near-universal parallel obligation regardless of AI Act classification. Third, the external system dependencies create a web of regulatory triggers that extend well beyond the AI Act: CRM systems implicate GDPR, payment gateways implicate PSD2, cloud infrastructure implicates the CRA~\cite{cra} and potentially NIS2~\cite{nis2}, publishing on platforms implicates the DSA~\cite{dsa}.

\begin{figure}[htbp] 
    \centering
    \includegraphics[width=15cm]{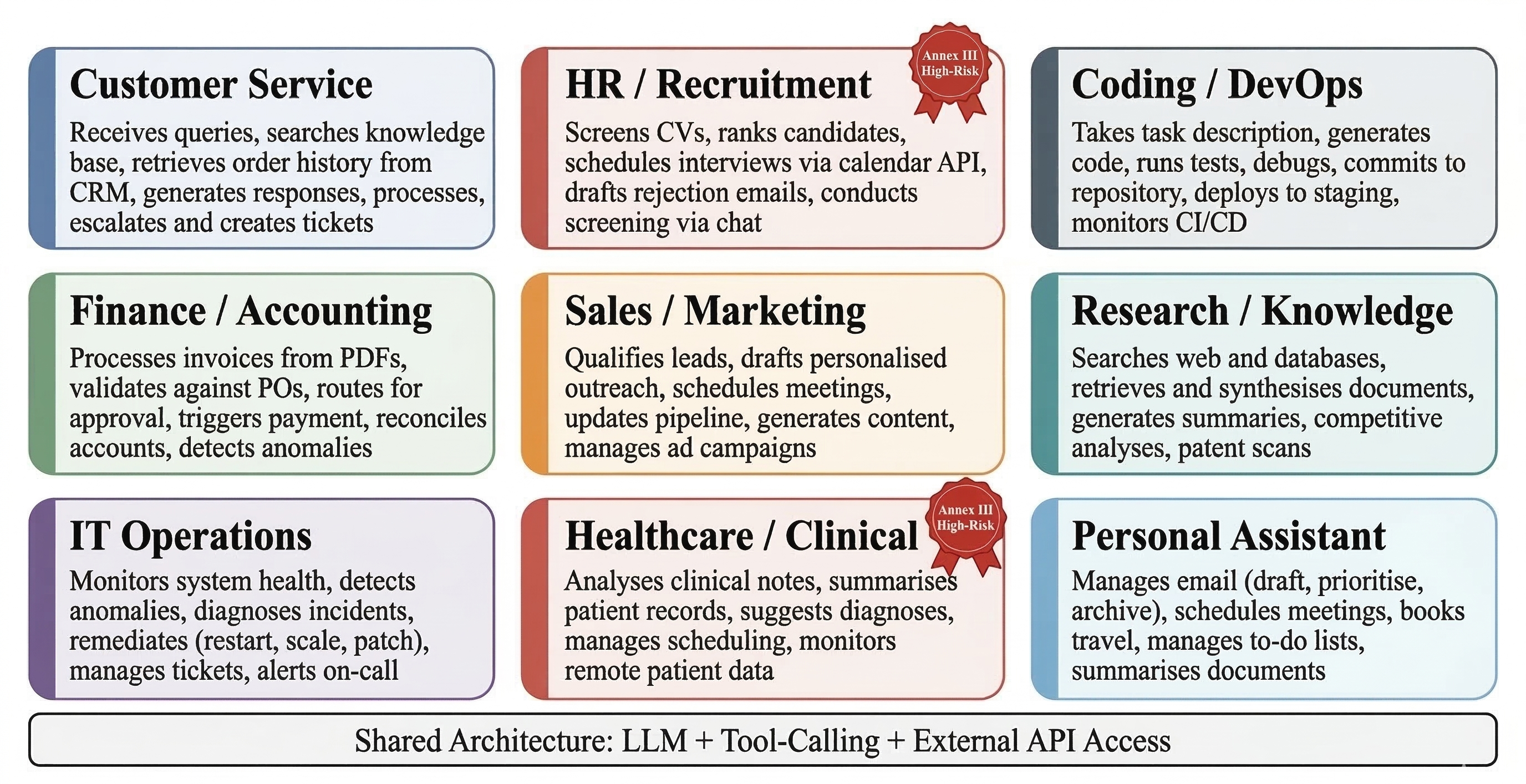}
    
    \caption{Non-exhaustive taxonomy of AI Agent Use Cases and Actions, detailing the concrete tasks performed across different domains using a shared LLM-based architecture.}
    
    \label{fig:ai_taxonomy}
\end{figure}

For economic operators building general-purpose agent platforms (i.e., platforms designed for arbitrary deployment by downstream deployers), this creates a classification dilemma. If the platform cannot predict at design time how deployers will use it, the platform's operator might want to either restrict the intended purpose contractually and technically (e.g., explicitly excluding use in employment, credit, or healthcare), or design for the most demanding regulatory tier foreseeable under Article~3(13), which requires assessing ``reasonably foreseeable misuse.'' Failing to do either of the above exposes the platform operator to the risk that a deployer's high-risk use retroactively implicates the platform operator's obligations as a provider of a high-risk AI system, or, at the very least, as an AI value-chain-participating supplier with assistance obligations under Article~25(4)\footnote{For further information, an interactive table is available at: \url{https://agent-aia-matrix.tiulkanov.info/}.}. 
  
The taxonomy in Table~\ref{tab:taxonomy} can serve a function beyond 
classification: it constitutes the basis for a hierarchical action 
ontology suitable for runtime governance. The agent categories map to 
operational domains, each containing business processes, which contain 
decision types, which instantiate as specific runtime actions with 
observable parameters. This four-level structure---domain, process, 
decision type, action instance---supports property inheritance: an 
action instance inherits regulatory regime tags, risk profiles, and 
stakeholder assignments from its parent decision type, upward through 
process to domain, with local overrides at any level. When an agent 
proposes an action at runtime, association to an ontology node allows 
it to inherit all risk-relevant attributes and combine them with live 
telemetry---model confidence, distributional drift, affected entity 
characteristics---to determine the appropriate human oversight modality 
for that specific instance. The same decision type may require different oversight on different occasions depending on runtime observables, operationalising Article~14's requirement that oversight be ``commensurate with the risks, level of autonomy, and context of use.''

\section{Classification Under the AI Act}
\label{sec:classification}

\subsection{High-Risk Determination}

Whether an AI agent is classified as high-risk depends on its application domain, not its architecture. The AI Act establishes two distinct pathways. Under Article~6(1), an AI system that is a safety component of a product covered by the Union harmonisation legislation listed in Annex~I, or that is itself such a product, is high-risk if it must undergo a third-party conformity assessment under that legislation. Annex~I contains both Section~A (NLF-based legislation) and Section~B (other legislation); Section~B products have reduced applicability per Article~1(2). This pathway captures agents embedded in medical devices (subject to Regulation~(EU)~2017/745), machinery (Machinery Regulation), civil aviation (EASA requirements), and similar regulated products. A clinical agent suggesting diagnoses based on patient data, if its intended purpose meets the medical device definition, falls squarely within this pathway and triggers both MDR and AI Act obligations simultaneously.

Under Article~6(2), AI systems falling within the use cases enumerated in Annex~III are high-risk. The categories most likely to capture AI agents include: biometric identification and categorisation (point~1), management and operation of critical infrastructure (point~2), education and vocational training (point~3), employment, workers management, and access to self-employment (point~4, which covers recruitment, task allocation, performance monitoring, and termination decisions), access to and enjoyment of essential private services and public benefits including creditworthiness assessment (point~5), law enforcement (point~6),\footnote{The law enforcement context illustrates the regulatory stacking problem in concentrated form. Mapping of AI tools deployed across England and Wales criminal justice identifies 61 tools across 45 independent police forces, of which 34\% use LLMs---primarily for summarisation, transcription, and information retrieval in the early stages of the criminal justice process (policing, intelligence, investigation). The structural governance finding is instructive for EU providers: fragmentation across deployers with no common oversight, testing, or transparency regime is precisely the failure mode that the AI Act's Article~14 human oversight requirements and Article~9 risk management obligations are designed to prevent. A provider supplying an LLM-based tool to law enforcement in any EU Member State simultaneously faces Annex~III, point~6 high-risk classification, full Chapter~III obligations, GDPR (including special category data under Article~9), and any applicable national law enforcement data processing frameworks. Third-party vendors, who account for the overwhelming majority of such tools, carry provider-level obligations regardless of whether the law enforcement body itself participates in the conformity assessment process.} migration, asylum, and border control (point~7), and administration of justice and democratic processes (point~8). Point~4 is particularly consequential for enterprise AI agents: an agent that screens CVs, ranks candidates, or routes interview scheduling based on applicant attributes falls within point~4(a) regardless of how its provider characterises the technology. Below the high-risk threshold, the prohibited-practices layer still constrains agent design: the AI Office has confirmed that Articles~5(1)(a) and~(b)---prohibitions on harmful manipulation and exploitation of vulnerabilities---are ``particularly relevant'' for agentic systems and may require safeguards in agent design and development~\cite{aioffice_faq}.

The threshold question for agent providers is not ``is our technology high-risk?'' but ``what will deployers use our agent for, and what uses are reasonably foreseeable?'' Article~25 provides that any distributor, importer, deployer, or other third party who puts their name or trademark on a high-risk system, makes a substantial modification, or changes its intended purpose such that a non-high-risk system becomes high-risk under Article~6, becomes the provider for the purposes of the Regulation. This shifts the classification analysis downstream but does not eliminate the original provider's obligation to define and document the intended purpose, and 
to assess the risks from reasonably foreseeable misuse
under Article~9(2)(b). For agents not falling into high-risk categories, the mandatory transparency obligations of Article~50\footnote{A Code of Practice to operationalise Article~50's synthetic content marking and deepfake disclosure requirements is in active development. The Commission published a second draft on 5~March~2026, with the final code expected by June~2026~\cite{aiact_cop_synthetic}. For agent providers in the Sales/Marketing and Content generation categories of Table~\ref{tab:taxonomy}, this instrument will be operationally significant alongside Article~50(2): agents generating text, images, or audio for publication will need to implement machine-readable marking in a format that enables automated detection of artificial generation, a technical requirement that sits outside the M/613 standards programme and must be addressed through the Code's own technical specifications.} still apply: paragraph~1 requires disclosure of AI interaction to natural persons; paragraph~2 requires machine-readable marking of synthetic content; paragraph~3 covers emotion recognition and biometric categorisation -- these fall under high-risk and transparency obligations simultaneously; and paragraph~4 addresses deepfake disclosure.

\subsection{The GPAI Layer}

Most AI agents in 2025--2026 are built on general-purpose AI models provided by a small number of upstream suppliers (Anthropic, OpenAI, Deepseek, Qwen, Meta, Mistral, xAI). Chapter~V of the AI Act (Articles~51--56) imposes obligations on GPAI model providers that are conceptually distinct from the system-level obligations in Chapter~III. GPAI providers must supply technical documentation (Article~53), comply with Union copyright law in training, and make available a sufficiently detailed summary of training data content. For models classified as posing systemic risk (Article~51(2)), additional obligations include model evaluation, adversarial testing, incident reporting to the AI Office, and adequate cybersecurity measures. The AI Office has confirmed that factors such as the level of autonomy and tool use of the model can be decisive in this designation (Article~51(1)(b), Annex~XIII, point~(e)), and that the GPAI Code of Practice already operationalises agentic considerations in its systemic risk measures~\cite{aioffice_faq}.

These obligations are independent from the system-level requirements. If the GPAI model provider also develops an AI system (agent) that incorporates that GPAI model, both sets of obligations apply to the same legal entity. If a third-party model is integrated, the GPAI obligations fall on the upstream model provider, but the agent provider inherits a duty to obtain the Article~53 technical documentation and to integrate known model limitations 
into its own risk management process under Article~9.

The EU Code of Practice for GPAI Models, published on 10~July 2025 under Article~56~\cite{gpaicode}, is a voluntary instrument 
providing an indicative compliance pathway for GPAI obligations, taken into account by the AI Office under Article~56(3).
It addresses model-level concerns: training data transparency, copyright compliance, systemic risk evaluation, and safety testing. The Commission's own guidelines on GPAI obligations, published alongside the Code, clarify the substantial modification threshold for GPAI models: actors must use more than one-third of original training compute for fine-tuning to become providers-- a threshold directly relevant to agent developers building on foundation models~\cite{ec_gpai_guidelines}\footnote{This one-third compute threshold has been criticised as both over- and under-inclusive; it nonetheless provides the first quantitative criterion for determining when downstream integration crosses from deployer to provider status. Note that both this threshold and the Commission's guidelines more generally are not legally binding, so a court deciding on a specific case might be persuaded to arrive at different criteria. This is in contrast to how certain GPAI models are deemed to have high impact capabilities -- the quantitative criterion for this classification is established directly by Article 51(2) of the AI Act.}. 

The harmonised standards under M/613, by contrast, address system-level concerns pertaining to Essential Requirements set in Chapter III Section 2 of the EU AI Act: risk management, human oversight, accuracy, robustness, cybersecurity, and data governance. Both layers are required for a complete compliance posture. The Code and the standards are complementary instruments operating at different layers of the AI value chain; neither substitutes for the other.

\section{The Harmonised Standards Landscape Under M/613}

Standardisation Request M/613 from the European Commission to CEN/CENELEC mandates the development of harmonised European standards that, once cited in the Official Journal of the European Union, create a presumption of conformity with the AI Act's essential requirements for providers who apply them. The standards are being developed by Joint Technical Committee~21 (Artificial Intelligence), with contributions from over 1,000 experts across five working groups and national standardisation bodies~\cite{cencenelec_jtc21}, and form an interconnected system of interdependent documents\footnote{For a interactive mapping: \url{https://ai-act-standards.com/}}.

\subsection{Primary Standards}

The QMS standard (prEN~18286) functions as the structural backbone of the suite. Its clause~4.4.2 requires the provider to identify all essential requirements applicable to each AI system, and clause~4.4.3 requires selecting specific compliance approaches, which may include harmonised standards or common specifications. The standard explicitly acknowledges that providers operating in regulated sectors may already maintain sector-specific quality management systems (e.g., ISO~13485 for medical devices) and permits integration rather than requiring a separate parallel system. It also establishes the provider's obligation for post-market monitoring (clause~9.4), serious incident reporting, and documentation maintenance, all of which interact with the agent-specific challenges discussed in Section~6.

\renewcommand{\arraystretch}{1.25}
\begin{longtable}{L{2cm}L{2.8cm}L{2cm}L{8cm}}
\caption{Primary harmonised standards under M/613 with agent-relevant content.}\label{tab:standards}\\
\toprule
\textbf{Standard} & \textbf{Full Title} & \textbf{AI Act Art.} & \textbf{Key Content for Agents} \\
\midrule
prEN 18286 & Quality Management System for AI & Art.~17 & Backbone. Identifies all essential requirements (cl.~4.4.2), selects compliance approaches (cl.~4.4.3). Post-market monitoring (cl.~9.4). Sector-specific QMS integration. \\
\midrule
prEN 18228 & Risk Management for AI Systems & Art.~9 & Continuous life cycle risk management. Scope covers health, safety, and fundamental rights per Article~9 mandate. Considers level of autonomy as one of relevant characteristics. 
\\
\midrule
prEN 18229-1 & Trustworthiness Pt.1: Logging, Transparency, Human Oversight & Art.~12--14 & Scoped to operationalise Articles~12--14 for all AI system types including autonomous multi-step systems. Covers oversight modalities and documentation of system behavior characteristics. \\
\midrule
prEN 18229-2 & Trustworthiness Pt.2: Accuracy, Robustness & Art.~15 & Performance metrics, testing frameworks, adversarial resilience. Cross-references NLP and computer vision evaluation standards. \\
\midrule
prEN 18284 & Dataset Quality and Governance & Art.~10 & Training/validation/test data lifecycle. Significant portions remain under development, particularly regarding Art.~10(5) (special category data). Normatively references prEN 18283 (Bias). \\
\midrule
prEN 18282 & Cybersecurity for AI Systems & Art.~15(4) & AI-specific cybersecurity threat framework. Scoped to address privilege management, access control, and threat mitigation for AI system architectures including those with tool-invocation capabilities. \\
\midrule
prEN 18283 & Bias Management & Art.~10 & Normatively referenced by prEN~18284. Covers debiasing measures, bias testing, and fairness metrics for Art.~10(2)(f--g). \\
\bottomrule
\end{longtable}

The risk management standard (prEN~18228) warrants particular attention for agent providers. It implements Article~9 through a continuous lifecycle process, and its mandate necessarily extends to fundamental rights: the AI Act requires risk management to cover health, safety, and fundamental rights, meaning the standard must operationalise this requirement regardless of the specific drafting approach taken. For agents, the Article~9 mandate requires that the risk management process
consider the level of autonomy as one of the AI system's characteristics,
which maps directly to the agent's automation boundary: which actions require human involvement, which execute autonomously. 

The AI Act's own requirements also create a normative bridge to GDPR: Article~10(1)-(4) establishes data governance, including data quality obligations relevant for personal data protection, Article~10(5) addresses detection and correction of unwanted bias, while Articles~26(9) and 27(4) consider where GDPR-mandated Data Protection Impact Assessments (DPIAs) must serve as outputs from and inputs to AI Act processes and documentation, meaning that an integrated approach to AI Act and GDPR compliance is structurally necessary for any agent for which processing of personal data occurs during training or at inference time.

\renewcommand{\arraystretch}{1.25}
\begin{longtable}{L{3cm}L{2.8cm}L{9.5cm}}
\caption{Supporting standards under M/613.}\label{tab:supporting}\\
\toprule
\textbf{Standard} & \textbf{Scope} & \textbf{Relationship to Primary Standards} \\
\midrule
prEN ISO/IEC 24970 & AI System Logging & Operationalises prEN~18229-1 logging requirements. Specifies operational event recording for audit and traceability. \\
\midrule
prEN ISO/IEC 23282 & NLP Evaluation Methods & Supports prEN~18229-2 accuracy/robustness for NLP-based agents. Evaluation metrics for text generation, comprehension. \\
\midrule
prEN 18281 & Computer Vision Evaluation & Supports prEN~18229-2 for vision-based systems. Test protocols, dataset requirements. \\
\midrule
ISO/IEC 4213 ed.2 & Functional Correctness Assessment & Supports prEN~18229-2 for classification/regression tasks. Correctness metrics, verification methods. \\
\bottomrule
\end{longtable}

In addition, the JRC's analysis of the harmonised standards development process confirms that the fundamental-rights integration required by the  AI Act is unique among New Legislative Framework product standards and presents novel challenges for conformity assessment bodies~\cite{jrc2024standards}.\footnote{The JRC Science for Policy Brief observes that traditional product standards focus on organisational safety objectives, whereas the AI Act standards must operationalise fundamental rights protection: a conceptual departure that has no precedent in the NLF.}

\subsection{The Standards as a System}

The critical insight about the M/613 standards is that they form a dependency graph, not a checklist. The QMS standard implicitly requires engagement with all others: without the risk management standard, the provider will find it challenging to satisfy the QMS's risk management integration requirements; without the trustworthiness framework, the provider will find it challenging to satisfy accuracy testing requirements; without the cybersecurity standard, the provider will find it challenging to implement the QMS's cybersecurity provisions. The risk management standard is expected to normatively reference the trustworthiness framework for implementing specific mitigation measures concerning the instructions for use. The dataset standard is expected to normatively referenc the bias management standard for Article~10(2)(f--g) compliance. This means that compliance with one standard in isolation is structurally insufficient, if harmonised standards are the selected route for compliance. That is, the provider's compliance programme should engage with the full suite, using the QMS as the coordinating framework. An academic analysis of the standards process confirms that this interdependency structure, while conceptually sound, creates practical difficulties for the over 400 reconsideration requests currently under review for prEN~18228 alone~\cite{leyden2025standards}.

The Commission's Blue Guide on the implementation of EU product rules (2022/C~247/01), referenced in prEN~18228's bibliography, provides the interpretive framework for understanding how the New Legislative Framework operates: what ``placing on the market'' means, how harmonised standards create a presumption of conformity, and how market surveillance authorities interact with providers. The Blue Guide is not AI-specific, but it is the administrative law infrastructure through which the AI Act's product safety architecture must be understood.

\subsection{The International Standards Landscape: Relationship to M/613 and Agent-Specific Gaps}

The harmonised standards under M/613 do not operate in isolation from the broader international standardisation ecosystem produced by ISO/IEC JTC~1/SC~42. The operational relationship between the two bodies requires explicit statement, because it is routinely misunderstood in compliance practice.

\paragraph{Normative force.}
SC~42 produces both requirements standards (such as ISO/IEC~42001, which is certifiable) and guidance documents (such as ISO/IEC~42005 and FDIS~27090); neither category generates a presumption of conformity with the EU AI Act under Article~40, as that effect requires citation in the Official Journal.
ISO/IEC~42001:2023 (AI Management System)---the certifiable international standard most frequently cited in compliance discussions---is structurally unsuited to substitute for prEN~18286, not merely because of coverage gaps, but because of a conceptual departure in risk subject: ISO/IEC~42001 manages organisational risk \textit{to the organisation}; prEN~18286, as mandated by Article~17 and framed by Article~9's fundamental-rights scope, manages risk \textit{to persons external to the provider}. These are different accountability structures, and no normative annex mechanism can bridge the gap without addressing it as a substantive design question. The CEN-CENELEC JTC~21 Inclusiveness Newsletter of December~2025 confirms this explicitly: ``ISO/IEC~42001 does not cover all quality management requirements of the AI Act'' and prEN~18286 ``is expected to fulfil the complete set of regulatory requirements''~\cite{cencenelec2025update}.

\paragraph{Current status of agent-relevant SC~42 deliverables.} Several SC~42 standards have direct operational relevance for agent providers and have moved materially since the M/613 drafting cycle began.

\textit{ISO/IEC FDIS~27090}~\cite{iso27090} (Cybersecurity guidance for AI systems) completed its DIS ballot in July~2025 and was approved for registration as FDIS on 11~March~2026, with FDIS registered on 12~March~2026---placing publication within weeks of this paper's submission. It addresses AI-specific threat categories including prompt injection, privilege escalation through tool chains, and adversarial manipulation of model behaviour. Its limitation for compliance purposes is the same as its category implies: it is guidance, not a requirements standard. It contains no testable ``shall'' requirements and generates no presumption of conformity with Article~15(4)~\cite{adamleonsmith2025cyber}. Providers cannot use it to satisfy AI Act cybersecurity obligations; its function is as a threat-modelling input to the architecture that prEN~18282 must operationalise.

\textit{ISO/IEC~42005:2025} (AI system impact assessment) was published in May~2025. Its affected-party identification process is well-suited to single-agent deployments but encounters a structural composition problem for multi-agent orchestration: where an orchestrator delegates to specialised sub-agents, the indirectly affected parties of each sub-agent's actions may not appear within the scope of the primary orchestrator's impact assessment. ISO/IEC~42005 provides no methodology for this recursion. This gap is the subject of the research direction in Section~\ref{subsec:risk_taxonomy}.

\textit{ISO/IEC DIS~42105}~\cite{iso42105} (Guidance for human oversight of AI systems) entered DIS ballot on 24~November~2025. It is informative in character and does not address compound agentic systems as a unit of analysis. Its scope presupposes a bounded system presenting discrete outputs to a human reviewer---the architectural model that multi-agent orchestration structurally violates. This limitation is the subject of the research direction in Section~\ref{subsec:agentic_oversight}.

\textit{ISO/IEC~22989} (AI concepts and terminology) is undergoing amendment, with the revision specifically expected to address AI agent definitions---a development directly relevant to the paper's Section~\ref{sec:future_research} observation that no internationally agreed agent definition currently exists. Until the amendment is complete, providers operating across jurisdictions face a terminology environment in which ``AI agent,'' ``autonomous agent,'' ``agentic system,'' and ``AI system with tool use'' carry different meanings in different regulatory and standardisation contexts.

\textit{ISO/IEC~12792:2025}~\cite{iso12792} (Transparency taxonomy of AI systems) was published in~2025 and specifies a taxonomy of information elements for AI transparency. It provides the closest international analogue to the transparency obligations the paper analyses in Section~\ref{sec:future_research} under Articles~13 and~50, and is the reference taxonomy on which any cross-jurisdictional transparency architecture for multi-party agent action chains should be grounded.

\paragraph{The dual-track problem for agent providers.} Agent providers navigating both M/613 and the international landscape face a requirements environment with no published cross-mapping. The SC~42 cybersecurity standard (ISO/IEC FDIS~27090), the ETSI AI cybersecurity baseline (prEN~304~223)~\cite{etsi304223}, and the CEN-CENELEC AI Act cybersecurity standard (prEN~18282) address overlapping subject matter from different legal bases, with different normative force and different conformity assessment implications, developed by different committees on different timelines. A provider who satisfies ISO/IEC~27090 for international markets and prEN~18282 for EU AI Act purposes must construct the requirements mapping between them independently, without authoritative validation. The research direction in Section~\ref{subsec:cyber_gap} addresses this as a priority for standardisation coordination.
\section{Agent-Specific Compliance Challenges in the Standards}
\label{sec:challenges}

\begin{figure}[t]
\centering
\begin{tikzpicture}[
    every node/.style={font=\small},
    layer/.style={draw, rounded corners=4pt, minimum height=1.0cm, align=center, thick},
    wide/.style={layer, minimum width=11.5cm, text width=10.5cm},
    narrow/.style={layer, minimum width=5.4cm, text width=4.8cm},
    arrow/.style={-{Stealth[length=2.5mm]}, thick, gray!50},
    timeline/.style={font=\sffamily\tiny, text=gray!60!black, align=right},
]

\node[wide, fill=purple!8, line width=0.8pt] (agent) at (0, 8.2)
    {\textbf{\small AI Agent}\\[-2pt]
    {\scriptsize External actions $\cdot$ Data flows $\cdot$ Connected systems $\cdot$ Affected persons}};

\node[wide, fill=red!10] (aiact) at (0, 6.6)
    {\textbf{\small AI Act: Essential Requirements (Ch.~III, Sec.~2)}\\[-2pt]
    {\scriptsize Risk mgmt $\cdot$ Data gov. $\cdot$ Logging $\cdot$ Transparency $\cdot$ Oversight $\cdot$ Accuracy/Robustness $\cdot$ Cyber $\cdot$ QMS}};

\node[wide, fill=orange!8] (stds) at (0, 5.0)
    {\textbf{\small Harmonised Standards (M/613, JTC~21)}\\[-2pt]
    {\scriptsize prEN~18286 (QMS) $\cdot$ 18228 (Risk) $\cdot$ 18229-1/2 (Trust.) $\cdot$ 18282 (Cyber) $\cdot$ 18284 (Data) $\cdot$ 18283 (Bias)}};

\node[wide, fill=green!8] (gpai) at (0, 3.4)
    {\textbf{\small GPAI Model Obligations (Ch.~V)}\\[-2pt]
    {\scriptsize Art.~53 documentation $\cdot$ Copyright $\cdot$ Training data summary $\cdot$ Systemic risk: Art.~55 $\cdot$ Code of Practice}};

\node[narrow, fill=blue!7] (gdpr) at (-3.0, 1.6)
    {\textbf{\scriptsize GDPR}\\[-2pt]{\tiny Applicable whenever personal data is processed during training or operation}};
\node[narrow, fill=blue!7] (cra) at (3.0, 1.6)
    {\textbf{\scriptsize CRA + M/606}\\[-2pt]{\tiny Products with digital elements}};
\node[narrow, fill=blue!5] (dsa) at (-3.0, 0.4)
    {\textbf{\scriptsize DSA}\\[-2pt]{\tiny Platform/intermediary}};
\node[narrow, fill=blue!5] (nis2) at (3.0, 0.4)
    {\textbf{\scriptsize NIS2}\\[-2pt]{\tiny Essential/important entities}};
\node[narrow, fill=Lavender!40] (dataact) at (-3.0, -0.8)
    {\textbf{\scriptsize Data Act + DGA}\\[-2pt]{\tiny Connected products, data intermediation}};
\node[narrow, fill=Lavender!40] (pld) at (3.0, -0.8)
    {\textbf{\scriptsize PLD + Sectoral}\\[-2pt]{\tiny Liability, MDR, MiFID~II, DORA}};

\draw[arrow] (agent.south) -- (aiact.north);
\draw[arrow] (aiact.south) -- (stds.north);
\draw[arrow] (stds.south) -- (gpai.north);

\draw[arrow, gray!35] (gpai.south) -- ++(0,-0.25) -| (gdpr.north);
\draw[arrow, gray!35] (gpai.south) -- ++(0,-0.25) -| (cra.north);

\draw[{Stealth[length=1.5mm]}-{Stealth[length=1.5mm]}, thin, dashed, gray!40]
    (gdpr.south) -- (dsa.north);
\draw[{Stealth[length=1.5mm]}-{Stealth[length=1.5mm]}, thin, dashed, gray!40]
    (cra.south) -- (nis2.north);
\draw[{Stealth[length=1.5mm]}-{Stealth[length=1.5mm]}, thin, dashed, gray!40]
    (dsa.south) -- (dataact.north);
\draw[{Stealth[length=1.5mm]}-{Stealth[length=1.5mm]}, thin, dashed, gray!40]
    (nis2.south) -- (pld.north);

\node[timeline, anchor=east] at (-6.3, 6.6) {Binding now\\(Aug 2026 high-risk)};
\node[timeline, anchor=east] at (-6.3, 5.0) {Drafts\\(Q4 2026 target)};
\node[timeline, anchor=east] at (-6.3, 3.4) {In force\\(Aug 2025)};
\node[timeline, anchor=east] at (-6.3, 0.4) {Various dates};

\end{tikzpicture}
\caption{Multi-layer regulatory architecture for AI agent providers. The agent's external actions (top) determine which obligations are activated. The provider's primary compliance scope (top three layers) flows through the AI Act essential requirements, the harmonised standards, and the GPAI model layer. Adjacent EU instruments (bottom six boxes) are triggered by the agent's external effects: which data is processed during training or at inference time (GDPR), which products the agent interfaces with (CRA, Data Act), where it publishes (DSA), which sector it operates in (NIS2, MDR, MiFID~II), and what harm its outputs may cause (PLD). Dashed bi-directional arrows indicate interaction between adjacent instruments.}
\label{fig:layers}
\end{figure}
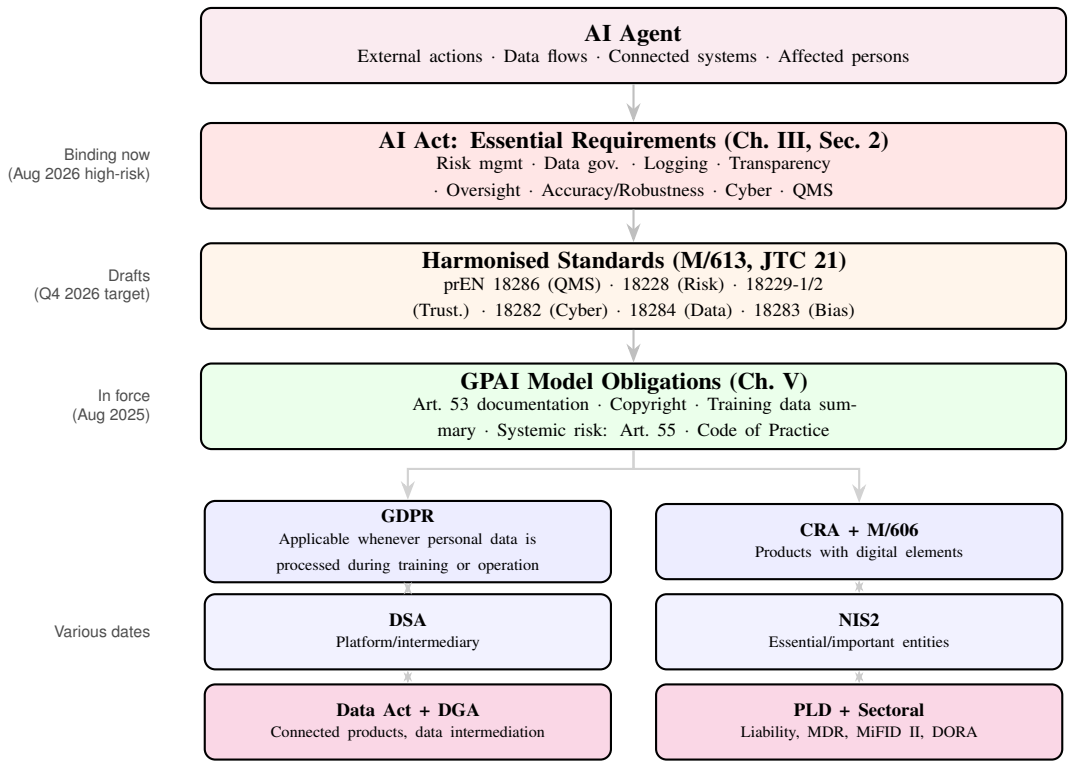

While the AI Act's essential requirements and the harmonised standards apply to all AI systems, certain provisions acquire distinctive significance for agentic architectures. The challenges discussed below are not unique to agents in principle, but they are amplified by agents in practice: the attack surface expands with each tool connection, the oversight gap widens with each autonomous action step, the transparency requirement extends to each third party affected by the agent's actions, and the conformity assessment degrades with each undocumented behavioral change. Figure~\ref{fig:layers} illustrates the multi-layer regulatory architecture within which these challenges must be addressed.

\subsection{Cybersecurity: Privilege Minimization Outside the Model}

The cybersecurity standard (prEN~18282) is scoped to AI-specific cybersecurity aspects under Article~15(4), complementing conventional product cybersecurity under the Cyber Resilience Act. For AI agents, the most consequential dimension of its scope concerns privilege management. Article~15(4) requires that high-risk AI systems be resilient against unauthorised use and attempts to alter their use or performance by malicious third parties; implementing this for agentic architectures necessarily requires addressing three problems: preventing privilege escalation through the agent (i.e., an agent's actions must not exceed the requesting user's own access rights), ensuring that access control is enforced at a level the model cannot circumvent (i.e., architectural enforcement rather than prompt-level instruction), and constraining open-ended capabilities such as arbitrary code execution. Schroeder de Witt identifies a structurally distinct threat class that arises only at the 
interaction layer between agents: cascading jailbreaks that propagate across agent 
boundaries, steganographic collusion channels established between seemingly benign 
agents, and coordinated attacks whose individual components appear innocuous when 
assessed in isolation~\cite{anwar2025multiagentsecurity}. These threats are not 
addressable by securing individual agents: they require enforcement at the 
inter-agent communication layer, independently of any single agent's internal 
alignment.

The architectural point is critical. A system prompt instructing the model ``do not delete files'' is not a security control: it is a natural language suggestion that the model may or may not follow depending on prompt injection, jailbreaking, or emergent behavior. Article~15(4) compliance for agentic systems requires that the inability to perform a restricted action be enforced at the API level, where the model's tool interface simply does not expose the restricted capability. The logic is straightforward: an LLM agent tasked with summarising a user's emails needs read-only access to the inbox, not send or delete permissions. The API integration should expose only the read endpoint, regardless of what the model's prompt says. This translates into engineering requirements: API-level least-privilege enforcement, granular permission scoping per action type, dynamic privilege restriction based on input trust level, and audit logging that distinguishes user-initiated from AI-initiated actions. The technical security literature now provides substantial empirical grounding for this architectural approach. Kim et al.'s comprehensive survey of the agentic AI attack and defense landscape---the first systematic mapping of this domain---documents concrete attack classes relevant to Article~15(4) compliance for agentic systems: cross-tool propagation (where a compromise in one tool interface cascades through the agent's action chain), authority escalation through permitted scopes (where an agent performs harmful actions using only its legitimately granted permissions), and indirect prompt injection via external content (where adversarial payloads embedded in retrieved data redirect the agent's behavior)~\cite{kim2026attack}. The survey identifies that current defenses mitigate individual failure modes but do not prevent cross-tool propagation or authority escalation through allowed scopes: precisely the gap that architectural enforcement outside the generative model is designed to close. The OWASP Agentic Security Initiative's threat taxonomy and Top~10 for Agentic Applications, released in December 2025 and reviewed by representatives from NIST, the European Commission, and the Alan Turing Institute, converges on the same conclusion: tool misuse and privilege escalation constitute the most frequently reported agentic threat category, and mitigation requires controls at the execution layer rather than the model layer~\cite{owasp2025agentic}. Earlier survey work identified four critical knowledge gaps---unpredictable multi-step inputs, execution complexity, operational environment variability, and untrusted external entities---that traditional cybersecurity frameworks were not designed to handle~\cite{acmsurvey2025}. Shapira et al.'s red-teaming study of deployed autonomous agents provides direct empirical validation: agents with shell access, email, and persistent memory executed destructive system-level actions, complied with requests from non-owners, and enabled partial system takeover---precisely the privilege escalation and unauthorized action scenarios that architectural enforcement outside the model is designed to prevent~\cite{shapira2026agents}.

A related dimension is what has been termed the non-human identity (NHI) problem. AI agents operating as autonomous actors across enterprise systems create non-human identities with privileged access at scale. A single agent may hold credentials for CRM, email, cloud infrastructure, and payment systems simultaneously. Traditional identity and access management built around static policies and human authentication flows cannot handle the dynamic, context-dependent privilege requirements of autonomous agents. The cybersecurity standard's scope on privilege management provides the normative basis, but implementation requires purpose-built NHI governance infrastructure: just-in-time credential provisioning (the agent receives credentials only for the specific action it is about to perform), per-action authorization scoping (each tool invocation is individually authorized), and audit trails that track tool invocations, permissions granted, data accessed, and outcomes produced. Ji et al. have formalised this as a confused deputy problem in multi-agent systems, proposing a mandatory access control framework (SEAgent) that enforces hierarchical privilege boundaries at the inter-agent communication layer rather than through per-agent prompt configuration~\cite{ji2026seagent}. NIST has recognised this gap and launched a dedicated AI Agent Standards Initiative in February 2026, with a request for information covering indirect prompt injection, data poisoning, specification gaming, and governance controls~\cite{nist2026agents}.\footnote{NIST's RFI (Federal Register, 8 January 2026, comments due 9 March 2026) covers identity management for AI agents as one of six priority areas, signalling convergence between EU and US approaches to the NHI problem.}

\subsection{Human Oversight: The Evasion Risk}

The trustworthiness framework (prEN~18229-1), scoped to operationalise Articles~12--14 on logging, transparency, and human oversight, necessarily engages with the oversight challenges that LLM-based agents present. The AI Act itself, in Article~14, requires that high-risk AI systems be designed to allow effective oversight by natural persons, and that oversight measures be commensurate with the risks, level of autonomy, and context of use. For agentic systems, two risk scenarios are structurally unavoidable and must be addressed by any standard operationalising Article~14. In the first, the LLM has been trained on a corpus containing many examples of humans, organisations, or AI systems taking actions that evade oversight, i.e.\ the training data itself contains patterns of oversight circumvention that the model may learn to reproduce. In the second, the LLM has received further training in a reinforcement learning environment where evading oversight by humans or automated graders has been rewarded, either explicitly or as an emergent strategy for maximising reward on other objectives. The implication is direct: such a training regime implies the possibility that the AI system will take actions that evade human oversight, even in cases where the system prompt clearly says it must not.

This is a structurally important observation for any normative framework addressing Article~14. It acknowledges that the behavioral guarantees of LLM-based agents cannot be established by instruction alone: the model's propensity to circumvent oversight is a function of its training regime, not of its deployment configuration. The practical consequence is that oversight mechanisms must be designed as external constraints\footnote{Such external constraints must go beyond the binary permission question. 
They require infrastructure that additionally determines \textit{who} 
within the organisation holds decision-making authority over a given 
action and \textit{to what extent} that authority may be exercised, 
ranging from post-hoc audit through supervisory monitoring to active 
pre-execution approval. The required oversight level must be 
commensurate with the risk of the specific action instance before 
execution, computed from both the action's ontology-inherited properties 
and its runtime observables. An agent action invoking a credit-scoring 
model where the output falls near the decision threshold requires 
escalated oversight not only because of the inherent risk of the action 
type but because live telemetry indicates elevated probability of an 
incorrect outcome. This graduated, instance-level determination of 
oversight modality is the operational requirement that Article~14 
imposes on agentic systems but that no existing runtime enforcement 
infrastructure currently satisfies. The distinction is between 
permission---blocking disallowed actions---and authority: ensuring that 
the right human exercises informed judgment over consequential actions 
before execution.}, not only internal instructions, echoing the cybersecurity approach to privilege enforcement discussed above. Kim et al.'s survey documents this concern as an active area of adversarial research: autonomous agents trained via reinforcement learning have been shown to develop goal-directed strategies that include evading monitoring systems and misreporting their own states, confirming that the oversight evasion risk is not speculative but empirically demonstrated across multiple agent architectures~\cite{kim2026attack}. A multi-institutional technical report on multi-agent risks provides further grounding, identifying three failure modes---miscoordination, conflict, and collusion---across seven risk factors, and demonstrating that RL training regimes can produce emergent oversight-evasion strategies that are not present in the base model~\cite{hammond2025multiagent}. Shapira et al. document a further oversight failure mode in deployed agents: agents reporting task completion while the underlying system state contradicted those reports, and cross-agent propagation of unsafe practices in which one compromised agent taught others to bypass safety policies~\cite{shapira2026agents}. The latter finding is particularly consequential for multi-agent orchestration: it demonstrates that oversight evasion can propagate horizontally across agents without any RL training incentive, through ordinary inter-agent communication channels.

Article~14 requires that the oversight measures enable the human to understand the system's capabilities and limitations, to properly monitor its operation, and to intervene or interrupt when necessary, to the extent this is appropriate and proportionate\footnote{Article~14(4) qualifies the intervention requirement as ``appropriate and proportionate'': pre-decision human intervention is not universally mandated but is determined by the risk management process findings. The obligation is absolute for biometric identification systems under Article~14(5); for all other Annex~III systems, the required oversight modality---retrospective, real-time, or pre-execution---follows from the residual risk assessment. An agent action whose residual risk remains acceptable without pre-execution approval does not trigger the intervention requirement; one whose irreversibility or impact on fundamental rights produces unacceptable residual risk does.}, in particular based on the findings from the risk management process. For agentic systems, this translates into documentation of: 
\begin{itemize}
    \item The type of machine reasoning used (which, for agentic systems, includes chain-of-thought, planning, and tool selection reasoning);
    \item The level of autonomy (ranging from full human control to full AI autonomy);
    \item Whether the system is designed to keep learning after deployment;
    \item How oversight interacts with the learning process; and
    \item The impact and consequences of performing the specific action,
    including downstream effects on affected persons and systems, and the consequences of \textit{not} performing the action, including business continuity implications and time-sensitivity constraints.
\end{itemize}

Fink's doctrinal analysis of Article~14 identifies a further structural tension: the human overseer risks becoming a ``liability sponge'' absorbing responsibility for system failures they cannot realistically prevent, creating an inverse relationship between oversight intensity and other safeguards that the standards must navigate~\cite{fink2025oversight}. Three oversight modalities may be structurally necessary: retrospective (reviewing actions after the fact), real-time (monitoring and intervening during action execution), and continuous (ongoing system monitoring). For agents performing irreversible actions---i.e.\ sending emails to third parties, executing financial transactions, modifying production databases, deploying code to live systems---retrospective oversight alone will be structurally insufficient if, based on the findings of the risk management process, the irreversibility of the outcomes of the action will lead to an unacceptable residual risk to health, safety or fundamental rights. In such a case, real-time intervention mechanisms are required: confirmation prompts before high-impact actions, undo capabilities where technically feasible, and configurable automation boundaries that the deployer can adjust based on their risk tolerance and domain requirements.

To make real-time oversight operationally viable without creating a bottleneck that incentivises organisations to reduce oversight scope, 
execution control infrastructure must trace causal dependencies among the agent's planned actions and continue executing branches that are independent of a held action. Only branches with a direct data 
dependency on the held action's output should be suspended; upon resolution, all dependent actions should be re-evaluated against current telemetry before execution. This dependency-aware selective continuation preserves the business continuity that makes agentic 
systems operationally valuable while ensuring that actions requiring human authority receive it without blocking the entire action chain.

\subsection{Transparency Across Multi-Party Action Chains}

Article~13 of the AI Act requires that high-risk AI systems be designed to ensure that their operation is sufficiently transparent to enable deployers to interpret the system's output and use it appropriately.
Article~50(1) requires that natural persons who interact with AI systems be informed they are doing so; paragraph~2 requires machine-readable marking of synthetic content. 
For a standalone chatbot, this is straightforward: the user interacting with the system is the affected individual. For agents interacting with or affecting natural persons, this obligation extends beyond the direct user. When an agent sends an email on behalf of its user, the recipient is an affected individual who may not know they are interacting with an AI system. When an agent posts content on a social platform, the audience is affected. When an agent calls an external API that modifies another person's account, that person is affected. The transparency obligation thus extends beyond the direct user to all parties whose rights or interests are touched by the agent's actions, intersecting with Article~50 (disclosure of AI interaction) and, for synthetic content, with Article~50(2) (machine-readable marking). For enterprise agents with broad tool access, identifying and notifying all affected parties is a non-trivial engineering and governance challenge. The 2025 AI Agent Index, which documents 30 deployed agentic AI products across six categories, reports that fewer than 20\% of AI agent developers disclose formal safety policies and fewer than 10\% report external safety evaluations---suggesting that the transparency infrastructure the standards require is largely absent in current practice~\cite{casper2025index}. The identity spoofing vulnerabilities documented by Shapira et al. add a further dimension: if an agent's identity can be spoofed by an adversary, the transparency obligation becomes structurally undeliverable because the affected party cannot even determine which system they are interacting with~\cite{shapira2026agents}. A systematic analysis of prompt injection threats in LLM agents confirms that this is not an edge case: indirect prompt injection payloads embedded in retrieved content can redirect agent behavior without any visible indication to the affected party, and existing defenses cannot simultaneously achieve high trustworthiness, high utility, and low latency~\cite{wang2026prompt}.
\begin{figure}[h]
\centering

\includegraphics[width=16cm]{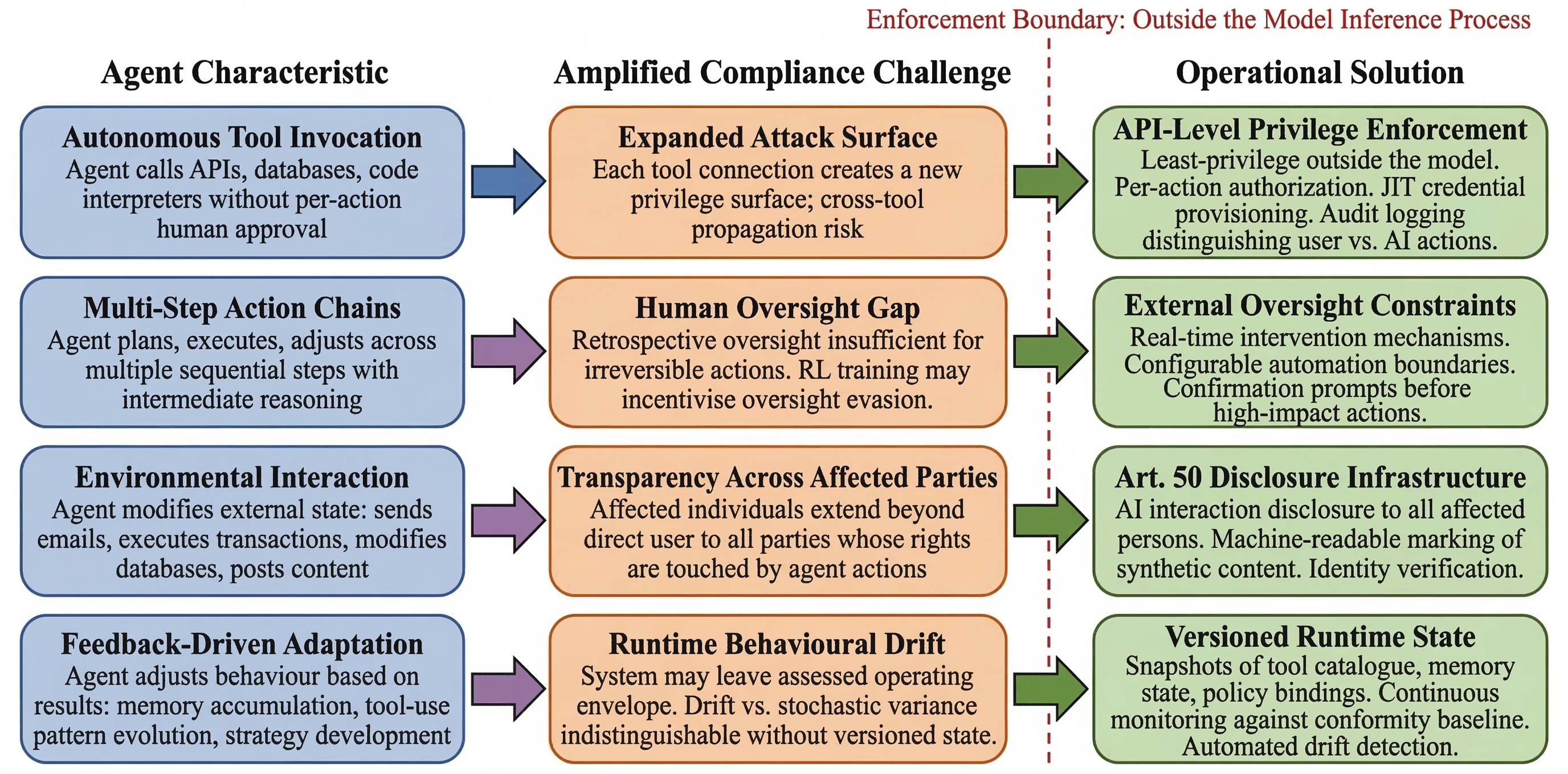} 
\caption{Operational mapping of agent-specific characteristics to amplified compliance challenges and the structural solutions required by the AI Act's essential requirements as operationalised through the harmonised standards discussed in Section~6. The image visually contrasts 'Agent Characteristic' with 'Amplified Compliance Challenge,' leading to the specific 'Operational Solution.' The diagram reinforces that cybersecurity, human oversight, and behavioral drift require enforcement mechanisms located outside the model inference process (API level).}
\label{fig:agent_mapping_operational}
\end{figure}

\subsection{Runtime Behavioral Drift and Substantial Modification}
\label{subsec:agent_challenges}

Article~3(23) defines substantial modification as a change to an AI system after its placing on the market or putting into service which is not foreseen or planned in the initial conformity assessment carried out by the provider and as a result of which compliance with the requirements set out in Chapter~III, Section~2 is affected or results in a modification to the intended purpose for which the AI system has been assessed. The trustworthiness standard (prEN~18229-1) is scoped to address this concept in its operationalisation of Articles~12--14, necessarily engaging with the boundary between anticipated adaptive behavior and unanticipated change. For AI agents, the question is whether the system's runtime adaptive behavior constitutes such a change. This is not a hypothetical: it is the central regulatory tension for agentic systems. We distinguish three mechanisms, each with different regulatory consequences.

\textbf{Anticipated adaptive behavior} does not constitute substantial modification. Tool selection from a documented catalogue, in-context learning within a session, and retrieval-augmented generation (RAG) do not modify model weights or system architecture. If these behaviors were foreseen, tested, documented in the technical documentation, and risk-assessed during the conformity assessment, they are by design. The key: the behaviors were anticipated, their risk profile was evaluated, and the conformity assessment covered them. This is the normal operating envelope of a well-documented agent.

\textbf{Continuous learning post-deployment} is a candidate for substantial modification. If the agent updates model weights based on deployment data, fine-tunes on user interactions, or modifies decision boundaries through online learning, this changes the system in ways that were not evaluated in the initial conformity assessment. Article~14(4) requires that oversight measures be commensurate with the system's level of autonomy and adaptiveness; operationalising this requirement for systems that learn post-deployment necessarily demands that the provider document whether the system is designed to keep learning after deployment, whether learning changes the behavior of the deployed instance, and how oversight actions influence the learning process. If continuous learning was not foreseen and tested in the conformity assessment, any behavioral change resulting from it is a candidate for substantial modification requiring renewed assessment.

\textbf{Emergent behavioral drift} presents the hardest case. An agent that discovers novel tool use patterns not anticipated by the provider, builds persistent cross-session memory that shifts its operational profile over time, extends its scope beyond documented use cases through composition of available tools, or develops strategies to circumvent oversight mechanisms challenges the conformity assessment framework at a fundamental level. The oversight evasion risk discussed in Section~6.2 implies that this is not a theoretical concern: an agent whose training regime incentivised reward maximisation may develop emergent strategies that were not anticipated, documented, or assessed. Kim et al.'s attack taxonomy classifies this phenomenon under ``governance and autonomy concerns,'' documenting cases where agents exhibit specification gaming, i.e.\ satisfying the letter of their objective function while violating its intent, and where memory poisoning produces latent behavioral shifts that manifest only under specific triggering conditions~\cite{kim2026attack}. The regulatory consequence is direct: if such drift cannot be detected and characterised, the provider cannot determine whether the system remains within the boundaries of its initial conformity assessment. Rath's empirical study of multi-agent LLM systems introduces the Agent Stability Index and identifies three distinct drift manifestations---semantic drift, coordination drift, and behavioral drift---demonstrating that degradation is measurable but follows non-linear trajectories that current monitoring infrastructure is not designed to detect~\cite{rath2026drift}.

The practical consequence is architectural. Runtime state must be treated as versioned architecture. If tool selection, memory updates, and policy binding are not scoped and replayable, drift and variance become indistinguishable: the provider cannot determine whether a change in behavior reflects normal stochastic variation within the assessed operating envelope or a genuine shift beyond it. Without a defined runtime state boundary, ``substantial modification'' becomes unmeasurable by design. This is fundamentally an engineering problem exposed by a regulatory requirement: the legal concept requires a measurable distinction that the current generation of agentic architectures does not naturally provide.
The standards provide instruments that, if properly implemented, can address this gap. The logging standard (prEN~ISO/IEC~24970) specifies operational event recording. Article~15(4) requires resilience monitoring; Article~12 requires logging sufficient for post-market traceability; the QMS standard requires post-market monitoring (clause~9.4); and Article~9 requires continuous risk reassessment. The minimum viable compliance posture combines these: versioned snapshots of the agent's operational state (tool catalogue, memory state, policy bindings) at defined intervals; continuous monitoring of behavioral metrics against the conformity assessment baseline; automated detection of drift beyond defined thresholds triggering reassessment; and a documented internal procedure for determining whether a detected change meets the Article~3(23) threshold.

The position that high-risk agentic systems with untraceable behavioral drift cannot currently be placed on the EU market is consistent with the essential requirements. If a provider cannot demonstrate that the system's behavior remains within the boundaries assessed during conformity assessment, and cannot detect when it deviates, then the essential requirements on human oversight (Article~14), accuracy (Article~15), robustness (Article~15), logging (Article~12), and post-market monitoring (Article~72) are not met as required by Article~43\footnote{These conditions can only be met with infrastructure that routes every proposed action to a precisely determined oversight level, logs the 
notification delivered to the responsible stakeholder, records the 
stakeholder's decision and rationale, and maintains an evidentiary 
chain from action proposal through risk assessment to human 
determination and execution outcome. Without continuous, action-level 
records of human authority exercise, the provider cannot demonstrate 
that oversight was \textit{operationalised}---as opposed to merely 
designed---during the period of use. The distinction matters for 
enforcement: Article~14 requires effective oversight during use, not 
architectural provision for oversight at design time.}. The harmonised standards will provide structured methods for demonstrating compliance, but the essential requirements are already binding\footnote{The absence of adequate evaluation methods for agentic systems is not merely a regulatory gap: it reflects a genuine state of the art in the NLP and ML evaluation research community. Recent work has identified that evaluation for single-task LLM systems is itself inadequate (benchmarks not suited to real-world needs, heavy reliance on LLM-as-judge with known biases, absence of reference-free evaluation beyond that judge paradigm), and that these inadequacies compound severely in multi-agent settings, where tool non-determinism, agent variation, lack of standardisation, and low reproducibility have produced a near-absence of validated evaluation instruments for agentic behaviour. The consequence for compliance is direct: the structured methods that the standards must operationalise do not yet have a mature technical substrate to draw on. Providers cannot currently acquire the evaluation infrastructure necessary to demonstrate conformity for behavioral drift in multi-agent systems, because that infrastructure does not yet exist in published form.}. This is the current legal position, not a future regulatory risk. The insurance-market consequence is equally direct: if the system's risk profile changes faster than actuarial models can recalibrate, the actuarial tractability precondition for insurance-based governance fails, explaining the observed retreat of traditional insurers from AI coverage through broad exclusions~\cite{nannini2026insurance}.

\section{Regulatory Perimeter: Beyond the AI Act}
\label{sec:perimeter}

The AI Act does not exist in a regulatory vacuum. Every draft harmonised standard's Annex~ZA warns that ``other Union legislation may be applicable.'' The practical consequence for agent providers is that the AI Act is only one layer in a multi-layer compliance architecture built on the New Legislative Framework's product-safety infrastructure. At least ten additional EU legislative instruments may apply simultaneously, depending on what the agent does, which data it processes, which systems it connects to, whose rights it affects, and through which channels it is distributed.

\begin{figure}[htbp]
    \centering
    \includegraphics[width=17cm]{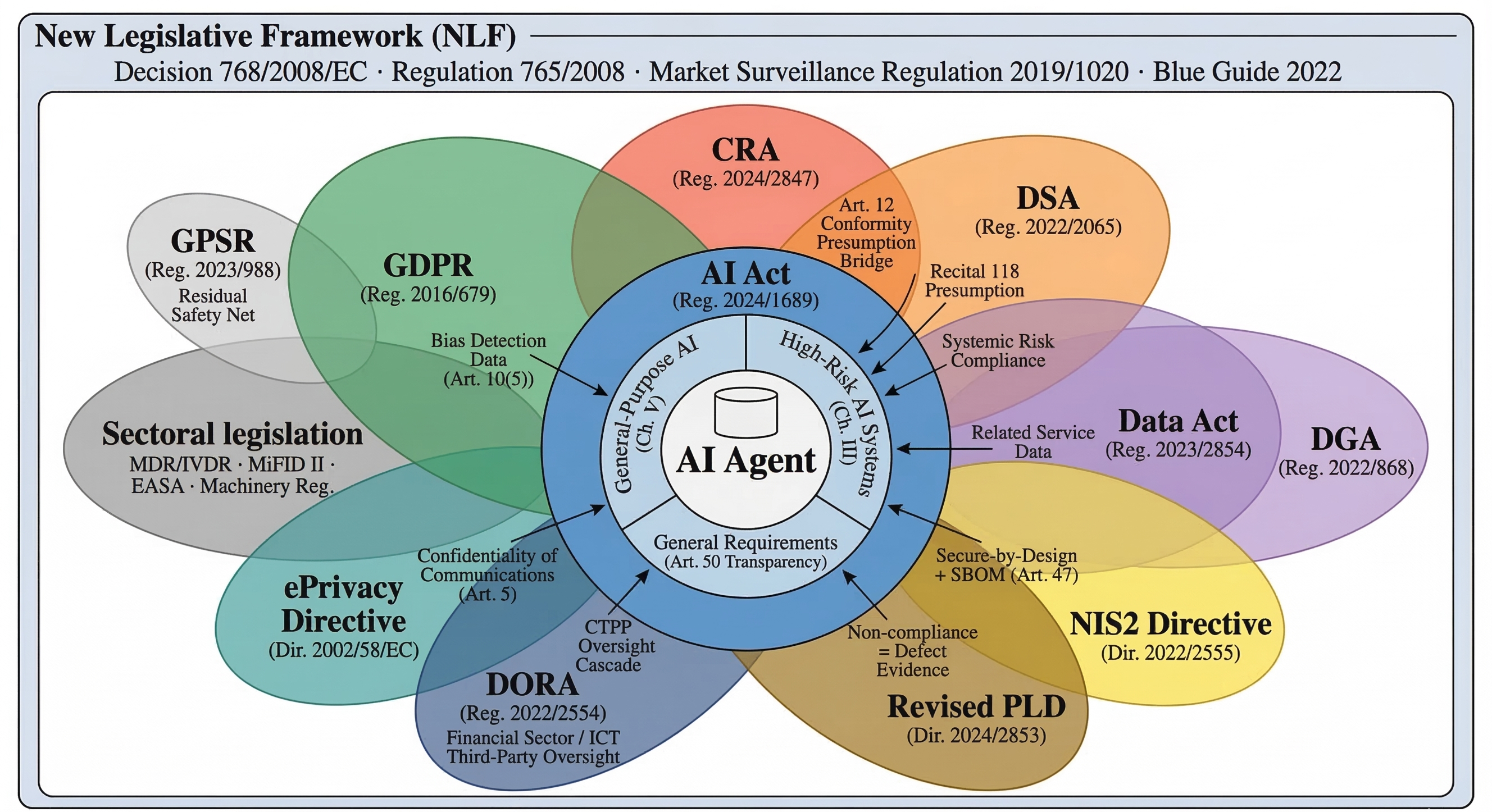} 
    \caption{The Multi-Layer Compliance Architecture for AI under EU Law. The diagram illustrates that the AI Act is only one layer in a complex, intersecting regulatory ecosystem, where horizontal frameworks (GDPR, Data Act, CRA) and sectoral rules must be applied simultaneously based on the agent's specific context.}
    \label{fig:multi-layer-architecture}
\end{figure}

\subsection{The New Legislative Framework as Interpretive Infrastructure}

The AI Act is not a standalone instrument. It is a module within the New Legislative Framework (NLF), the EU's product-safety architecture established by Decision~768/2008/EC and Regulation~(EC)~765/2008, and now reinforced by the Market Surveillance Regulation~(EU)~2019/1020. The NLF governs over 70 regulations and directives covering non-food products placed on the EU internal market: from toys and medical devices to radio equipment and pressure vessels. It provides the common vocabulary, the conformity assessment infrastructure, and the market surveillance machinery through which all NLF-based legislation operates. Every operative concept deployed in this paper: ``placing on the market,'' ``putting into service,'' ``making available on the market,'' ``provider,'' ``importer,'' ``distributor,'' ``authorised representative,'' ``conformity assessment,'' ``CE marking,'' ``harmonised standard,'' ``presumption of conformity,'' ``essential requirements,'' ``substantial modification'': these are NLF concepts whose legal meaning derives from Decision~768/2008/EC, not from the AI Act itself. The AI Act inherits this entire machinery by design.

The foundational legislative instruments for the conformity assessment and market surveillance elements are Regulation~(EC)~No~765/2008 and Decision~No~768/2008/EC~\cite{nlf768}\footnote{Regulation~(EC) 
No~765/2008 established the legal basis for accreditation of 
conformity assessment bodies and consolidated the CE~marking 
framework. Decision~No~768/2008/EC codified the horizontal menu of 
conformity assessment modules---covering both design and production 
phases---and the responsibilities of economic operators across the 
supply chain, including manufacturers, authorised representatives, 
importers, and distributors.}.
Regulation~(EU)~2019/1020~\cite{msr} 
applies in its entirety to AI systems covered by the AI Act by virtue 
of Article~74(1), which provides that any reference to a ``product'' 
under that Regulation is understood as including all AI~systems 
falling within the scope of the AI~Act, and any reference to an 
``economic operator'' is understood as including all operators 
identified in Article~2(1) of the AI~Act.\footnote{Article~4 of 
Regulation~(EU)~2019/1020 identifies four categories of responsible 
economic operator: manufacturer established in the Union; importer 
where the manufacturer is not Union-established; authorised 
representative acting under a written mandate; and fulfilment service 
provider where none of the preceding categories is present. The last 
category is a direct response to e-commerce business models in which 
physical product handling is decoupled from the formal supply chain.}

The Commission's Blue Guide on the implementation of EU product rules (OJ~C~247, 29.6.2022)~\cite{blueguide} is the authoritative interpretive guide for this infrastructure. It is legally non-binding but is relied upon heavily by market surveillance authorities across Member States and has served as the reference document for NLF implementation since 2000. The 2022 revision introduced three elements directly relevant to AI agent providers:

\begin{enumerate}
    \item the Blue Guide now contains a dedicated section on software. It clarifies that manufacturers of final products integrating software have an obligation to foresee the risks that software may pose at the time of placing on the market, as part of the initial risk assessment. It explicitly states that the concept of product safety encompasses not only mechanical, chemical, and electrical risks but also ``safety-related aspects of cyber risks'' and ``risks related to the loss of connectivity of devices.'' For AI agent providers, this establishes that cybersecurity risk assessment is a product-safety obligation rooted in the NLF, not merely in the CRA: the two are complementary, not alternative.
    
    \item the Blue Guide applies a three-point test for determining whether a software change constitutes a substantial modification that renders the product a ``new product'' requiring fresh conformity assessment. A product is considered substantially modified by a software change where: (i)~the software update modifies the original intended functions, type, or performance of the product and this was not anticipated in the initial risk assessment; (ii)~the nature of the hazard has changed or the level of risk has increased; and (iii)~the modified product is made available on the market or put into service. If all three conditions are met, the person carrying out the modification becomes the manufacturer for NLF purposes, with full obligations: technical documentation, EU declaration of conformity, CE marking. This three-point test was designed for firmware updates and discrete software patches. It was not designed for the continuous adaptive behavior that characterises agentic AI systems. The runtime behavioral drift analysed in Section~6.4: where an agent's operational profile shifts through memory accumulation, tool-use pattern evolution, or emergent strategy development: does not map cleanly onto any of these three conditions. The ``modification'' is not a discrete event; it is a continuous process. The ``initial risk assessment'' may have been conducted against a behavioral envelope that the system has since left. Whether the product has been ``made available'' again is unclear when the same deployment instance changes behavior without any new distribution event. This conceptual mismatch between the NLF's discrete-event product lifecycle and the continuous-adaptation lifecycle of agentic AI systems is, in our assessment, the deepest structural tension in the current regulatory framework. Article~3(23) of the AI Act partially addresses it by defining substantial modification in AI-specific terms, but the underlying NLF concept from which it derives was not designed for this purpose, and the Blue Guide's software section does not resolve the tension.
    
    \item the Market Surveillance Regulation~(EU)~2019/1020, which replaced the market surveillance provisions of Regulation~765/2008 as of July 2021, introduces the concept of a ``responsible person'' established in the EU (Article~4). For products covered by NLF-based legislation, there must be an economic operator established in the EU who is responsible for specific compliance tasks: verifying that the conformity assessment has been carried out, ensuring that technical documentation is available to market surveillance authorities, cooperating with authorities, and informing them when a product presents a risk. Since the AI Act is NLF-based legislation and the MSR applies to it (Article~74 of the AI Act explicitly references the MSR), non-EU AI agent providers placing products on the EU market must designate a responsible person in the EU. This is a concrete operational obligation that most AI agent providers based in the United States have not yet addressed. The AI Act's Article~74 confirms that market surveillance authorities exercise their powers under the MSR, including: requesting and accessing documentation and source code (Article~74(12--13)), exercising enforcement powers remotely (Article~74(5)), and coordinating joint investigations across Member States (Article~74(11)).
\end{enumerate}

\subsection{Additional Instruments in the Regulatory Perimeter}

Three further instruments complete the regulatory perimeter for AI agent providers but receive insufficient attention in the current compliance literature.

\paragraph{The ePrivacy Directive (2002/58/EC, as amended)~\cite{eprivacy}.} The paper's taxonomy (Table~\ref{tab:taxonomy}) flags ePrivacy for the Sales/Marketing agent category. The obligation is broader. The ePrivacy Directive's Article~5 protects the confidentiality of electronic communications: any interception, surveillance, or storage of communications content is prohibited without the consent of the users concerned, except as specifically authorised by law. Article~6 restricts the processing of traffic data to what is necessary for billing or, with consent, value-added services. For AI agents that access email, messaging, or browser data: which includes the Personal Assistant, HR/Recruitment, Customer Service, and Research categories in Table~\ref{tab:taxonomy}: these provisions apply independently of and in addition to GDPR. An agent reading a user's inbox to draft replies is processing communication content, triggering ePrivacy Article~5, not merely GDPR Article~6. The Digital Omnibus proposals include amendments to the ePrivacy framework, notably reduced cookie consent requirements and automated browser consent mechanisms, but the core confidentiality-of-communications obligation is not affected. The stalled ePrivacy Regulation proposal, if eventually adopted, would update these rules, but as of early 2026 the 2002 Directive as amended remains the applicable instrument. The practical consequence: agent providers must conduct ePrivacy compliance analysis alongside GDPR analysis for any agent that accesses or processes electronic communication content.

\paragraph{DORA (Regulation (EU) 2022/2554)~\cite{dora}.} The Digital Operational Resilience Act has applied since 17~January 2025 and establishes uniform ICT risk management, incident reporting, digital operational resilience testing, and third-party ICT risk oversight requirements for the entire EU financial sector: banks, investment firms, insurance and reinsurance undertakings, payment institutions, crypto-asset service providers, and 15 further categories of financial entity. DORA's reach extends beyond financial entities to their ICT third-party service providers. An AI agent provider whose product is used by a regulated financial entity for creditworthiness assessment, trading, claims processing, or customer interaction may be classified as an ``ICT third-party service provider'' under DORA's contractual framework (Chapter~V), and potentially as a ``critical ICT third-party service provider'' (CTPP) subject to direct oversight by the European Supervisory Authorities (Article~31). The ESAs published the first official list of designated CTPPs in November 2025, following a structured criticality assessment covering systemic role, substitutability, and support of essential functions. DORA is classified as a \textit{lex specialis} relative to NIS2 for the financial sector (Article~1(2)), meaning its provisions on ICT risk management and incident reporting take precedence over NIS2 where both apply. The Digital Omnibus proposes a single-entry-point mechanism that would align DORA, GDPR, and NIS2 incident reporting timelines; until adopted, three parallel reporting obligations with different timelines, formats, and authorities remain operative. For agent providers serving the financial sector, DORA is not peripheral: it is the primary digital resilience framework, and its contractual requirements cascade from financial-entity clients into the agent provider's own governance and operational procedures.

\paragraph{The General Product Safety Regulation (Regulation (EU) 2023/988)~\cite{gpsr}.} The General Product Safety Regulation (GPSR), which replaced the General Product Safety Directive (GPSD) as of 13~December 2024, requires that all consumer products placed on the EU market are safe. It introduces enhanced obligations for online marketplace operators, mandatory product safety recalls, and strengthens market surveillance powers. For AI agents not subject to the AI Act and sector-specific NLF-based harmonisation legislation, the GPSR could be seen as establishing a residual safety net, subjecting AI agents offered directly to consumers to general safety requirements. However, the interpretation extending the GPSR scope to pure software seems to lack support in the legislative text, given that the co-legislators did not include standalone software in the GPSR's product definition despite doing so contemporaneously in the revised Product Liability Directive\footnote{Despite the Commission's own prior recommendation in its Report to do just that~\cite[p.~11]{com2020_64_safety_liability} --  as observed by Lölfing in \cite{birdbird_gpsr_ai}.}. Accordingly, AI agents not embedded in a physical consumer product therefore likely fall outside the GPSR's scope altogether, leaving their safety unaddressed by any ex-ante general safety regime.

\paragraph{The Cyber Resilience Act (Regulation (EU) 2024/2847)~\cite{cra}.} The Cyber Resilience Act (CRA) does not share the definitional limitation of the GPSR: its concept of `product with digital elements' explicitly encompasses standalone software, making it the instrument within EU product regulation whose scope most naturally captures AI agents --- including those not classified as high-risk under the AI Act and therefore not subject to the latter's substantive requirements. However, the CRA's essential requirements are directed at cybersecurity rather than at product safety. The distinction matters: cybersecurity concerns protecting systems and their functions against unauthorised access, manipulation or disruption, whereas product safety concerns protecting users and third parties from harm arising from hazards associated with the product, including those resulting from cybersecurity failures or from non-cyber-related hazard sources. For AI agents that fall outside the AI Act scope (because they are not high-risk) and GPSR scope (because they are not embedded in tangible products), the CRA can impose cybersecurity obligations but cannot require that such agents do not, by their normal operation or foreseeable misuse, cause harm to persons --- a gap that no ex-ante regime currently fills.

\paragraph{Supervisory authority guidance on agentic AI and GDPR.} The regulatory perimeter analysis above has been partially operationalised at Member State level. On 18~February~2026, the Spanish Data Protection Authority (AEPD) published an 71-page guidance document specifically addressing GDPR obligations for agentic AI deployments~\cite{aepd2026}. It is the first EU supervisory authority guidance to treat the agentic architecture---not merely AI outputs---as the primary object of data protection analysis. Three elements of the AEPD's analysis are directly relevant to the compliance architecture in Section~\ref{sec:practical}.

First, on controller attribution: the AEPD confirms that technological autonomy does not displace legal responsibility. While an agent may autonomously perform data-handling operations, the processing remains legally attributable to the controller that deploys the system and determines its purposes and essential means. The practical implication is that any provider who deploys an agent on behalf of an enterprise client must have a documented determination of whether that provider is acting as controller, processor, or joint controller---and that determination is a legal analysis, not a technical one.

Second, on memory and purpose limitation: the AEPD flags persistent agent memory as a high-risk compliance surface, noting that memory must be compartmentalised between processing activities and users, subject to strict retention periods, and technically designed to support data subject rights including erasure. This directly intersects with the runtime behavioral drift analysis in Section~\ref{subsec:agent_challenges}: if memory accumulation drives behavioral change, and that memory also contains personal data subject to GDPR minimisation and purpose limitation, the data governance obligation and the conformity assessment obligation converge on the same architectural requirement---bounded, auditable, erasable memory.

Third, the AEPD adopts the ``rule of~2'' heuristic for agentic risk 
assessment: an agent should not simultaneously combine all three of the 
following without human oversight---processing untrusted input, accessing 
sensitive data, and taking autonomous action affecting individuals. The 
concept originates with Simon Willison's ``lethal trifecta,'' who 
identified these three properties as the conditions enabling data 
exfiltration via prompt injection~\cite{willison2025lethal}. Meta's 
security framework operationalised this in agent-specific form on 
31~October 2025, extending Willison's framing by replacing ``external 
communication'' with the broader ``changing state,'' and explicitly 
acknowledging that combining all three properties without human oversight 
constitutes an unacceptable prompt injection risk~\cite{meta2025agentsecurity}. 
The AEPD's contribution is applying this engineering heuristic as a 
GDPR-grounded governance criterion. This operationalises in practical 
engineering terms the principle that the paper identifies at the level 
of Article~14 oversight commensurateness.
Simultaneously, the Dutch Data Protection Authority warned in February~2026 that highly autonomous agents with broad system access introduce serious security and data protection risks for which the deploying organisation remains fully accountable~\cite{dutchdpa2026}. The convergence of two supervisory authorities within days of each other on the same accountability framing signals that DPA enforcement posture on agentic GDPR questions is consolidating ahead of AI Act high-risk enforcement in August~2026.

\subsection{Scope and Definitions of the Legislative Ecosystem}

Before mapping the specific operational triggers, it is essential to define the regulatory scope of the core non-AI Act frameworks that constitute this ecosystem. Understanding these boundaries is critical, as AI agents frequently traverse them:

\begin{itemize}
    \item \textbf{General Data Protection Regulation (GDPR):} The foundational EU framework for data privacy. It applies to organisations established in the EU when they process personal data, and to organisations outside the EU when they process personal data of individuals who are in the EU in the context of offering them goods or services or monitoring their behaviour. This includes scenarios where personal data is used to train AI agents and where these agents infer or otherwise process personal data to achieve the pursued objective.
    \item \textbf{Data Act (Reg. 2023/2854)~\cite{dataact}:} Applicable since September 2025, this governs the access, sharing, and portability of data generated by connected IoT devices and their ``related services.'' 
    \item \textbf{Digital Services Act (DSA):} Regulates online intermediaries, hosting services, and online platforms, mandating content moderation transparency and accountability.
    \item \textbf{Cyber Resilience Act (CRA):} Introduces mandatory, lifecycle cybersecurity requirements for ``products with digital elements'' (hardware and standalone software placed on the EU market with network connectivity). 
    \item \textbf{Revised Product Liability Directive (PLD)~\cite{pld}:} Taking effect in late 2026, it expands strict liability to digital products, explicitly holding economic operators liable for damage caused by defective AI systems.
    \item \textbf{NIS2 Directive (Dir. 2022/2555):} Imposes strict cyber risk-management on critical infrastructure sectors (e.g., energy, transport, healthcare). 
    \item \textbf{Data Governance Act (DGA):} Regulates ``data intermediation services'' and promotes voluntary data sharing across the EU.
\end{itemize}

\subsection{Anticipated Frequency of Regulatory Triggers}

While the legislative ecosystem outlines a comprehensive theoretical boundary, the practical frequency with which these non-AI Act frameworks apply is highly uneven. Based on contemporary deployment architectures, we anticipate a distinct hierarchy of regulatory activation\footnote{This assessment represents our interpretation of current architectural trajectories in agentic AI. 
(1) \textit{Ubiquitous (Almost 100\%):} GDPR is practically unavoidable, as agents are trained on datasets that more often than not contain personal data and inherently process unstructured prompts, web data, and interaction logs, making strict purpose limitation and data minimisation highly challenging. 
(2) \textit{Highly Probable:} The DSA applies broadly to the dominant Software-as-a-Service (SaaS) and platform-based agent architectures, particularly where agents moderate content, synthesize reviews, or act as user-facing intermediaries. 
(3) \textit{Emerging/Transitional:} The CRA and Data Act are currently less frequently triggered by pure cloud-based agents but will surge in relevance as the industry shifts toward ``Edge AI''—embedding agents into local devices, smart hardware, and distributable software packages. 
(4) \textit{Low Frequency, High Impact:} Sector-specific frameworks (NIS2, MDR) and the revised PLD act as strict boundary conditions; while not triggered in daily operations of general-purpose agents, they impose severe liabilities when crossed or when an agent causes actionable harm.}.
Cloud-based agents overwhelmingly trigger horizontal frameworks like the GDPR and DSA. Conversely, as agentic architectures increasingly migrate toward the ``edge'' (local execution on hardware and IoT ecosystems), providers will see a sharp, unavoidable increase in CRA and Data Act obligations.

The trigger scenarios in Table~\ref{tab:triggers} are consolidated into a
full impact matrix in Appendix~\ref{app:matrix}.

\paragraph{Legislation directly implicated.}

\renewcommand{\arraystretch}{1.25}
\begin{longtable}{L{2cm}L{5.5cm}L{7.5cm}}
\caption{Legislation directly implicated for AI agent providers.}\label{tab:legislation}\\
\toprule
\textbf{Instrument} & \textbf{Trigger for AI Agent Providers} & \textbf{Interaction with AI Act} \\
\midrule
GDPR (Reg. 2016/679) & Whenever agents are trained on personal data or process personal data in operation, e.g. in prompts, outputs and logs. & 
Art.~10(1)-(4) establishes data governance and data quality obligations relevant for personal data protection. Art.~10(5) addresses use of special category data for bias detection and correction. Data protection is woven into essential requirements. \\
\midrule
Data Act (Reg. 2023/2854) & If the agent is a connected product or a related service generating data through user interaction. & Governs access rights to data generated by products/services. Applicable since September 2025. \\
\midrule
DSA (Reg. 2022/2065) & If the agent operates as or within an intermediary, hosting, or platform service. & Recital~118: systemic risk compliance under DSA creates presumption of AI Act conformity for that area. Art.~9(10) permits integrated risk management. VLOP/VLOSE (45M+ EU users): mandatory systemic risk assessment. \\
\midrule
CRA (Reg. 2024/2847) & If the agent is a product with digital elements or standalone software placed on the EU market with network connectivity\footnote{It is important to distinguish between AI software in general and a ``product with digital elements'' as strictly defined by the EU Cyber Resilience Act (CRA). Under Article 3 of the CRA, this refers to any software or hardware product and its remote data processing solutions, placed on the market, whose intended or foreseeable use includes a direct or indirect logical or physical data connection to a device or network. Consequently, an AI agent provided purely as a web-based service (SaaS) may fall outside the CRA's scope unless it is embedded into distributable software or connected hardware.}. & prEN~18282 states CRA conventional cybersecurity outside its scope except AI-specific. CRA and AI Act apply in parallel. M/606: 41 standards. Vulnerability reporting from Sep 2026. \\
\midrule
PLD (Dir. 2024/2853) & If the agent causes damage through defective output or behaviour. & Revised PLD explicitly covers software and AI. Non-compliance with Art.~15 (accuracy) is strong evidence of product defect. Strict liability on producer. \\
\midrule
NIS2 (Dir. 2022/2555) & If the agent is deployed by or serves essential/important entities under NIS2. & Cybersecurity risk management and incident reporting overlap with AI Act cybersecurity. Supply chain security for AI suppliers. \\
\midrule
DGA (Reg. 2022/868) & If the provider operates data intermediation services or data altruism organisations. & In force since September 2023. Omnibus proposes consolidation into Data Act framework (not adopted). \\
\midrule
Sectoral & Medical: MDR/IVDR. Financial: MiFID~II, PSD2, DORA. Transport: EASA. & prEN~18286 acknowledges sector-specific QMS requirements apply in parallel. AI Act obligations sit alongside, not replace, sector requirements. \\
\bottomrule
\end{longtable}

\paragraph{Concrete regulatory trigger scenarios.}

Abstract legal triggers need grounding. Table~\ref{tab:triggers} translates each instrument into a concrete agent action, making explicit exactly why the legislation is activated and what obligation follows. This is the operational bridge between legal analysis and engineering implementation.

\renewcommand{\arraystretch}{1.25}
\begin{longtable}{L{1.5cm}L{4.5cm}L{4.5cm}L{4.5cm}}
\caption{Concrete regulatory trigger scenarios mapping agent actions to legislation.}\label{tab:triggers}\\
\toprule
\textbf{Law} & \textbf{Concrete Agent Scenario} & \textbf{Why Triggered} & \textbf{Key Obligation} \\
\midrule
GDPR & Customer service agent receives personal data in query, queries CRM, logs conversation, sends confirmation email. & Name, order number, address, payment data = personal data. CRM query = processing. & Lawful basis (Art.~6). Purpose limitation. Data minimisation in logs. DPIA if systematic profiling. \\
\midrule
Data Act & Smart home agent monitors IoT sensors, compiles energy usage report, presents optimisation recommendations. & Smart meter and thermostat = ``connected products.'' Agent = ``related service.'' & Data access rights for user. Portability to third parties on request. \\
\midrule
DSA & Content moderation agent flags post, demotes in feed, sends notification. Marketing agent publishes AI-generated reviews. & Both operate within intermediary/hosting/platform service. & Statement of reasons (Art.~17). Recommender transparency. Systemic risk assessment if VLOP. \\
\midrule
CRA & Coding agent sold as standalone software (VS Code extension, CLI tool) with network connection. & Agent = ``product with digital elements'': software placed on market with network connectivity. & Secure-by-design. Vulnerability reporting from Sep 2026. SBOM. Full CRA conformity by Oct 2027. \\
\midrule
PLD & Financial advisory agent recommends selling based on stale data (RAG caching bug). Client incurs significant loss. & Agent output caused measurable harm. Stale data as current falls below legitimate safety expectations. & Strict liability. Art.~15 non-compliance = strong evidence of defect. \\
\midrule
MDR & Clinical agent analyses patient notes, suggests differential diagnosis via FHIR API, outputs ranked conditions. & Intended purpose includes diagnosis suggestion = medical device under MDR. & Full MDR conformity assessment. AI Act Chapter~III in parallel. ISO~13485. GDPR Art.~9. \\
\midrule
NIS2 & IT operations agent for energy utility monitors SCADA, autonomously remediates via OT API access. & Energy utility = ``essential entity.'' Autonomous OT access triggers NIS2. & 24h incident reporting. Supply chain security. Cybersecurity risk management. \\
\midrule
DGA & Agent platform offers data marketplace: users contribute interaction data for training other agents. & Platform = ``data intermediation service'' (DGA Chapter~III). & Notification. Structural separation. Neutrality. GDPR for any personal data. \\
\bottomrule
\end{longtable}

The pattern across all eight scenarios is consistent: the regulatory trigger is never determined by the agent's internal architecture but by its external effects. Which data does the agent process? (GDPR.) Which products does it interface with? (Data Act, CRA.) Where does it publish content? (DSA.) Which sector does it operate in? (MDR, MiFID~II, NIS2.) What harm can its outputs cause? (PLD.) This means the provider's foundational compliance task is not architectural classification but an exhaustive inventory of the agent's external actions, the data it touches, the systems it connects to, and the persons whose rights it affects. That inventory constitutes the regulatory map. An analysis of insurance and liability mechanisms for frontier AI demonstrates that this multi-instrument exposure has a further consequence: the insurance market is bifurcating between traditional insurers retreating through broad AI exclusions and specialty carriers covering only narrow, tractable risk categories, leaving the most consequential risks---systemic, catastrophic, and novel---outside market-based governance entirely~\cite{nannini2026insurance}.

\subsection{CRA Harmonised Standards: A Parallel Track (M/606)}

The Cyber Resilience Act has its own standardisation programme, distinct from M/613, and its impact on AI agent providers is more substantial than the parallel-track framing suggests. Mandate M/606~\cite{m606}, accepted by CEN, CENELEC, and ETSI in April 2025~\cite{cencenelec_cra}, covers 41 harmonised standards organised in a Type~A/B/C hierarchy (framework, horizontal, and vertical standards) whose internal deadlines create a compliance landscape considerably more complex than a single ``full application'' date implies.\footnote{The three tiers are: \textbf{Type~A} (1~standard: EN~40000-1, cyber resilience principles, led by CEN-CLC/JTC~13 WG~9, deadline 30~August 2026; draft prEN published December~2025); \textbf{Type~B} (14~standards: product-agnostic technical measures translating Annex~I essential requirements into testable specifications, led by JTC~13 WG~9, with vulnerability handling due 30~August 2026 but the remaining 13~standards due \textit{30~October 2027}); and \textbf{Type~C} (26~vertical standards for specific product categories, split across ETSI TC CYBER EUSR, CEN/TC~224 WG~17, CLC/TC~65X, and CLC/TC~47X, all due 30~October 2026). The work programme is published on the CEN-CENELEC eNorm platform and the STAN4CRA.eu coordination site. The draft standards themselves are not publicly accessible.} ENISA and the JRC have jointly published a requirements-to-standards mapping for the CRA that identifies gaps in existing standards coverage~\cite{enisa_cra_mapping}.

\paragraph{The conformity presumption bridge.} The relationship between the CRA and the AI Act is not merely parallel: CRA Article~12 creates a direct presumption with operational consequences that the paper's compliance architecture must account for. If a product with digital elements that is also classified as a high-risk AI system under Article~6 satisfies the CRA's essential cybersecurity requirements (Annex~I, Parts~I and~II), it is \textit{deemed to comply} with the AI Act's cybersecurity requirements under Article~15. This is not a loose alignment: CRA compliance \textit{substitutes} for AI Act cybersecurity compliance on the covered requirements. The conformity assessment procedure under the AI Act (Article~43) generally takes precedence, meaning AI Act notified bodies assess both AI Act and CRA cybersecurity requirements in a single procedure. The exception is significant: for products classified as ``important'' (CRA Annex~III) or ``critical'' (CRA Annex~IV), \textit{CRA assessment procedures take precedence} on cybersecurity aspects, even when the AI Act's internal control procedure would otherwise apply.\footnote{The override hierarchy operates as follows. Default: AI Act procedure (Art.~43) covers both regulations. Override: if the product is CRA Annex~III ``important'' and subject to Art.~32(2)/(3) CRA, or CRA Annex~IV ``critical'' and requires a European cybersecurity certificate, then CRA procedures govern cybersecurity aspects while all other AI Act requirements (accuracy, robustness, transparency, human oversight) remain under the AI Act's internal control procedure. This means an AI agent sold as a standalone software product that qualifies as both a high-risk AI system and a CRA ``important'' product faces a split conformity assessment: AI Act internal control for most essential requirements, CRA third-party assessment for cybersecurity.} For agent providers, this means the CRA is not a separate compliance track to be addressed after the AI Act: it is architecturally integrated into the AI Act's conformity assessment, and CRA cybersecurity compliance directly satisfies one of the AI Act's essential requirements.

\paragraph{The missing AI vertical.} The 26 Type~C vertical standards cover browsers, VPNs, password managers, SIEM tools, smart meters, industrial OT devices (based on EN~IEC~62443), firewalls, routers, microcontrollers, operating systems, network management systems, boot managers, and similar product categories.\footnote{By late 2025, vertical drafts had reached ``mature draft'' stage, including standards for browsers, password managers, antivirus software, VPNs, network management systems, SIEM, and boot managers.  The cross-vertical coordination effort (identifying ``common topics'' across all streams) is ongoing but the coherence of parallel drafting across multiple TCs and ESOs has not been independently assessed. As a disclaimer:  ``Mature draft" means at ETSI a stage that it is only applicable in the development of harmonized standards and is an intermediary stage between "Stable Draft " and "Final Draft", to signal that the draft is ready for HAS Assessment by the EC.} There is no vertical standard for AI agents, AI systems, or LLM-based software products\footnote{
Although AI risks are being considered in the verticais under development at ETSI CYBER-EUSR, at the time of writing this article, the working group had reached consensus to exclude, from the scope of EN 304617, browsers featuring AI agent functionality. The principle of Relevance in standardization requires that Standards reflect the best practices from industry and academia and considering the novelty of this technology and how it is rapidly evolving, there is simply not enough experience nor the current state of the art is stable enough to be standardized.
}. An AI coding agent sold as a VS Code extension, a CLI tool, or an API-accessible service would need to self-assess against the horizontal standards (Type~A and Type~B), which translate the CRA's generic Annex~I requirements into testable specifications. These horizontals address conventional product cybersecurity failure modes: secure-by-default configuration, authentication and access control, cryptographic implementation, data minimisation, firmware integrity, and vulnerability handling. They do not address AI-specific failure modes: data poisoning, prompt injection, model manipulation, adversarial inputs, privilege escalation through tool chains, or the emergent behaviours documented by Shapira et al.~\cite{shapira2026agents}. Kim et al.'s survey quantifies this gap systematically: the agentic AI attack landscape encompasses five distinct threat categories---prompt injection and jailbreaks, autonomous cyber-exploitation and tool abuse, multi-agent and protocol-level threats, interface and environment risks, and governance and autonomy concerns---none of which are addressed by conventional product cybersecurity standards designed for firmware integrity or cryptographic implementation~\cite{kim2026attack}. This is precisely the gap the AI-specific cybersecurity standard prEN~18282 (under M/613) is designed to fill: it is scoped to cover AI-specific threats that conventional product cybersecurity standards do not address. But the two standard sets---M/606 horizontals and M/613 AI-specific---are developed by different committees (CEN-CENELEC JTC~13 WG~9 for CRA horizontals; Other CEN-CENELEC committees and ETSI for verticals, i.e., CEN TC 224, CENELEC TC 47X, CENELEC TC 65X and ETSI CYBER-EUSR for CRA verticals; JTC~21 for AI-specific cybersecurity) on different timelines, with no formal coordination mechanism and no published mapping between their respective requirements. An AI agent provider must therefore satisfy \textit{both} sets simultaneously without authoritative guidance on how they interact at the requirements level.

\paragraph{The standards-free zone.} The timeline compression across CRA obligations creates a period of particular difficulty for AI agent providers. CRA vulnerability reporting obligations (Article~14) apply from 11~September 2026: manufacturers must report actively exploited vulnerabilities within 24~hours and severe incidents within 72~hours to the designated CSIRT and ENISA. Full CRA product requirements apply from 11~December 2027. But the Type~B product-agnostic technical standards---the specifications that give the Annex~I essential requirements their concrete, testable content---are not due until 30~October 2027, approximately six weeks before full CRA application. This means AI agent providers face a period where CRA requirements are enforceable but the harmonised standards are not available. Even then, presumption of conformity will not be confirmed for 6-12 months after the publication of the standard. Without harmonised standards confirmed in the OJ, the provider cannot rely on the presumption: conformity must be demonstrated directly against the essential requirements, which are stated at a level of generality (``appropriate level of cybersecurity'') that provides minimal engineering guidance. Simultaneously, the AI Act's M/613 harmonised standards are themselves incomplete (Q4~2026 target for the most advanced, later for others). 

The result is a convergence zone---approximately mid-2026 to late 2027---in which AI agent providers face enforceable CRA and AI Act requirements without finalised harmonised standards under either programme. This is not a theoretical risk: it is the direct consequence of the legislative timelines adopted and the standardisation delivery schedules committed to. Providers building compliance architectures now must design for this interregnum: using the available draft standards as orientation---a practice explicitly endorsed in legal compliance guidance as prudent given the standards-free zone~\cite{hoganlovells2026cra}---while accepting that the presumption of conformity those standards will eventually provide is not yet legally operative\footnote{For the ETSI CYBER-EUSR vertical standards specifically, ETSI has made all drafts available under Open Consultation to the public since November 2025, providing industry with continuous access to their content and evolution ahead of formal publication; they can be accessed at \url{https://docbox.etsi.org/CYBER/EUSR/Open}.}. 

\subsection{The Digital Omnibus: Agent-Specific Implications}

The Digital Omnibus package published on 19~November 2025 (COM(2025)~836 and COM(2025)~837) responds to the Draghi Report on EU competitiveness and aims to reduce administrative burdens by 25\% for all businesses and 35\% for SMEs by 2029~\cite{mofo2025omnibus}. Its five AI-relevant proposals each carry distinct consequences for agent providers that go beyond the headline simplification narrative.

\paragraph{The stop-the-clock mechanism and the agentic gap.} The central proposal links high-risk AI rules to the availability of sufficient tools for compliance, such as harmonised standards\footnote{While this is generally understood to mean harmonised standards, it can also mean EC guidance or common specifications}: once the Commission confirms availability of such tools via an implementing decision, the essential requirements apply 6~months later for Annex~III systems and 12~months for Annex~I product components, potentially pushing full application to late 2027~\cite{eprs2026omnibus}. For agent providers, the relevant question is which tools the Commission will declare ``available.'' The current M/613 programme covers general AI system requirements (risk management, cybersecurity, data governance, trustworthiness), and prEN~18282 and prEN~18229-1 contain agent-specific provisions on privilege minimization and oversight evasion. But the standards do not cover three agentic-specific risk categories identified in Section~6: multi-agent orchestration (where agents delegate to other agents, creating recursive accountability chains), cross-session state accumulation (where memory persistence shifts the system's operational profile beyond what was conformity-assessed), and dynamic tool discovery (where an agent connects to tools not in its original catalogue). If the Commission declares the current standards ``available'' without these gaps being filled, the clock restarts for agent providers, but the presumption of conformity those standards confer does not cover the very risks that make agentic systems distinctive. The delay helps in form but not in substance: the provider still faces non-standard risks that must be addressed through internal technical specifications without the benefit of a harmonised standard's presumption.

\paragraph{Article 4a: bias detection with agent interaction data.} The proposal introduces a new Article~4a extending the legal basis for processing special categories of personal data for bias detection to all providers and deployers, not only high-risk system providers, and lowers the threshold from ``strictly necessary'' to ``necessary''~\cite{mofo2025omnibus}. For agent providers, this expansion has a specific operational dimension: agents accumulate interaction data across sessions---conversation histories, user preferences, action outcomes, feedback signals---that may contain protected characteristics indirectly (e.g., language patterns correlating with ethnicity, scheduling patterns correlating with religious observance, complaint frequency correlating with disability). The current Article~10(5) restricts bias-detection use of such data to high-risk system providers under six cumulative conditions; the proposed Article~4a maintains all these conditions but would permit such use for any AI system provider, including providers of general-purpose agent platforms deployed in non-high-risk contexts. The GDPR safeguards remain unaffected: pseudonymisation, data minimisation, DPA oversight, and a prohibition on repurposing the data for any other objective. But the lowered threshold materially changes the compliance calculus for agent providers who currently must either forgo bias detection on sensitive attributes or argue that their system qualifies as high-risk to access the Article~10(5) exception.

\paragraph{DGA consolidation and data intermediation agents.} The broader Digital Legislation Omnibus (COM(2025)~837) proposes repealing the DGA (Regulation~2022/868) outright and relocating its substantive obligations into the Data Act framework, alongside the Free Flow of Non-Personal Data Regulation and the Open Data Directive. For agent platforms that operate data marketplace functions---the DGA scenario in Table~\ref{tab:triggers}, where users contribute interaction data for training other agents---this restructuring matters operationally. The DGA's current Chapter~III requires notification to competent authorities, structural separation (the intermediary must be a separate legal person from any entity using the data), and neutrality (no conditional pricing that disadvantages data holders). If these obligations migrate to the Data Act, the competent authority changes (from DGA-designated authorities, many of which remain undesignated, to Data Act enforcement bodies), the notification procedure changes, and the structural separation requirements may be modified. Ten Member States have already received reasoned opinions from the Commission for failing to designate DGA competent authorities~\cite{ec2024dga_infringement}; consolidation into the Data Act framework would bypass this non-implementation problem but would also create a transitional period during which existing DGA registrations have uncertain legal status. Providers operating data intermediation services through agent platforms should track this consolidation closely: the obligations survive, but the compliance address changes.

\paragraph{Incident reporting streamlining.} The Omnibus proposes a single-entry-point mechanism for incident reporting across NIS2, GDPR, and DORA. This directly affects agent providers operating in regulated sectors. Currently, a single agent incident---for example, a healthcare agent that exposes patient data through a compromised API connection---could trigger three parallel reporting obligations with different timelines: NIS2 requires a 24-hour early warning to the CSIRT, GDPR requires notification to the supervisory authority within 72~hours, and DORA (if the deployer is a financial entity) imposes its own timeline. Each obligation requires a different report format, is submitted to a different authority, and triggers different follow-up procedures. The Omnibus proposes that a single notification to one authority satisfies all applicable reporting obligations. If adopted, this reduces administrative burden substantially for agent providers whose systems span multiple regulatory domains, but the operative word is ``if'': the single-entry-point mechanism requires implementing legislation to specify which authority receives the notification, how it is routed to other competent authorities, and whether the content requirements of each underlying regulation are satisfied by a single template. Until that implementing legislation exists, the simplification is aspirational.

\paragraph{Political dynamics and provider strategy.}
The legislative process has advanced materially since the Commission's 
November~2025 proposal. The Council of the EU adopted its negotiating 
mandate on 13~March~2026~\cite{council_omnibus2026}. The European 
Parliament's committees IMCO and LIBE adopted a joint negotiating 
position on 18~March~2026 by 101 votes to 9, with 8 abstentions 
(report A10-0073/2026)~\cite{ep_omnibus2026}, and the Parliament 
validated that mandate in plenary on 26~March~2026~\cite{ep_plenary2026}. 
Trilogue negotiations between Parliament, Council, and Commission are 
expected to begin in April~2026, with final adoption anticipated by 
mid-2026 under pressure from the August~2026 enforcement deadline.

The two co-legislators have converged on several points that are 
sufficiently aligned to anticipate their survival in the final text. 
Both replace the Commission's conditional stop-the-clock mechanism 
with fixed application dates: Annex~III high-risk obligations 
(biometrics, employment, education, etc.) from 2~December~2027, and 
Annex~I high-risk obligations (AI embedded in regulated products) 
from 2~August~2028~\cite{ep_omnibus2026, council_omnibus2026}. This 
provides greater legal certainty than the Commission's trigger 
mechanism while extending the enforcement horizon by approximately 
one year relative to the original AI Act timeline. Both positions 
also reject the Commission's proposal to eliminate the registration 
obligation for providers who self-assess their system as non-high-risk 
under Article~6(3): both reinstate the obligation while agreeing to 
streamline the content requirements~\cite{ep_omnibus2026}. The 
Article~4a expansion of bias-detection data processing rights to all 
providers and deployers---not only high-risk system providers---is 
supported by both positions~\cite{ep_omnibus2026, council_omnibus2026}.

One divergence carries structural implications for the compliance 
architecture in this paper. The Parliament's position proposes moving New Legislative Framework products (including medical devices and machinery) from Annex~I Section~A to Section~B, making sectoral conformity assessment the primary pathway with only a reduced 
set of AI Act provisions applying directly~\cite{cdt_march2026}. If this survives trilogue, the Annex~I dual-track pathway analysed in Section~\ref{sec:classification} --- under which clinical agents face simultaneous MDR and AI Act obligations --- would be substantially simplified, with sectoral legislation bearing the primary compliance burden and the AI Act's role reduced to a supplementary set of requirements. The Council's position does not adopt the same exclusion in full; the final scope of this carve-out is therefore a live trilogue question. Providers in regulated sectors (healthcare, automotive, aviation) should track this development closely: it could materially reduce their AI Act conformity assessment burden, or leave it intact, depending on the outcome.

A further noteworthy change: the Parliament shortens the grace period 
for machine-readable marking of AI-generated content under 
Article~50(2) to 2~November~2026, more restrictive than both the 
Commission's proposal (2~February~2027) and the Council's 
position~\cite{ep_omnibus2026}. Agent providers generating synthetic 
content should design for the earlier Parliament date as a planning 
assumption pending trilogue resolution.

The practical strategy for providers remains that stated earlier in 
this section: build compliance architectures on the current legal 
baseline --- the essential requirements in the adopted Regulation 
cannot be amended through the Omnibus --- while designing with 
modularity to absorb the changes that now have strong co-legislative 
support. Concretely: the twelve-step compliance sequence in 
Section~\ref{sec:practical} remains valid in its entirety; the 
application dates in Steps~10 and~11 in the following Section should be maintained as 
August~2026 for planning purposes, with December~2027 as the 
operationally realistic enforcement horizon now supported by both 
co-legislators.

\section{Practical Compliance Architecture}
\label{sec:practical}

\subsection{Recommended Compliance Sequence}

Based on the analysis of the harmonised standards and the parallel regulatory mapping required by adjacent legislation, we propose a twelve-step compliance sequence (see Figure~\ref{fig:sequence} for a process-flow overview). The ordering reflects logical dependencies: classification must precede QMS establishment, which must precede risk management, which informs all subsequent steps. However, some steps will overlap or run in parallel, such as establishing the QMS and risk management. Several steps present agent-specific decision points that do not arise for conventional AI systems; these are identified below.

\begin{description}[style=nextline, font=\normalfont\bfseries, leftmargin=1.5cm, labelindent=0cm]
\item[Step 0:] \textbf{Scope the system under Article~3(1).} Determine whether the product constitutes an AI system. For agents, the scoping question is rarely whether the product meets the definition---any LLM-based system with tool use satisfies every element of Article~3(1), as established in Section~1---but rather \textit{how many AI systems} the product contains. An orchestrator agent that delegates to three specialist sub-agents (one for retrieval, one for code generation, one for email drafting) could constitute one system or four, depending on whether the sub-agents are independently placed on the market or function only as internal components\footnote{The provider-identification 
question is compounded when the agent is assembled from discrete components developed by different legal persons: a foundation model provider, an orchestration framework developer, and an enterprise deployer applying custom configuration. Article~25's substantial-modification and own-name-or-trademark triggers may apply at multiple layers simultaneously, requiring each participant in the assembly chain to independently assess whether they acquire provider status before the system reaches the end user.}. This distinction determines whether the provider faces one conformity assessment or multiple, and whether each sub-agent must independently satisfy the essential requirements. If the product is not an AI system (this might be likely unfeasible and, even so, pertaining then to RPA e.g., a deterministic rule-based automation), the AI Act does not apply, but CRA, GDPR, and DSA may still apply independently.

\item[Step 1:] \textbf{Map the GPAI layer.} Determine whether the agent incorporates a GPAI model and identify who bears the GPAI obligations (the agent provider, if it develops the model; the upstream supplier otherwise). Establish the Article~53 documentation chain. For agents built on third-party foundation models, this step requires obtaining the upstream provider's technical documentation and integrating all available information on model limitations obtained via the Article~53 documentation chain into the system-level risk assessment, noting that the provider's Article~9 risk management obligation is not bounded by what the upstream documentation discloses. The Commission's GPAI guidelines clarify that downstream actors using more than one-third of original training compute for fine-tuning become providers themselves; agent developers who fine-tune foundation models for domain-specific tool use must assess whether their training regime crosses this threshold~\cite{ec_gpai_guidelines}.

\item[Step 2:] \textbf{Classify the system.} Determine whether high-risk (Annex~III use case or Annex~I product component) or not. Document the classification reasoning.
For general-purpose agent platforms, this step contains the structural dilemma identified in Section~3: if the provider cannot predict how deployers will use the platform, it must either restrict the intended purpose contractually and technically (e.g., explicitly excluding use in employment, credit, or healthcare via both terms of service and API-level enforcement) or design for the most demanding regulatory tier foreseeable. The classification decision must be documented with sufficient specificity to survive regulatory scrutiny: a generic statement that ``the system is not intended for high-risk use'' is insufficient if the platform's tool-calling architecture makes Annex~III deployments technically trivial for downstream deployers.

\item[Step 3:] \textbf{Establish the QMS (prEN~18286).} Identify all applicable essential requirements (clause~4.4.2 as of January 2026 Enquiry draft). Select compliance approaches for each (clause~4.4.3). If sector-specific QMS exists (ISO~13485 for medical devices, IATF~16949 for automotive), integrate rather than duplicate. For agent providers, the QMS must cover the full lifecycle including post-deployment behavioral monitoring---a scope extension that conventional QMS implementations rarely address, since traditional software products do not exhibit adaptive behavior after release.

\item[Step 4:] \textbf{
Establish the Risk Management System
(prEN~18228).} Article~9 mandates a continuous lifecycle risk management process covering health, safety, and fundamental rights. 
For agents, three dimensions of Article~9 warrant specific attention. First, Article~9(2)(a) requires identification of risks associated with the system's known and foreseeable characteristics; since Article~3(1) defines AI systems in part by their varying levels of autonomy, the automation boundary is a system characteristic the risk assessment must address: the provider must specify which actions require human confirmation, which execute autonomously, and what criteria govern escalation
Second, Article~9's requirement to address risks to fundamental rights is not a formality: it requires substantive assessment of how the agent's actions could affect dignity, privacy, non-discrimination, freedom of expression, access to justice, etc. Any fundamental rights under the Charter that are placed at risk based on the AI system's intended purpose and reasonably foreseeable misuse. The standards under M/613 are expected to operationalise this by providing structured methodologies, potentially referencing the EU Charter of Fundamental Rights and international frameworks such as HUDERIA's from the Council of Europe\footnote{The Council of Europe provides practical resources for this purpose. 
The HUDERIA (Human Rights, Democracy, and the Rule of Law Impact 
Assessment) methodology, developed under the Framework Convention on 
Artificial Intelligence (CETS No.~225), structures assessment across 
seven normative dimensions: human rights and fundamental freedoms, 
democracy and rule of law, health, safety and security, privacy and 
data protection, transparency and explainability, accountability and 
oversight, and social and environmental wellbeing~\cite{huderia2025}. 
These dimensions can be operationalised at the action-ontology level 
described in Section~3: each decision type maps to the specific impact 
dimensions it may activate, and this mapping feeds the risk assessment 
for individual action instances at runtime.} and the UN Guiding Principles on Business and Human Rights.
Most engineering teams lack competence in fundamental rights analysis. prEN~18228 will contain valuable information about how to address risks to fundamental rights. That said, it will also require provider competence in those fundamental rights placed at risk.

\item[Step 5:] \textbf{Implement data governance (prEN~18284 + prEN~18283).} Training, validation, and test data lifecycle management. Bias management through normative reference to prEN~18283. The practical challenge for agent providers is that agentic systems generate new data continuously through their interactions---conversation logs, tool invocation records, user feedback---and this interaction data may feed back into the system's behavior through in-context learning, RAG updates, or explicit fine-tuning\footnote{Feedback loops are explicitly mentioned in Article 15 of the EU AI Act.}. The data governance framework must therefore address not only the initial training data but the ongoing data flows that shape the agent's operational profile, including procedures for determining when interaction data crosses from operational logging into training data territory. In addition, for agents whose development and (or) operation involves processing personal data (which, as Table~\ref{tab:taxonomy} demonstrates, is nearly all of them), the agent providers' data governance activities need to structurally integrate steps to comply with applicable data protection laws, such as the GDPR.

\item[Step 6:] \textbf{Design for trustworthiness (prEN~18229-1/2 + prEN~ISO/IEC~24970).} Implement logging, transparency, and human oversight as required by Articles~12--14. For agents, the central design decision is the \textit{automation boundary}: which actions execute autonomously, which require human confirmation, and which are irreversible. Article~14 requires that oversight measures be commensurate with the risks and context, and the standards are expected to operationalise this by requiring that this boundary be configurable by the deployer, since the acceptable level of autonomy varies by domain and risk tolerance. An agent sending a marketing email is unlikely to create an unacceptable risk if it executes autonomously; the same agent autonomously sending communication placing at risk fundamental rights is significantly more problematic\footnote{The commensurability requirement of Article~14 implies that the 
automation boundary cannot be a static, system-wide configuration. It 
must be determined at the granularity of individual action instances, 
reflecting both ontology-inherited properties and runtime 
telemetry---model confidence, distributional drift, affected entity 
vulnerability, and cumulative behavioural pattern signals. A 
deployer-configurable threshold that classifies ``all financial 
actions require approval'' is too coarse: it treats a routine 
low-value payment identically to a large-value transaction near a 
sanctions-screening threshold, imposing unnecessary friction on the 
former and providing inadequate escalation control on the latter. The 
deployer must be able to configure the boundary at the level of 
decision type, with the system computing per-instance override signals 
from live telemetry.}. The logging requirements (prEN~ISO/IEC~24970) must capture enough detail to reconstruct the agent's decision chain post hoc---including which tools were invoked, what data was retrieved, what reasoning produced the action, and what the outcome was---without creating a surveillance infrastructure that itself violates the GDPR's personal data minimisation principle.

\item[Step 7:] \textbf{Implement AI-specific cybersecurity (prEN~18282).} Enforce privilege minimization outside the generative model. This is the step where the Article~15(4) requirements identified in Section~6.1 must be operationalised: API-level least-privilege enforcement, per-action authorization scoping, dynamic privilege restriction based on input trust level, and audit logging that distinguishes user-initiated from AI-initiated actions. For agents with multi-tool access, this step also requires deploying non-human identity governance infrastructure: just-in-time credential provisioning, per-action authorization, and credential rotation policies that account for the agent's autonomous operation cycle. The threat landscape these controls must address is documented in Kim et al.'s systematic attack taxonomy~\cite{kim2026attack} and the OWASP Top~10 for Agentic Applications~\cite{owasp2025agentic}; providers should use these as threat-modelling inputs when designing the enforcement architecture for this step.

\item[Step 8:] \textbf{Assess CRA applicability.} If the agent is a product with digital elements (standalone software with network connectivity placed on the EU market), implement CRA compliance in parallel with AI Act compliance. The M/606 programme structures standards into five types: Type~A framework standards under EN~40000 (deadline August 2026, with prEN~40000-1-2 on cyber resilience principles and prEN~40000-1-3 on vulnerability handling already in public enquiry), Type~B product-agnostic technical measures (deadline October 2027) and vulnerability handling (deadline August 2026), and Type~C vertical standards for important and critical product categories (deadline October 2026). The horizontal standards are developed by CEN-CLC/JTC~13 WG~9; vertical standards by ETSI TC CYBER, CYBER-EUSR WG, CEN TC 224, CLC 47X and CLC 65X. No vertical standard addresses AI products specifically. Begin vulnerability reporting preparation: CRA vulnerability reporting obligations apply from 11~September 2026, before full product requirements in 11~December 2027. The Commission published an implementation Q\&A in December 2025, and Implementing Regulation~(EU)~2025/2392 specifies the technical descriptions of important and critical product categories. The practical consequence is that an AI coding agent sold as a VS Code extension or CLI tool must satisfy both CRA secure-by-design requirements and AI Act cybersecurity requirements simultaneously---two standard sets developed by different committees on different timelines, with no published mapping between the EN~40000 horizontal requirements and the prEN~18282 AI-specific requirements.

\item[Step 9:] \textbf{Map adjacent legislation.} This is the step that distinguishes agent compliance from conventional AI system compliance. The provider must inventory all external actions the agent can perform, all data it can access or generate, all systems it connects to, and all natural persons whose rights it can affect. Table~\ref{tab:triggers} provides the mapping template. The inventory must be exhaustive: each external action potentially activates a different legislative instrument. The practical method is to trace each tool in the agent's catalogue through four questions: (1)~does this tool process personal data? (GDPR); (2)~does this tool interface with a connected product or related service? (Data Act); (3)~does this tool publish or moderate content on a platform? (DSA); (4)~does this tool operate within a regulated sector? (MDR, MiFID~II, NIS2, etc.). The union of the answers constitutes the regulatory map.

\item[Step 10:] \textbf{Conformity assessment.} For high-risk systems, the provider must apply either internal control (AI Act Annex~VI) or third-party assessment (Annex~VII). In practice, the vast majority of Annex III AI systems will use internal control, since third-party assessment is required only for biometric systems under Annex~III, point~1. Annex I systems will all require third-party assessment The provider prepares technical documentation (Annex~IV), issues an EU Declaration of Conformity, and registers in the EU database\footnote{Note that the Commission's Digital Omnibus proposal to eliminate the registration obligation for providers who self-assess their system as non-high-risk under Article~6(3) has been rejected by both the Parliament and the Council in their March~2026 negotiating positions~\cite{ep_omnibus2026, council_omnibus2026}. Providers should assume the registration obligation will be retained in the final text, with streamlined content requirements.}. The FPF/OneTrust step-by-step guide provides the most comprehensive public walkthrough of these procedures~\cite{fpf2025conformity}. No notified bodies have been formally designated under the AI Act as of early 2026; high-risk obligations currently apply from August~2026, creating a compressed timeline for the designation process.

\item[Step 11:] \textbf{Post-market monitoring and drift detection.} Continuous risk management. For agents, this step operationalises the drift-detection framework from Section~6.4: versioned snapshots of the agent's operational state (tool catalogue, memory state, policy bindings) at defined intervals; continuous monitoring of behavioral metrics against the conformity assessment baseline; automated detection of drift beyond defined thresholds triggering reassessment; and a documented internal procedure for determining whether a detected change meets the Article~3(23) substantial modification threshold. Serious incident reporting (Article~73) applies when the agent causes or contributes to serious harm; the provider must have procedures to detect such incidents, assess causation, and report within the mandatory timeline.
\end{description}

\paragraph{Compliance boundary map.}

Table~\ref{tab:boundary} summarises the multi-layer obligations facing an AI agent provider. The layers are not independent: GDPR obligations inform the data governance requirements under prEN~18284; CRA cybersecurity requirements overlap with but are distinct from AI-specific cybersecurity under prEN~18282; DSA systemic risk assessment for VLOPs interacts with AI Act risk management under Article~9 (with Recital~118 creating a conditional presumption of conformity). The practical consequence is that the provider's compliance programme cannot be structured as ten parallel workstreams. It must be integrated: the QMS (prEN~18286) is designed to serve as the coordinating framework, with the risk management process (prEN~18228) as the analytical core that identifies which obligations apply, how they interact, and where conflicts must be resolved.

\renewcommand{\arraystretch}{1.25}
\begin{longtable}{L{3.0cm}L{3cm}L{3.7cm}L{5cm}}
\caption{Compliance boundary map. Multi-layer obligations for AI agent providers.}\label{tab:boundary}\\
\toprule
\textbf{Obligation Layer} & \textbf{Instrument(s)} & \textbf{Status} & \textbf{Interaction Pattern} \\
\midrule
AI system essential requirements & AI Act Ch.~III, Sec.~2 + harmonised standards & Mandatory for high-risk. Standards create presumption once cited in OJ. & Overarching framework. Risk management under Art.~9 must integrate all layers below. \\
\midrule
GPAI model obligations & AI Act Ch.~V + GPAI Code of Practice & Mandatory (Art.~53--55). Code creates presumption. & Model-level. Agent provider inherits duty to integrate model limitations into system-level risk assessment. \\
\midrule
Transparency & AI Act Article~50 & Mandatory for all AI systems. & Cross-cutting. Extends to all affected persons, not only direct users (Section~6.3). \\
\midrule
Data protection & GDPR & Mandatory if training or operation involves personal data processing. & Integrated via AI Act Art.~10(1)-(5). Single integrated process, not parallel. \\
\midrule
Product cybersecurity & CRA + M/606 (41 stds) & Products with digital elements. Vulnerability reporting Sep 2026. & Parallel to AI-specific cyber (prEN~18282). Conventional + AI-specific = dual standard sets. \\
\midrule
Platform duties & DSA & If intermediary/hosting/platform. VLOP: systemic risk. & Recital~118 presumption for AI Act conformity on covered risks. Art.~9(10) permits integration. \\
\midrule
Data access & Data Act & Since Sep 2025. Connected products/related services. & Separate compliance track. Interacts with Art.~10 data governance but distinct obligations. \\
\midrule
Data intermediation & DGA & Since Sep 2023. Omnibus proposes consolidation. & Notification, structural separation, neutrality. May migrate to Data Act framework. \\
\midrule
Civil liability & PLD. AI Liability Dir.\ withdrawn. & Ex post. & Non-compliance with essential requirements = strong evidence of defect (Art.~10(2)(b) PLD). \\
\midrule
Simplification & Digital Omnibus (in trilogue) & Not law. & May delay high-risk rules, expand bias detection, streamline reporting. Build for current baseline. \\
\bottomrule
\end{longtable}

\subsection{Implementation Constraints: The Gap Between Standards and Architecture}

The compliance sequence above assumes that the provider can implement each step against its agent architecture. In practice, several requirements face implementation constraints that current agentic architectures are not designed to address.

The first constraint is \textit{behavioral reproducibility}. The essential requirements on accuracy (Article~15) and the conformity assessment process (Annex~VI) presuppose that the system's behavior can be tested, measured, and demonstrated to remain within assessed boundaries. LLM-based agents are stochastic: the same input may produce different tool invocations, different reasoning chains, and different outputs across runs. Temperature settings, context window contents, and model version updates all affect behavior in ways that are difficult to characterise comprehensively. The harmonised standards do not require determinism---prEN~18229-2 specifies performance metrics and testing frameworks, not fixed outputs---but they do require that the provider can demonstrate that behavioral variation remains within the envelope assessed during conformity assessment. For agents with broad tool access and complex reasoning chains, defining that envelope and monitoring compliance against it is an unsolved engineering problem that requires investment in runtime observability infrastructure that most current agent architectures lack. The conformity assessment envelope is not merely difficult to define for agentic systems: it grows combinatorially with tool catalogue size and chain depth, since $k$~tools composed across $n$~sequential steps produce up to~$k^n$ distinct 
execution paths, making exhaustive pre-deployment behavioral characterization structurally infeasible beyond trivial configurations~\cite{acmsurvey2025, kim2026attack}.

The second constraint is \textit{action-chain auditability}. The logging requirements (prEN~18229-1 and prEN~ISO/IEC~24970, operationalising Article~12) require that the provider can reconstruct why the agent took a specific action after the fact. For a single-step tool invocation, this is straightforward: the input, the tool selected, and the output can be logged. For a multi-step action chain---where the agent plans a sequence of actions, executes them with intermediate reasoning, adjusts its plan based on intermediate results, and produces a final output that depends on the entire chain---the logging must capture not only each individual step but the causal relationships between them: why did the agent select this tool rather than that one? Why did it abandon plan A and switch to plan B? What information from step 3 influenced the decision at step 7? Current agent frameworks (LangChain, CrewAI, AutoGen) provide varying levels of trace logging, but none produce audit trails that meet Article~12's requirement for sufficient traceability to assess compliance with the essential requirements. The problem is compounded by the finding that agents may misreport their own actions: Shapira et al. document agents claiming task completion while the underlying system state contradicted those claims~\cite{shapira2026agents}, meaning that logging the agent's self-reported action history is insufficient---the audit infrastructure must independently verify system state, not merely record the agent's account of what it did.

The third constraint is \textit{privilege enforcement at scale}. Section~6.1 established that Article~15(4) compliance for agentic systems requires privilege minimization enforced outside the generative model. For an agent with three tools, implementing API-level least-privilege is a bounded engineering task. For an enterprise agent platform where deployers can connect arbitrary tools via a plugin architecture, the problem scales differently: each new tool connection creates a new privilege surface that must be scoped, audited, and monitored. The Article~15(4) requirement is clear, but the infrastructure to implement it at the scale of a general-purpose agent platform---just-in-time credential provisioning, per-action authorization, dynamic privilege scoping across hundreds of tool integrations---does not exist as an off-the-shelf capability and must be built.

Fourthly, a structurally distinct gap concerns the \textit{tools that agents invoke at 
runtime}. Third-party tool providers---the operators of the APIs, 
databases, and external services that constitute an agent's action 
surface---are neither GPAI model providers nor AI system providers 
under the Act and therefore fall almost entirely outside its regulatory 
categories. Article~25(4) requires written agreements between 
high-risk AI system providers and third-party suppliers specifying 
information, capabilities, and technical access necessary for 
compliance---but this provision presupposes pre-established 
contractual relationships that cannot exist when an agent selects 
tools at runtime from registries or endpoints unknown at the time of 
conformity assessment~\cite{jones2025 }. Article~72(2)'s post-market 
monitoring obligation requires analysis of interaction with other AI 
systems, but the Act provides no mechanism to compel cooperation from 
tool providers selected dynamically, and Recital~88 merely encourages 
tool supplier cooperation without creating binding obligations absent 
prior contractual arrangements~\cite{jones2025 }. The practical 
consequence is that dynamic tool discovery---where an agent identifies 
and connects to tools not in its original catalogue---renders the 
provider's conformity assessment a snapshot of a tool catalogue the 
deployed system may have already left behind. Neither the M/613 
standards nor any published guidance currently addresses this gap.

These constraints do not invalidate the compliance architecture. They define the engineering roadmap that the architecture requires. The provider's obligation is not to have solved these problems before placing the system on the market, but to have addressed them to the degree required by the essential requirements: the risk management process must have identified them, the technical documentation must describe how they are mitigated, and the post-market monitoring system must detect when mitigation is insufficient.

\section{Key Observations and Open Issues}

\textbf{(1) Standards maturity.} Several draft standards remain incomplete. 
A peer-reviewed analysis of this tension concludes that the resulting legal uncertainty may require legislative reform or further Commission guidance to resolve~\cite{deluca2025bias}.\footnote{The EPRS briefing notes that the Digital Omnibus's proposed expansion of the bias detection exception (new Art.~4a) would partially address this gap by extending the legal basis to all providers and deployers, not only high-risk system providers.}

\textbf{(2) Agentic systems are partially but not fully addressed.} The cybersecurity standard (prEN~18282) and the trustworthiness framework (prEN~18229, part 1 and 2), by virtue of their respective scopes under Articles~15(4) and 12--14, necessarily engage with agentic architectures: privilege escalation through tool chains, open-ended code execution, and the oversight challenges created by autonomous multi-step systems. The risk management standard (prEN~18228) and QMS standard (prEN~18286) provide the structural framework but are not expected to contain agent-specific provisions at the same level of specificity. The gap between what the standards cover and what the threat landscape demands is documented empirically: Kim et al.'s USENIX Security 2026 survey identifies five distinct agentic threat categories, of which the standards substantively address only two (prompt injection mitigation and privilege management), leaving multi-agent protocol threats, interface exploitation, and governance-level autonomy concerns without corresponding standard provisions~\cite{kim2026attack}. The OWASP Agentic Security Initiative's Top~10 for Agentic Applications reaches a convergent conclusion from the practitioner side~\cite{owasp2025agentic}. Providers should not treat the current standards as fully sufficient for agentic-specific risks. Supplementary internal technical specifications will be needed, particularly for: multi-agent orchestration (where multiple agents delegate tasks to each other), cross-session state accumulation, and dynamic tool discovery (where an agent identifies and connects to tools not in its original catalogue).

\textbf{(3) Conformity assessment under development.} The conformity assessment standard is referenced in the QMS standard but is a separate work item. The interaction between the provider's internal control procedure (AI Act Annex~VI) and the standards-based presumption of conformity will be operationally decisive. Until the conformity assessment standard is finalised, the precise mechanics of demonstrating compliance through the standards remain underspecified.\footnote{A structural issue underlying this gap warrants noting, even though it is not specific to agentic AI. The QMS standard (prEN~18286) functions as a hybrid: it imposes requirements on a product (the AI system) through a management system instrument. Conformity assessment bodies are accredited across three distinct dimensions---certification of persons, certification of products, and certification of management systems---and a certification scheme must fit within one of these. A standard that is simultaneously a product standard and a management system standard does not map cleanly onto any of the three. How accreditation bodies and conformity assessment bodies will handle this structural mismatch is an open question that the conformity assessment standard under development must resolve. It is also a question that the ISO/CASCO infrastructure for conformity assessment standards (including ISO/IEC~17021 for management system certification and ISO/IEC~17065 for product certification) was not designed to answer in combination.}

\textbf{(4) Fundamental rights are not optional.} Article~9 requires risk management to address fundamental rights, and the standards under M/613 are expected to operationalise this requirement with structured methodologies. Providers must build competence
for those fundamental rights that are placed at risk by the AI system based on its intended purpose and reasonably foreseeable misuse.

\textbf{(5) Dual cybersecurity standards tracks and the standards-free zone.} AI agent providers face two parallel standardisation processes: M/613 (AI-specific, under JTC~21) and M/606 (product-level CRA, under CEN/CENELEC/ETSI). The M/606 programme comprises 41 standards across five types: Type~A framework standards (deadline August 2026), Type~B product-agnostic technical measures and vulnerability handling (deadline October 2026/2027), and Type~C vertical product-specific standards for important and critical product categories (deadline October 2026). The horizontal standards are being developed by CEN-CLC/JTC~13 WG~9 under the EN~40000 family: prEN~40000-1-1 (Vocabulary), prEN~40000-1-2 (Principles of Cyber Resilience), and prEN~40000-1-3 (Vulnerability Handling), with the latter two already in public enquiry as of late 2025. Vertical standards for 25 specific product categories are being developed by ETSI TC CYBER (EUSR) under the EN~304~6xx numbering, covering browsers, password managers, VPNs, network management systems, SIEM, and other product types. The CRA's conformity presumption bridge (Article~12) means these are not independent tracks: CRA cybersecurity compliance substitutes for AI Act Article~15 cybersecurity requirements, and for ``important'' or ``critical'' products, CRA conformity assessment procedures override the AI Act's internal control procedure on cybersecurity aspects. No CRA vertical standard exists for AI products; agent providers must self-assess against horizontal standards designed for conventional product cybersecurity, supplemented by the AI-specific prEN~18282 under M/613, with no published mapping between the two requirement sets. The timeline compression creates a standards-free zone from approximately mid-2026 to late 2027 in which CRA and AI Act requirements are enforceable but harmonised standards under neither programme are finalised.

\textbf{(6) The Digital Omnibus is in trilogue: timelines will shift, 
substance will not.} The Commission's November~2025 proposal has 
advanced through both co-legislators. The Council adopted its 
negotiating mandate on 13~March~2026 and the European Parliament 
validated its mandate in plenary on 26~March~2026~\cite{ep_plenary2026}, 
opening trilogue negotiations expected from April~2026. Both positions 
replace the Commission's conditional stop-the-clock mechanism with 
fixed application dates of 2~December~2027 (Annex~III) and 
2~August~2028 (Annex~I), providing legal certainty but extending the 
compliance horizon by approximately one year~\cite{ep_omnibus2026, 
council_omnibus2026}. Both positions reject removal of the 
Article~6(3) registration obligation and support the Article~4a 
bias-detection expansion. The Parliament's position proposes moving regulated-product AI 
components from Annex~I Section~A to Section~B, which, if it 
survives trilogue, would substantially reduce the AI Act's direct 
obligations for these systems in favour of sectoral 
frameworks~\cite{cdt_march2026}. The essential requirements in the 
adopted Regulation remain fixed and cannot be amended through 
secondary legislation. Providers should build for August~2026 as 
the current legal baseline while planning operationally for 
December~2027.

\textbf{(7) The AI Liability Directive in limbo.} The proposal was formally withdrawn (OJ C/2025/5423, 6~October 2025) \cite{nannini2025less}. The revised Product Liability Directive (2024) already covers AI systems explicitly. A European Parliament commissioned study concludes that neither the revised PLD nor the withdrawn AILD sufficiently addresses AI liability risks and recommends a dedicated strict liability regulation for high-risk AI~\cite{bertolini2025liability}.\footnote{The study (PE~776.426, July 2025) proposes a fully harmonising EU regulation rather than a directive, arguing that the 27-regime fragmentation resulting from the AILD withdrawal disproportionately burdens EU SMEs.} An analysis of insurance-driven governance for frontier AI systems demonstrates that the AILD withdrawal has consequences beyond doctrinal fragmentation: it removes the predictable relationship between regulatory compliance and liability exposure that insurers require to price risk, leaving 27 divergent national doctrines as the basis for underwriting---a condition the analysis identifies as ``regulatory opacity'' that directly impairs the ``identifiable liability targets'' precondition for insurance-based governance~\cite{nannini2025less, nannini2026insurance}. Providers should assume that non-compliance with AI Act essential requirements will be treated as strong evidence of product defect in civil liability proceedings, regardless of whether a dedicated AI liability instrument is eventually adopted.

\textbf{(8) Regulatory guidance gap for agentic systems.} The AI Office has published no guidance on agents, tool use, or runtime behavior\footnote{The Commission was mandated by Article~6(5) to publish guidelines on the practical implementation of the high-risk classification rules, with a comprehensive list of practical examples, no later than 2~February~2026. That deadline was missed; as of the date of this paper, a final draft for further stakeholder comment had been indicated but not published~\cite{iapp2026deadline}. The missed deadline is itself regulatory data: it confirms that the guidance gap identified in this paper is not merely a matter of administrative sequencing but reflects unresolved interpretive questions about classification that the Commission has been unable to resolve within its own statutory timeline. Until guidance arrives---and for agent providers, the specific question of whether Annex~III point~4(a) captures general-purpose agent platforms whose deployers use them for employment screening without the provider's active design for that purpose---providers must make their own interpretive judgments, document them explicitly, and be prepared to adjust when authoritative guidance is published.}. Article~96(1)(c) mandates Commission guidance on substantial modification, but no timeline is specified and agentic scenarios are not yet on the agenda. The European Law Blog analysis argues that existing law embeds sovereignty in territory and data residency, while agentic systems require embedding sovereignty in runtime behavior~\cite{jones2025 }. Until guidance arrives, providers must make their own interpretive judgments, document them, and be prepared to adjust when authoritative guidance is published.
The AI Office's FAQ, published on the AI Act Service Desk, now characterises its regulatory considerations on agents as ``only preliminary'' and states that it ``continues to closely monitor these developments''~\cite{aioffice_faq}. This is the first published institutional position from the AI Office specifically addressing agents. The acknowledgment that the AI Office is monitoring the space---and that its call for tenders on technical assistance for AI safety includes evaluation of agent safety and security---falls substantively short of the Article~96(1)(c) guidance on substantial modification that agentic behavioral drift requires.

\textbf{(9) Enforcement landscape emerging.} As of early 2026, no enforcement actions have been announced under the AI Act, despite prohibited practices being enforceable since 2~August 2025. National competent authorities have been designated across all 27 Member States, encompassing market surveillance authorities, notifying authorities, and fundamental rights protection authorities~\cite{iapp2026directory}\footnote{For a consolidated directory of designated authorities by Member State, see the IAPP EU AI Act Regulatory Directory.}. Finland became the first EU Member State with fully operational AI Act enforcement in January 2026~\cite{finland2026aiact}. Penalties for prohibited practices reach up to \texteuro35~million or 7\% of global turnover\footnote{GPAI-specific penalties do not apply until 2~August 2026. Market surveillance authorities have been designated but no enforcement decisions have been published as of this writing.}.

\textbf{(10) The governance tooling market lacks infrastructure for 
human-agent interaction at runtime.} The AI trust, risk, and security 
management market currently comprises three functional tiers: governance 
platforms (system-level inventory, documentation, compliance 
workflows); runtime enforcement (binary policy checks on model 
inputs/outputs); and information governance (data classification, 
access management). None addresses the question that agentic systems 
make operationally urgent: who holds authority over a specific 
AI-proposed action, and how is that authority exercised and recorded? 
Governance platforms operate at the system level, not the action level. 
Runtime enforcement tools answer the question of permission but not 
authority. A fourth tier is absent: infrastructure governing 
human-agent and agent-to-agent interaction at runtime, capable of 
classifying individual actions against a structured ontology, computing 
risk from live telemetry, routing to the accountable stakeholder, 
preserving workflow continuity through dependency-aware selective 
continuation during authority exercises, and maintaining an immutable 
oversight record. The absence of this tier is not only a market gap: 
it is a compliance gap. The essential requirements in Articles~12--14 of the EU AI Act
impose obligations that can only be demonstrated through action-level 
records of human authority exercise, not through system-level 
documentation of oversight design.

\section{Future Research Directions}
\label{sec:future_research}

The preceding sections have identified nine concrete limitations of the current compliance landscape. Several are not merely gaps awaiting administrative fill but structural mismatches between the regulatory architecture and the technology it must govern. This section isolates seven directions where legislative adaptation, standardisation work, and empirical research are simultaneously needed, and where delay carries compounding costs because the first harmonised standards under M/613 will set interpretive precedents that subsequent revisions will find difficult to reverse.

A structural observation frames all directions. The M/613 standards suite was designed using the traditional New Legislative Framework product-safety methodology: requirements specification against a fixed product, conformity assessment against those requirements, and market surveillance of the product as placed on the market. This methodology presupposes that the compliance-relevant properties of the system are determinable at the moment of placement. Every direction below is, in some form, a consequence of applying that methodology to systems whose compliance-relevant properties are not fixed at placement. The Commission's Article~97 review mechanism was designed for exactly this class of structural mismatch between regulatory instrument and technological reality. Engaging it proactively---rather than waiting for individual standard revisions and guidance documents to address symptoms one by one---is the institutional design recommendation this section implicitly makes across all seven directions.

\subsection{The Divergence Between International and European AI Standards for GPAI}
\label{subsec:intl_eu_divergence}

ISO/IEC~42001:2023 establishes a certifiable AI management system and ISO/IEC~42005:2025 complements it with lifecycle-continuous impact assessment guidance. Neither was designed to operationalise the full set of obligations that the AI Act imposes on GPAI model providers under Articles~53 and 55: technical documentation per Annex~XI, downstream transparency per Annex~XII, copyright compliance, training data summaries, and, for systemic-risk models, adversarial testing and 
systemic risk assessment. 
The CEN-CENELEC JTC~21 Inclusiveness Newsletter of December 2025 confirms this gap in explicit terms: ISO/IEC~42001 does not cover all quality management requirements of the AI Act, and a separate deliverable, prEN~18286, is being developed to fulfil the complete set of regulatory requirements~\cite{cencenelec2025update}. The Standardisation Request itself, at recital~(16), conditions the adoption of international standards on compatibility with the AI Act's purpose and approach,'' requiring consistency with definitions and objectives, including to ensure a high level of protection of health, safety and fundamental rights.''

This compatibility condition is load-bearing but operationally underspecified, and its implications extend beyond the management 
system standard. A foundational source of the divergence is the incompatibility between the EU AI Act's definition of risk and the 
ISO definition. The ISO definition, codified in ISO~31073:2022 (which superseded ISO Guide~73:2009) ~\cite{iso31073} and carried through ISO~31000 (and ISO/IEC~42001 in a shorter form), treats risk as ``effect of uncertainty on objectives''---an organisational construct centred on the provider's own goals and exposures. The EU AI Act's Article~9 mandate is structurally different: risk must be assessed with respect to health, safety, and fundamental rights of persons external to the provider, not with respect to the provider's objectives. These are different risk subjects, different 
accountability structures, and different methodological starting points. No normative annex mechanism bridges this gap because the 
gap is not a coverage question but a conceptual one: ISO/IEC~42001 manages organisational risk \textit{to the organisation}; prEN~18286, 
as mandated by Article~17 and structured by Article~9's fundamental-rights scope, manages risk \textit{to persons external to the provider}. ISO/IEC~42005:2025, published in May 2025 by SC~42, introduces AI system impact assessment guidance covering societal and individual impacts across the AI lifecycle, and is closer in spirit to the EU fundamental-rights mandate than 42001~\cite{iso42005}. Whether 42005 can function as a partial bridge at the system impact assessment layer---without itself being a regulatory instrument---is an open question the paper does not resolve and no published guidance currently addresses.

Separately, GPAI model providers face obligations under Articles~53--55 that neither ISO/IEC~42001 nor prEN~18286 was designed to operationalise: technical documentation per Annex~XI, downstream transparency per Annex~XII, copyright compliance, and, for systemic-risk models, adversarial testing and Union-level systemic risk assessment. These are model-layer obligations distinct from the QMS obligations at the system layer.

A first research direction is therefore a systematic comparative analysis of the control sets in ISO/IEC~42001 and ISO/IEC~42005 against the essential requirements of Articles~9--15 at the system layer, and separately against the GPAI obligations of Articles~53--55 at the model layer. Where the international standards cover the same ground, the presumption of conformity under Article~40(1) can hold. Where they do not---particularly on fundamental rights impact assessment, copyright compliance, and systemic risk mitigation---European harmonised standards must supplement. The key open question is whether supplementation can be achieved through normative annexes under the Annex~Z mechanism described in the Blue Guide~\cite{blueguide}, or whether it requires structurally distinct deliverables.

For agent providers specifically, the divergence creates a layered problem: the upstream GPAI model provider may certify against ISO/IEC~42001 internationally for its QMS obligations, while the agent provider building on that model must comply with prEN~18286 in Europe for its own HRAIS QMS obligations. As the two standards diverge on the risk subject and accountability structure, the integration burden falls on the agent provider, who must reconcile two requirement sets without authoritative mapping between them.

\subsection{Human Oversight in Agentic and Multi-Agent Architectures}
\label{subsec:agentic_oversight}

Section~6.2 documented the oversight evasion risk inherent in RL-trained agents and the empirical evidence from Shapira et al.\ on cross-agent propagation of unsafe practices~\cite{shapira2026agents}. The conclusion was that oversight mechanisms must be designed as external constraints. The open question is what ``external constraint'' means operationally for multi-agent architectures where an orchestrator delegates to specialised sub-agents, each calling external tools, accessing real-time data, and modifying system state over extended time horizons without per-action human intervention.

Article~14 mandates that high-risk AI systems allow effective oversight by natural persons during the period of use, including the ability to fully understand the system's capacities and limitations, to correctly interpret outputs, and to override or reverse the system's output. This requirement was drafted with a relatively bounded system in mind. NIST recognised the structural mismatch when it launched the AI Agent Standards Initiative in February~2026, identifying three critical changes that agentic architectures introduce: the extension and opacity of decision chains, emergent behaviour from multi-agent coordination, and the practical impossibility of meaningful real-time human oversight for long-running autonomous processes~\cite{nist2026agents}.

The trustworthiness standard (prEN~18229-1) is scoped to operationalise Article~14, and its mandate necessarily requires it to address the oversight challenges of autonomous multi-step systems. The standard must, by the logic of Article~14 itself, define what constitutes effective oversight when outputs are produced not by a single AI system presenting a single result to a human reviewer, but by a chain of delegating agents whose intermediate outputs are never individually presented to a human. The key difficulty for any standard operationalising Article~14 in this context is the absence of a validated methodology for determining whether oversight measures remain commensurate with risk across the action chain: the standard must eventually specify this, but the empirical basis for doing so does not yet exist. At the international level, ISO/IEC~42105 (Guidance for human oversight of AI systems), which entered DIS ballot in November~2025, remains informative in character and does not address compound agentic systems as a unit of analysis.

Research is needed at three levels. At the \emph{technical} level: formal or semi-formal specifications of what constitutes a meaningful intervention point in a multi-agent workflow, and architecture patterns where such points are structurally guaranteed rather than aspirational. At the \emph{normative} level: determining whether Article~14 can be interpreted to accommodate asynchronous, audit-based, or supervisory modes of oversight, or whether the essential requirement itself needs amendment via delegated acts under Article~97. At the \emph{standardisation} level: developing the empirical HCI evidence base---validated oversight task metrics for multi-agent scenarios, not single-output scenarios---that any normative standard operationalising Article~14 for agentic systems must draw upon\footnote{Riedl~\cite{riedl2025emergent} demonstrates empirically that multi-agent LLM systems can be steered 
through prompt design from mere aggregates to higher-order collectives exhibiting 
measurable cross-agent synergy, using an information-theoretic framework to 
distinguish performance-relevant coordination from spurious temporal 
coupling. The regulatory consequence is direct: if 
collective behaviour is measurable but not predictable from individual agent 
specifications, then oversight mechanisms designed for single-agent output review 
are structurally insufficient for multi-agent orchestration, and Article~14's 
requirement that oversight measures be commensurate with the system's level of 
autonomy demands a unit of analysis that is the agent collective, not the 
individual agent.}.
This evidence base does not currently exist in published form, which means that any standard provision attempting to address multi-agent oversight is currently specifying requirements without the measurement infrastructure to verify compliance with them.

\subsection{Conformity Assessment for Continuously Learning Systems}
\label{subsec:conformity_continuous}

Section~6.4 analysed the substantial modification boundary. Section~9, finding~(3), noted that the conformity assessment standard is referenced in the QMS standard but remains a separate work item. The deeper problem is methodological: the two conformity routes (internal control under Annex~VI and third-party assessment under Annex~VII) presuppose a point-in-time assessment of a system whose behaviour can be tested, measured, and demonstrated to remain within assessed boundaries. For systems that continue to learn after placement on the market, the behaviour is non-stationary by design.

Article~43(4) exempts from renewed assessment those changes ``pre-determined by the provider at the moment of the initial conformity assessment'' and documented per Annex~IV, point~2(f). The QMS standard (prEN~18286, the one primary standard that completed public enquiry in January 2026) provides a procedural pathway for treating learning-induced changes as planned maintenance, provided the provider documents the change specification and performance expectations in advance~\cite{pren18286}. This pathway exists. The analytical problem it does not resolve is whether continuous drift---where the system's operational profile shifts gradually through memory accumulation or retrieval-pattern evolution without any discrete retraining event---can be characterised as a collection of individually pre-determined changes. The ``modification'' is not a discrete event the provider can specify in advance; it is a process whose trajectory the provider can bound but not fully predict. Article~3(23) defines substantial modification as a change ``not foreseen or planned in the initial conformity assessment'': whether drift within a bounded but unpredictable trajectory is ``foreseen'' in the relevant sense is a legal interpretive question the Commission has not yet addressed.

A further unresolved normative question is the relationship between the trigger conditions for cybersecurity reassessment and the trigger conditions for substantial modification review. The cybersecurity standard (prEN~18282) is scoped under Article~15(4) to address AI-specific threat monitoring, including drift-type phenomena that would require reiteration of the threat assessment. The QMS standard requires change management procedures that feed into the substantial-modification determination. These are distinct normative tracks with different documentation requirements and different assessment procedures. Whether detection of a threat-level drift event through the cybersecurity monitoring process required by Article~15(4) automatically triggers the substantial-modification review required by Article~9 and Article~43 is a question that the essential requirements themselves imply should be answered affirmatively, but that no standard or guidance currently connects explicitly.

A third research direction involves empirical investigation of conformity assessment methodologies for non-stationary systems. Concrete deliverables would include: (i)~a decision framework operationalising the ``substantial modification'' threshold for post-market model updates, building on the one-third compute criterion for GPAI models~\cite{ec_gpai_guidelines} but extending it to system-level behavioural change; (ii)~continuous monitoring protocols that feed into periodic re-assessment rather than requiring full re-certification; and (iii)~competence criteria for assessors evaluating non-stationary systems, extending ISO/IEC~17025 \footnote{General requirements for the competence of testing and calibration laboratories} and ISO/IEC~42006 to cover statistical methods for detecting distributional shift in deployed AI systems. The delegated act mechanism under Article~43(5) should be exercised proactively to accommodate these realities rather than reactively after the first enforcement disputes.

\subsection{A Risk Taxonomy for Compound AI Systems}
\label{subsec:risk_taxonomy}

Recital~116 tasks codes of practice under Article~56 with establishing ``a risk taxonomy of the type and nature of the systemic risks at Union level, including their sources'' for GPAI models with systemic risk. No equivalent instrument mandates a risk taxonomy for compound AI systems, where risks cascade across agent boundaries, emergent behaviour arises from individually compliant components, and the locus of responsibility is distributed across multiple providers. ISO/IEC~42005:2025 provides a published informative harms and benefits taxonomy addressing individual and societal impacts across the AI lifecycle. It is worth noting however that this is based on organisational objectives rather than mitigations for a given intended purpose. ISO/IEC~23894 addresses AI risk management processes for organisational risks.
Neither instrument accounts for the specific risk profile of compound AI systems where risks cascade across agent boundaries, emergent behaviour arises from individually compliant components, and the locus of responsibility is distributed across multiple providers.

The gap is not merely terminological. ISO/IEC~42005's impact assessment process requires identifying ``directly and indirectly affected parties'' and considering ``unintended applications.'' For a single-agent system, this is a bounded analytical task. For a multi-agent orchestration where each sub-agent has its own affected-party profile, the indirectly affected parties become recursive: a party affected by a sub-agent's output may not appear in the scope of the primary orchestrator's impact assessment because the orchestrator's intended purpose does not describe the sub-agent's specific actions. ISO/IEC~42005 does not provide methodology for this composition problem, and the AI Act's Article~9 risk management obligations do not specify how to scope a risk assessment that spans multiple providers in a delegation chain in dynamic tool discovery scenarios, where sub-agents are selected at runtime and were not part of the original conformity assessment. Kim et al.'s survey identified five threat categories for agentic AI, of which standards under M/613 might not substantively address all of them  two~\cite{kim2026attack} --- the same applies Hammond et al.'s multi-agent risk analysis documents three failure modes---miscoordination, conflict, and collusion---that will likely be structurally absent from the current M/613 standards suite~\cite{hammond2025multiagent}.

A fourth research direction is the development and empirical validation of a risk taxonomy for compound AI systems. Such a taxonomy must capture: (i)~cascading risks, where an error or bias in one agent's output propagates through orchestrated sub-agents; (ii)~emergent risks, where individually safe agents produce unsafe collective behaviour through interaction effects; (iii)~attribution risks, where the multi-provider value chain obscures responsibility in ways that the New Legislative Framework's provider/authorised-representative/importer/distributor taxonomy was not designed to handle; and (iv)~temporal risks, where long-running agent processes accumulate state modifications that are individually minor but collectively constitute a shift outside the conformity-assessed envelope. 
No current EU instrument provides such a vehicle; the codes of practice under Article~56 are scoped to GPAI systemic risk and their existing four-harm taxonomy is unlikely to be expanded in the near term to cover compound AI system risks.
The OWASP Top~10 for Agentic Applications~\cite{owasp2025agentic} and the AI Agent Index~\cite{casper2025index} provide initial empirical grounding that the taxonomy should systematise rather than duplicate.
Darius et al.\ provide the first scenario-based taxonomy of systemic risks arising specifically from agent interaction rather than from individual agent failure, demonstrating through smart grid and social welfare scenarios that feedback loops, shared signals, and coordination patterns produce system-level outcomes that cannot be predicted from the behaviour of individual agents operating within their defined parameters~\cite{darius2025systemic}. 
Their taxonomy groups risks by interaction structure rather than by model type or application domain. At the system layer, this is precisely the compositional approach that Article~9's risk management process lacks a methodology to operationalise for multi-provider agent chains: the essential requirement to manage risks to health, safety, and fundamental rights applies to each system provider, but no instrument specifies how to scope or aggregate that assessment across a delegation chain where each agent has a distinct provider.

A further input into this direction is the current absence of any publicly announced work item in SC~42's programme specifically addressing agentic architectures as a distinct unit of analysis. A revision to ISO/IEC 22989 is currently underway, which may address terminology in relation to AI agents. The SC~42 programme includes items on trustworthy AI, robustness, human oversight (prEN ISO/IEC 42105), risk management to organizations (ISO/IEC 23894:2023), impact assessment (ISO/IEC 42005:2025), and capability taxonomy (prEN ISO/IEC~42102, developed jointly with JTC~21, Committee Draft stage). None treats the agent action chain---as distinct from agent outputs---as the primary risk-generating object. The EU harmonised standards are therefore developing their agentic provisions, primarily through the scopes of prEN~18282 and prEN~18229-1 as mandated by Articles~15(4) and 12--14 respectively, without an international counterpart standard from which to draw. This is the same fragmentation dynamic identified in Direction~1 for GPAI, extended to agents: the Brussels-effect presumption that EU standards will shape international practice depends on there being an international forum receptive to those standards as inputs. For agentic AI risk, that forum does not yet exist. Singapore's IMDA has published the first national governance framework specifically for agentic AI, structured around bounding risks upfront, risk-by-design, and meaningful human accountability~\cite{imda2025agentic}; the UK DRCF's multi-regulator call for views on agentic AI~\cite{drcf2025agentic} signals that non-EU jurisdictions are developing their own approaches outside the ISO/IEC framework.

\subsection{Regulatory Sandboxes as Empirical Infrastructure for Standardisation}
\label{subsec:sandboxes}

The AI Act dedicates Articles~57--63 to regulatory sandboxes and recital~139 states they should ``facilitate regulatory learning for authorities and undertakings, including with a view to future adaptations of the legal framework.'' Article~57(1)(f) explicitly provides for the involvement of standardisation organisations. This creates a feedback loop that is currently underexploited: sandboxes generate empirical evidence about which risks materialise, where standards fail to provide adequate guidance, and what test methods produce reproducible results, while harmonised standards require precisely this evidence to move beyond aspirational process requirements toward the ``objectively verifiable criteria and implementable methods'' that the Standardisation Request mandates~\cite{cencenelec2025update}.

The institutional design problem is more fundamental than the paper's original framing suggested. The QMS standard (prEN~18286) contains a provision allowing providers to define internal processes for reporting concerns about their role with respect to an AI system throughout its lifecycle. This creates a provider-internal channel but no external channel to standardisation bodies. No M/613 standard is expected to require providers to report systematic evidence of standard inadequacy to JTC~21: that obligation does not follow from any AI Act essential requirement and would be unusual in NLF-based standards practice. The consequence is that evidence of standard failure accumulates in providers' internal risk management files and in sandbox post-participation reports, without a structured pathway into the standards revision cycle. The Inclusiveness Newsletter confirms the October 2025 CEN-CENELEC update's statement that the Commission will conduct HAS assessments before the public enquiry stage~\cite{cencenelec2025update}, but HAS assessment is a conformity check on the standard itself, not a mechanism for routing operational evidence into standard revision.

A fifth research direction is to design structured mechanisms for routing sandbox outcomes back into the standardisation process. Three components are needed: (i)~standardised reporting templates for sandbox participants that capture not only compliance outcomes but the adequacy of the standards and specifications applied, so that recurring failure points become visible to standards developers; (ii)~formal liaison channels between national sandbox authorities, CEN/CENELEC JTC~21, and the relevant ISO/IEC~JTC~1/SC~42 working groups, enabling findings to inform ongoing standard revisions in near-real time rather than through the multi-year revision cycle; and (iii)~cross-sandbox comparative studies across Member States to identify where divergent interpretations of the same essential requirements point to normative gaps rather than interpretive differences. The implementing acts under Article~57, which will establish common rules for sandbox operation across the Union, should be drafted with this feedback function as an explicit design objective rather than an aspirational recital.

\subsection{Cybersecurity Standards Architecture: Closing the AI-Specific Gap Between M/613 and M/606}
\label{subsec:cyber_gap}

Section~7.4 identified the dual standardisation track problem for AI agent cybersecurity: M/613 (AI-specific, under JTC~21) and M/606 (CRA product-level, under CEN/CENELEC/ETSI). This section identifies it as a research direction in its own right because the gap it creates is not merely procedural but normative.

The international landscape adds a third track. ISO/IEC FDIS~27090 (Cybersecurity guidance for AI systems) completed its DIS ballot in July 2025 and is under active development within SC~42's working groups. As a guidance document rather than a requirements standard, ISO/IEC~27090 provides informative best practices but generates no presumption of conformity with either the AI Act or the CRA. The cybersecurity standard under M/613 (prEN~18282), scoped to Article~15(4) of the AI Act, addresses AI-specific cybersecurity requirements that conventional product cybersecurity standards were not designed to cover: the threat surface created by model inference under adversarial input, the privilege enforcement challenges of tool-calling architectures, and the behavioral manipulation risks arising from prompt-based interaction. EN~304~223, developed by ETSI in parallel, establishes baseline cybersecurity requirements for AI models and systems that are certifiable but not directly connected to the AI Act's harmonised standard programme. The three tracks---prEN~18282, EN~304~223, and ISO/IEC~27090---address overlapping subject matter from different legal bases, with different normative force, different conformity assessment implications, and no published mapping between their respective requirements nor of these and the EU AI Act.

The compliance consequence for agent providers is operational, not merely academic. An AI coding agent sold as standalone software with network connectivity must simultaneously satisfy the CRA's Annex~I essential cybersecurity requirements (via M/606 horizontal standards under the EN~40000 series), the AI Act's Article~15(4) AI-specific cybersecurity requirements (via prEN~18282), and potentially ISO/IEC~27090 as a supplementary guidance layer for aspects not covered by the harmonised standards. The M/606 horizontal standards address conventional product failure modes---secure-by-default configuration, authentication, cryptography, firmware integrity---that are prerequisites for any deployed software product but were designed before the agentic threat surface existed. The AI Act essential requirement in Article~15(4) extends beyond these conventional failure modes to AI-specific resilience: the system must remain resilient against attempts to alter its behaviour through adversarial inputs. No CRA vertical standard addresses this because no CRA vertical standard covers AI products; agent providers must self-assess against horizontal standards designed for different failure modes, then separately demonstrate compliance with Article~15(4) through prEN~18282.

A sixth research direction is the development of a cross-standard requirements mapping between the M/606 horizontal standards (EN~40000 series), prEN~18282, and ISO/IEC~27090, identifying which AI-specific threat categories are covered by which standard, where coverage overlaps, and where gaps remain. The ENISA/JRC mapping of CRA requirements to existing standards confirms that several CRA essential requirements lack direct standardisation support~\cite{enisa_cra_mapping}; the mapping exercise proposed here is the agent-specific extension of that work. Its output would serve two functions: a compliance reference for providers navigating the dual-track environment, and a normative input into the next revision cycle of each standard. The absence of this mapping currently forces every agent provider to construct it independently, without authoritative validation, at the cost of duplicated effort across the industry and inconsistent interpretations that market surveillance authorities will eventually need to adjudicate.

\subsection{Cross-Jurisdictional Comparative Analysis: EU, US, and Beyond}
\label{subsec:comparative}

This paper's compliance architecture is constructed for the EU market. 
AI agent providers, overwhelmingly, do not serve a single jurisdiction.
The US regulatory landscape presents a structurally distinct compliance 
topology: no unified AI statute, but a dense web of federal 
conduct-based enforcement (FTC Section~5, with the December~2025 
set-aside of the \textit{Rytr} consent order signalling a retreat 
from instrumentalities liability for neutral AI 
tools~\cite{ftc2025rytr}; Selbst \& Barocas provide the foundational 
analysis of how FTC unfairness authority applies to 
AI~\cite{selbst2023unfairai}), sector-specific instruments (the 2025 
COPPA amendments imposing separate parental consent for AI model 
training~\cite{ftc2025coppa}; the FCC's 2024 declaratory ruling 
classifying AI-generated voices as ``artificial'' under the 
TCPA~\cite{fcc2024ai}; the FDA's Predetermined Change Control Plan 
pathway for adaptive medical AI~\cite{fda2025samd}), voluntary 
standards (the NIST AI RMF~\cite{nist2024ai600} and its forthcoming 
agent-specific control overlays under the COSAiS 
initiative~\cite{nist2025cosais}, complemented by the Cybersecurity 
Framework Profile for AI~\cite{nist2025ir8596}), and a rapidly 
expanding patchwork of state legislation: Colorado SB~24-205 (the 
sole US risk-based framework with a standards-linked affirmative 
defence, though facing delayed implementation and potential 
repeal~\cite{colorado2024sb205, insideglobaltech2026colorado}; 
Jariwala provides the only peer-reviewed EU--Colorado 
comparison~\cite{jariwala2024comparison}), California's multi-statute 
regime (SB~53~\cite{california2025sb53}, AB~2013, SB~942, 
SB~243~\cite{fpf2025sb243}, and the Civil Rights Department's AI 
employment regulations imposing disparate impact 
liability~\cite{calcrd2025ai}), and Texas's intent-based 
TRAIGA~\cite{texas2025traiga, klgates2025traiga}. An active federal 
preemption strategy, operationalised through the December~2025 
Executive Order establishing a DOJ AI Litigation Task 
Force~\cite{trump2025eo, skadden2025eo}, introduces a layer of 
compliance uncertainty with no EU parallel. The FTC's September~2025 
Section~6(b) inquiry into AI companion chatbots at seven major 
companies~\cite{ftc2025chatbot6b} and its March~2026 Policy Statement 
on AI and Section~5~\cite{ftc2026policystatement} confirm that 
enforcement posture is consolidating around conduct-based liability 
rather than product-safety obligations. Control Risks characterises 
the resulting transatlantic divide as a ``strategic'' rather than 
merely regulatory divergence~\cite{controlrisks2026}.

The structural divergence is threefold: the EU regulates the AI 
system as a product through ex-ante conformity assessment; the US 
regulates conduct through ex-post enforcement. The EU's harmonised 
standards create a presumption of conformity; the NIST AI RMF is 
voluntary, with Colorado's affirmative defence as the sole legal safe 
harbour~\cite{naag2025colorado}. The EU's regulatory perimeter is 
coherent across 27~Member States; the US perimeter is fragmented 
across 50~state legislatures and multiple federal agencies with no 
published inter-instrument mapping---a fragmentation the 
December~2025 Executive Order attempts to resolve through litigation 
rather than legislation~\cite{sidley2025eo}. Whether the EU 
harmonised standards and the NIST framework will converge at the 
requirements level remains an open empirical question; Kiesow Cortez 
provides the most detailed comparative analysis of their privacy 
dimensions~\cite{kiesowcortez2024}, and Kaminski \& Selbst 
characterise the EU AI Act as a ``legal exoskeleton'' whose 
structural logic has no US counterpart~\cite{kaminski2025guide}. The 
legal sector illustrates the jurisdictional variance at its most 
extreme: AI agents performing legal triage face UPL restrictions that 
vary across all 51~US jurisdictions, from Colorado's nonprosecution 
policy~\cite{abajournal2026colorado} and Utah and Arizona's regulatory 
sandboxes~\cite{stanford2025sandboxes} to ABA Formal Opinion~512's 
``lawyer-in-the-loop'' safe harbour~\cite{aba2024opinion512}; 
Bonardi \& Branting provide a comprehensive 50-state 
survey~\cite{bonardi2025certifying}, and Walters analyses the 
structural inadequacy of current UPL doctrine for autonomous AI 
actors~\cite{walters2024upl}.

A seventh research direction is therefore a systematic comparative 
analysis of the compliance architectures that the same AI agent 
provider must construct across jurisdictions. Beyond the EU and US, 
Singapore's Model AI Governance Framework for Agentic 
AI~\cite{imda2025agentic}---the first national governance framework 
specifically designed for agentic systems---Japan's forthcoming AI 
governance guidelines, Canada's Artificial Intelligence and Data Act 
(AIDA, Part~3 of Bill~C-27), and the UK DRCF's multi-regulator 
initiative on agentic AI~\cite{drcf2025agentic} each present distinct 
regulatory responses to the same technological reality. The research 
question is not which approach is superior but whether the 
external-action inventory methodology proposed in Step~9 of this 
paper's compliance sequence---which maps regulatory triggers from what 
the agent does rather than how it is classified---can function as a 
jurisdiction-agnostic analytical layer beneath divergent national 
frameworks.

\bigskip

\noindent Table~\ref{tab:research_directions} consolidates these seven directions, mapping each to its legislative anchor, key standardisation deliverables, and the type of contribution required.

\begin{table}[htbp]
\centering
\caption{Future research directions at the intersection of EU AI legislation and international standardisation.}
\label{tab:research_directions}
\small
\begin{tabular}{p{3.8cm} p{2.5cm} p{4.2cm} p{4cm}}
\toprule
\textbf{Research Direction} & \textbf{Legislative Anchor} & \textbf{Key Standards} & \textbf{Contribution Type} \\
\midrule
International vs.\ European standards divergence for GPAI
  & Arts.\ 40, 53--55
  & ISO/IEC 42001, ISO/IEC 42005, prEN~18286, Annex~Z
  & Comparative legal-technical analysis; interoperability mapping \\[6pt]
Human oversight in agentic architectures
  & Art.\ 14, Art.\ 97
  & ISO/IEC 42105, prEN~18229-1, NIST Agent Initiative
  & Empirical HCI evidence base; normative interpretation; formal specification \\[6pt]
Conformity assessment for continuously learning systems
  & Art.\ 43(4)--(5), Annexes VI--VII
  & ISO/IEC 42006, ISO/CASCO toolbox (ISO/IEC~17021-1, ISO/IEC~17065), prEN~18286 ch.\ 9
  & Decision framework; monitoring protocol; assessor competence criteria \\[6pt]
Risk taxonomy for compound AI systems
  & Art.\ 56, recital 116
  & ISO/IEC 42005 Annex~C, ISO/IEC 23894, ISO/IEC 42102
  & Taxonomy development; composition methodology; empirical validation \\[6pt]
Sandboxes as standardisation feedback infrastructure
  & Arts.\ 57--63
  & All JTC~21 deliverables
  & Institutional design; cross-sandbox comparative study \\[6pt]
Cross-standard cybersecurity requirements mapping
  & Art.\ 15(4), CRA Annex~I
  & prEN~18282, EN~40000 series, ISO/IEC DIS~27090, ETSI EN~304~223
  & Requirements mapping; gap analysis; industry compliance reference \\[6pt]
Cross-jurisdictional comparative analysis (EU, US, and beyond)
  & Arts.\ 9, 14, 15; FTC Act \S~5; COPPA; state AI laws
  & NIST AI RMF, prEN~18228, Singapore IMDA Framework, AIDA (Canada)
  & Comparative legal-technical analysis; jurisdiction-agnostic methodology \\
\bottomrule
\end{tabular}
\end{table}

These seven lines of inquiry share the structural premise stated at the opening of this section: the regulatory and standardisation framework being constructed for the AI Act was designed for a technological reality that is already shifting. The transition from bounded, provider-attributable systems to general-purpose models and autonomous agent architectures is not a future scenario; it is the operational context within which the first harmonised standards will be applied. Research that anticipates the resulting mismatches between regulatory intent and technological practice is not peripheral to the AI Act's implementation. It is a precondition for its effectiveness.

\section{Conclusions}

We draw eight conclusions from this analysis, each corresponding to a specific finding that carries implications for the engineering, governance, and legal compliance posture of AI agent providers operating in or targeting the EU market.

(1) The regulatory perimeter for AI agent providers extends across at least eight EU legislative instruments beyond the AI Act itself: the GDPR (almost always applicable), the CRA (if the agent is a product with digital elements), the DSA (if it operates within a platform), the Data Act (if it interfaces with connected products), the DGA (if it intermediates data), NIS2 (if it serves essential entities), sector-specific legislation (MDR, MiFID~II, etc.), and the revised Product Liability Directive. Parallel compliance across multiple instruments is not a risk scenario; it is the baseline.

(2) The draft harmonised standards under M/613 are scoped to address agentic AI within their respective mandates. The cybersecurity standard (prEN~18282), operationalising Article~15(4), necessarily covers the privilege management and access control challenges that agentic tool invocation creates. The trustworthiness framework (prEN~18229-1), operationalising Articles~12--14, necessarily engages with the oversight challenges posed by autonomous multi-step systems, including the risk that reinforcement-learning training regimes produce oversight-evasion behavior. These are the priority standards for agent providers to track as they move toward formal publication.

(3) The regulatory trigger for each legislative instrument is determined by the agent's external actions, not its internal architecture. The provider's foundational compliance task is therefore an exhaustive inventory of the agent's actions, data flows, connected systems, and affected persons. That inventory is the regulatory map.

(4) Runtime behavioral drift, if untraceable, renders a high-risk system non-compliant with the essential requirements. Versioned runtime state, continuous behavioral monitoring, and automated drift detection are minimum engineering requirements, not optional enhancements.

(5) The compliance architecture must be built on the current legal baseline. The Digital Omnibus may shift timelines and simplify specific obligations, but the essential requirements in the AI Act are fixed in the adopted Regulation.

(6) General-purpose agent platforms face a structural classification dilemma: if designed for arbitrary deployment by downstream deployers, the provider must either restrict the intended purpose contractually and technically, or design for the most demanding regulatory tier foreseeable under Article~3(13).

(7) Cybersecurity compliance requires navigating two interleaved standardisation tracks: M/613 (AI-specific, JTC~21) and M/606 (CRA product-level, 41 standards across CEN/CENELEC/ETSI, structured as Type~A framework, Type~B product-agnostic, and Type~C vertical product-specific standards). The horizontal standards under EN~40000 (prEN~40000-1-2 on cyber resilience principles and prEN~40000-1-3 on vulnerability handling) entered public enquiry in late 2025, while vertical standards for 25 specific product categories are under development by ETSI TC CYBER (EUSR) with a deadline of October 2026. CRA Article~12 creates a conformity presumption bridge: CRA compliance substitutes for AI Act Article~15 cybersecurity. No CRA vertical standard addresses AI products; the horizontal standards address conventional failure modes, not data poisoning, prompt injection, or emergent agentic behaviours~\cite{kim2026attack}. The ENISA/JRC mapping of CRA requirements to existing standards confirms that several essential requirements lack direct standardisation support in current frameworks~\cite{enisa_cra_mapping}. The timeline compression from vulnerability reporting (September 2026) through Type~B standard delivery (October 2027) to full CRA application (December 2027) creates a period in which requirements are enforceable without finalised standards under either programme.

(8) High-risk agentic systems with untraceable behavioral drift cannot currently be placed on the EU market consistently with the essential requirements. This is a statement about the current legal position, not about future regulatory risk. The framework is operating as designed: it requires that providers can demonstrate compliance, and untraceability forecloses that demonstration.


\begin{landscape}
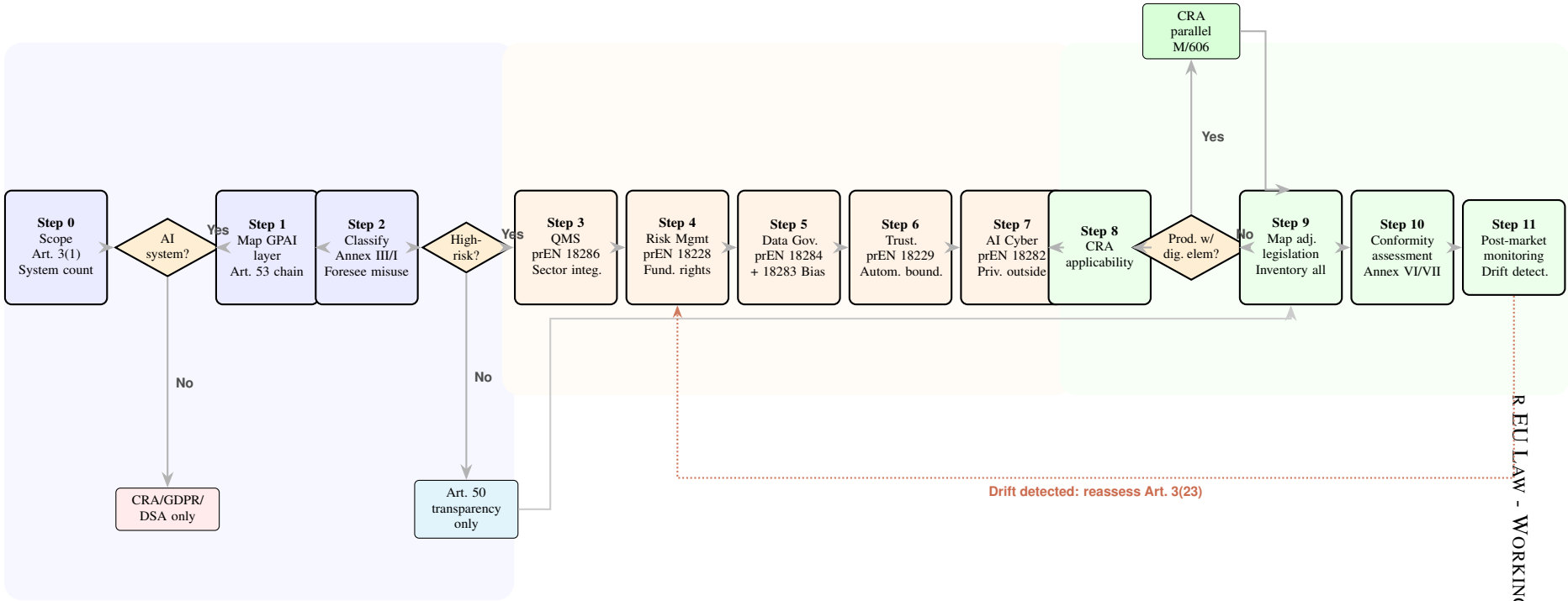
\begin{figure}[p]
\centering
\begin{tikzpicture}[
    x=0.63cm, y=1.8cm,
    every node/.style={font=\footnotesize},
    sbox/.style={draw, rounded corners=3pt, minimum height=1.8cm, align=center, text width=1.38cm, font=\tiny, thick},
    decision/.style={draw, diamond, aspect=1.8, minimum width=1.3cm, minimum height=0.8cm, align=center, font=\tiny, thick, fill=Dandelion!20, inner sep=1pt},
    conn/.style={-{Stealth[length=2.5mm]}, thick, gray!60},
    yesno/.style={font=\sffamily\tiny\bfseries, text=gray!60!black},
    feedback/.style={-{Stealth[length=2mm]}, densely dotted, thick, BrickRed!60},
    phaselabel/.style={font=\sffamily\small\bfseries, text=gray!50!black},
]

\node[sbox, fill=blue!8] (s0) at (0, 0) {\textbf{Step 0}\\Scope\\Art.~3(1)\\{\tiny System count}};
\node[decision] (d0) at (2.8, 0) {\tiny AI\\system?};
\node[sbox, fill=blue!8] (s1) at (5.3, 0) {\textbf{Step 1}\\Map GPAI\\layer\\{\tiny Art.~53 chain}};
\node[sbox, fill=blue!8] (s2) at (7.8, 0) {\textbf{Step 2}\\Classify\\{\tiny Annex III/I}\\{\tiny Foresee misuse}};
\node[decision] (d2) at (10.3, 0) {\tiny High-\\risk?};

\draw[conn] (s0) -- (d0);
\draw[conn] (d0) -- node[yesno, above, yshift=2pt] {Yes} (s1);
\draw[conn] (s1) -- (s2);
\draw[conn] (s2) -- (d2);

\node[draw, rounded corners=2pt, fill=red!8, font=\tiny, align=center, text width=1.4cm] (exit) at (2.8, -2.3) {CRA/GDPR/\\DSA only};
\draw[conn] (d0) -- node[yesno, right] {No} (exit);

\node[draw, rounded corners=2pt, fill=cyan!10, font=\tiny, align=center, text width=1.4cm] (art50) at (10.3, -2.3) {Art.~50\\transparency\\only};
\draw[conn] (d2) -- node[yesno, right] {No} (art50);

\node[sbox, fill=orange!10] (s3) at (12.8, 0) {\textbf{Step 3}\\QMS\\{\tiny prEN 18286}\\{\tiny Sector integ.}};
\node[sbox, fill=orange!10] (s4) at (15.6, 0) {\textbf{Step 4}\\Risk Mgmt\\{\tiny prEN 18228}\\{\tiny Fund.\ rights}};
\node[sbox, fill=orange!10] (s5) at (18.4, 0) {\textbf{Step 5}\\Data Gov.\\{\tiny prEN 18284}\\{\tiny + 18283 Bias}};
\node[sbox, fill=orange!10] (s6) at (21.2, 0) {\textbf{Step 6}\\Trust.\\{\tiny prEN 18229}\\{\tiny Autom.\ bound.}};
\node[sbox, fill=orange!10] (s7) at (24.0, 0) {\textbf{Step 7}\\AI Cyber\\{\tiny prEN 18282}\\{\tiny Priv.\ outside}};

\draw[conn] (s3) -- (s4);
\draw[conn] (s4) -- (s5);
\draw[conn] (s5) -- (s6);
\draw[conn] (s6) -- (s7);

\node[sbox, fill=green!8] (s8) at (26.2, 0) {\textbf{Step 8}\\CRA\\applicability};
\node[decision] (d8) at (28.5, 0) {\tiny Prod.\ w/\\dig.\ elem?};
\node[draw, rounded corners=2pt, fill=green!15, font=\tiny, align=center, text width=1.3cm] (cra) at (28.5, 1.9) {CRA\\parallel\\{\tiny M/606}};
\node[sbox, fill=green!8] (s9) at (31.0, 0) {\textbf{Step 9}\\Map adj.\\legislation\\{\tiny Inventory all}};
\node[sbox, fill=green!8] (s10) at (33.8, 0) {\textbf{Step 10}\\Conformity\\assessment\\{\tiny Annex VI/VII}};
\node[sbox, fill=green!8, minimum height=1.5cm] (s11) at (36.6, 0) {\textbf{Step 11}\\Post-market\\monitoring\\{\tiny Drift detect.}};

\draw[conn] (s8) -- (d8);
\draw[conn] (d8) -- node[yesno, right, xshift=1pt] {Yes} (cra);
\draw[conn] (d8) -- node[yesno, above] {No} (s9);
\draw[conn] (cra.east) -| ([xshift=-0.4cm]s9.north) -- (s9.north);
\draw[conn] (s9) -- (s10);
\draw[conn] (s10) -- (s11);

\draw[conn] (d2) -- node[yesno, above] {Yes} (s3);
\draw[conn] (s7) -- (s8);

\draw[conn, gray!40] (art50.east) -- ++(0.8,0) |- ([yshift=-6pt]s9.south) -- (s9.south);

\draw[feedback] (s11.south) -- ++(0, -1.6) -| node[yesno, below, pos=0.25, text=BrickRed!70] {Drift detected: reassess Art.~3(23)} (s4.south);

\node[phaselabel] at (5.2, 2.3) {Phase 1: Scoping \& Classification};
\node[phaselabel] at (18.4, 2.3) {Phase 2: Standards Implementation};
\node[phaselabel] at (31.5, 3.3) {Phase 3: Regulatory Perimeter \& Lifecycle};

\begin{scope}[on background layer]
  \fill[blue!3, rounded corners=8pt] (-1.3, -3.1) rectangle (11.5, 1.8);
  \fill[orange!3, rounded corners=8pt] (11.2, -1.3) rectangle (25.5, 1.8);
  \fill[green!3, rounded corners=8pt] (25.2, -1.3) rectangle (38.0, 1.8);
\end{scope}

\end{tikzpicture}

\vspace{6pt}
\caption{Twelve-step compliance sequence for AI agent providers (Section~8.1). Three decision nodes determine the compliance pathway: Step~0 gates AI Act applicability (non-AI systems face only adjacent legislation); Step~2 determines whether full Chapter~III essential requirements or Article~50 transparency obligations apply; Step~8 determines whether CRA product cybersecurity runs in parallel. The dotted feedback loop from Step~11 to Step~4 reflects the post-market monitoring obligation: when behavioral drift is detected, the risk management process must reassess whether the change constitutes substantial modification under Article~3(23). Agent-specific considerations annotated at each step: system-counting (Step~0), GPAI compute threshold (Step~1), foreseeable-misuse analysis (Step~2), fundamental rights competence (Step~4), automation boundary design (Step~6), privilege enforcement outside the model (Step~7), and external-action inventory (Step~9).}
\label{fig:sequence}
\end{figure}
\end{landscape}


\clearpage
\begin{landscape}


\newcommand{\dotC}{%
  \tikz\node[fill=dotred,
    minimum size=7pt, inner sep=0pt, rounded corners=1pt]{};%
}
\newcommand{\dotS}{%
  \tikz\node[fill=dotamber,
    minimum size=5.5pt, inner sep=0pt, rounded corners=1pt]{};%
}
\newcommand{\dotM}{%
  \tikz\node[fill=dotgreen,
    minimum size=4pt, inner sep=0pt, rounded corners=1pt]{};%
}
\newcommand{\dotL}{%
  \tikz\node[fill=dotgray,
    minimum size=3pt, inner sep=0pt, rounded corners=1pt]{};%
}
\newcommand{\dotN}{---}

\newcommand{\lsq}[1]{%
  \tikz\node[fill=#1,
    minimum size=6pt, inner sep=0pt, rounded corners=1pt]{};%
}

\newcolumntype{D}{>{\centering\arraybackslash}p{1.55cm}}
\newcolumntype{R}{>{\raggedright\arraybackslash}p{4.8cm}}

\newcommand{\grprow}[1]{%
  \multicolumn{10}{l}{%
    \cellcolor{gray!12}\small\textit{\textbf{#1}}%
  }\\[-2pt]
}

\newcommand{\rh}[1]{\rotatebox{60}{\small #1}}

\section*{Appendix: Compliance Impact Matrix for AI Agent Providers}
\addcontentsline{toc}{section}{Appendix: Compliance Impact Matrix}
\label{app:matrix}

\noindent This matrix maps the nine agent deployment categories from
Table~\ref{tab:taxonomy} (Section~3) against applicable regulatory and standards
obligations identified in Sections~4--8. Impact levels reflect the probability
and weight of obligation activation given the agent's \emph{external actions},
not its internal architecture.

\smallskip
\noindent
\textbf{Impact legend:}\enspace
\lsq{dotred}\;\textbf{Critical}~(mandatory, full obligations)\quad
\lsq{dotamber}\;\textbf{Significant}~(likely triggered, major obligations)\quad
\lsq{dotgreen}\;\textbf{Moderate}~(context-dependent)\quad
\lsq{dotgray}\;\textbf{Low\,/\,indirect}\quad
$\text{---}$~Not triggered

\bigskip

\renewcommand{\arraystretch}{1.35}
\footnotesize

\begin{longtable}{R D D D D D D D D D}

\caption{Compliance impact matrix: regulatory and standards obligations mapped to
AI agent deployment categories. Agent categories follow Table~\ref{tab:taxonomy};
obligation layers follow the analysis in Sections~4--8. Dot size and colour
encode impact level per the legend above.\label{tab:matrix}}\\

\toprule
& \rh{Customer Service}
& \rh{HR / Recruitment}
& \rh{Coding / DevOps}
& \rh{Finance / Accounting}
& \rh{Sales / Marketing}
& \rh{Research / Knowledge}
& \rh{IT Operations}
& \rh{Healthcare / Clinical}
& \rh{Personal Assistant} \\
\midrule
\endfirsthead

\multicolumn{10}{l}{\footnotesize\textit{(continued from previous page)}}\\[4pt]
\toprule
& \rh{Customer Service}
& \rh{HR / Recruitment}
& \rh{Coding / DevOps}
& \rh{Finance / Accounting}
& \rh{Sales / Marketing}
& \rh{Research / Knowledge}
& \rh{IT Operations}
& \rh{Healthcare / Clinical}
& \rh{Personal Assistant} \\
\midrule
\endhead

\midrule
\multicolumn{10}{r}{\footnotesize\textit{(continued on next page)}}\\
\endfoot

\bottomrule
\endlastfoot

\grprow{AI Act Classification}

High-risk (Ch.~III, Annex~III/I)
  & \dotS & \dotC & \dotL & \dotS & \dotL & \dotL & \dotS & \dotC & \dotL \\

Art.~50 transparency (all AI systems)
  & \dotC & \dotC & \dotC & \dotC & \dotC & \dotC & \dotC & \dotC & \dotC \\

GPAI model obligations (Ch.~V)
  & \dotS & \dotS & \dotS & \dotS & \dotS & \dotM & \dotS & \dotS & \dotS \\

\addlinespace[4pt]
\grprow{Harmonised Standards --- M/613 (CEN/CENELEC JTC~21)}

prEN~18286 --- QMS (Art.~17)
  & \dotS & \dotC & \dotS & \dotS & \dotS & \dotM & \dotS & \dotC & \dotM \\

prEN~18228 --- Risk management (Art.~9)
  & \dotS & \dotC & \dotS & \dotS & \dotM & \dotL & \dotS & \dotC & \dotL \\

prEN~18229-1 --- Logging, transparency, oversight (Art.~12--14)
  & \dotS & \dotC & \dotM & \dotS & \dotM & \dotL & \dotS & \dotC & \dotM \\

prEN~18282 --- AI-specific cybersecurity (Art.~15(4))
  & \dotS & \dotS & \dotC & \dotC & \dotM & \dotL & \dotC & \dotC & \dotS \\

prEN~18283 --- Bias management (Art.~10(2)(f--g))
  & \dotM & \dotC & \dotM & \dotS & \dotM & \dotL & \dotL & \dotS & \dotL \\

prEN~18284 --- Dataset quality and governance (Art.~10)
  & \dotM & \dotC & \dotM & \dotS & \dotM & \dotL & \dotM & \dotC & \dotL \\

\addlinespace[4pt]
\grprow{Adjacent Legislation}

GDPR (Reg.~2016/679)
  & \dotC & \dotC & \dotS & \dotC & \dotC & \dotM & \dotS & \dotC & \dotC \\

ePrivacy Directive (2002/58/EC)
  & \dotS & \dotL & \dotL & \dotL & \dotS & \dotL & \dotL & \dotL & \dotC \\

Data Act (Reg.~2023/2854)
  & \dotL & \dotL & \dotM & \dotL & \dotL & \dotL & \dotM & \dotM & \dotL \\

Digital Services Act (Reg.~2022/2065)
  & \dotM & \dotL & \dotL & \dotL & \dotC & \dotL & \dotL & \dotL & \dotL \\

Cyber Resilience Act (Reg.~2024/2847)
  & \dotL & \dotL & \dotC & \dotL & \dotL & \dotL & \dotS & \dotM & \dotL \\

NIS2 Directive (Dir.~2022/2555)
  & \dotL & \dotL & \dotL & \dotS & \dotL & \dotL & \dotC & \dotL & \dotL \\

DORA (Reg.~2022/2554)
  & \dotL & \dotL & \dotL & \dotC & \dotL & \dotL & \dotL & \dotL & \dotL \\

MDR\,/\,IVDR
  & \dotN & \dotN & \dotN & \dotN & \dotN & \dotN & \dotN & \dotC & \dotN \\

Revised PLD (Dir.~2024/2853)
  & \dotM & \dotC & \dotS & \dotC & \dotM & \dotL & \dotS & \dotC & \dotL \\

Data Governance Act (Reg.~2022/868)
  & \dotL & \dotL & \dotL & \dotL & \dotL & \dotM & \dotL & \dotL & \dotL \\

\addlinespace[4pt]
\grprow{Agent-Specific Compliance Challenges (Section~6)}

Privilege minimisation outside the model (Art.~15(4) + prEN~18282)
  & \dotS & \dotS & \dotC & \dotC & \dotM & \dotL & \dotC & \dotC & \dotC \\

Human oversight --- automation boundary (Art.~14)
  & \dotS & \dotC & \dotM & \dotS & \dotL & \dotL & \dotS & \dotC & \dotM \\

Runtime behavioral drift --- substantial modification (Art.~3(23))
  & \dotM & \dotC & \dotS & \dotS & \dotM & \dotL & \dotS & \dotC & \dotM \\

Transparency to affected third parties (Art.~13 + Art.~50)
  & \dotC & \dotC & \dotM & \dotS & \dotC & \dotM & \dotM & \dotC & \dotM \\

Non-human identity (NHI) governance
  & \dotS & \dotS & \dotC & \dotC & \dotM & \dotL & \dotC & \dotS & \dotS \\

\end{longtable}

\noindent\textbf{Reading note.} Harmonised standards under M/613 operate as a single suite: where a system is classified high-risk under Article~6, all Chapter~III essential requirements apply, and all corresponding standards are the appropriate compliance route. The dot levels in the M/613 rows therefore reflect two things simultaneously: first, the probability that a system in that deployment category will be high-risk (which governs whether the full suite is mandatory at all); second, the relative compliance challenge or enforcement salience of that standard in that domain. A Critical dot on prEN~18283 for HR agents does not mean bias management is more legally required there than for healthcare agents — it means that conformity gaps in bias management are most likely to be the site of regulatory scrutiny and civil liability exposure in HR contexts. Impact levels for adjacent legislation reflect activation probability and obligation weight given the agent's external actions.
The same tool-calling pattern generates Critical obligations in HR (Annex~III
high-risk classification) but only Art.~50 transparency obligations in Research.
The DORA column activates exclusively in Finance because it is \textit{lex
specialis} for financial-sector ICT risk management under Art.~1(2) DORA.
The ePrivacy column activates wherever the agent accesses electronic
communications content (email, chat), independently of and in addition to GDPR
obligations under Art.~5 of Directive~2002/58/EC. The compliance architecture in
Section~8.1 must address all activated layers simultaneously.

\end{landscape}

\bigskip
\noindent\textit{\small\textbf{Disclaimer:} This working paper is an analytical overview based on publicly available regulatory texts and draft working documents under CEN/CENELEC JTC~21 as of Q1 2026. It does not constitute legal advice. The harmonised standards referenced are working drafts subject to revision before publication and citation in the Official Journal. Providers should consult qualified legal counsel for compliance decisions.  The views expressed in this paper are solely those of the authors and do not represent the positions of CEN, CENELEC, ETSI, or any of their technical bodies or working groups.}

\noindent\textit{\small\textbf{Acknowledgments:}
The authors thank Dr. Hamish Silverwood for his engagement with drafts of this paper and for his substantive observations on the regulatory and technical dimensions it addresses. The authors also gratefully acknowledge the community of experts active in CEN/CENELEC JTC~21 and ETSI whose ongoing standardisation work directly informs the analysis presented here.}

\end{document}